\documentclass[transmag]{IEEEtran}

\usepackage{latexsym}
\usepackage{amsfonts,amssymb,amsmath}
\usepackage{hyperref}
\usepackage{cite}
\usepackage{graphicx}
\usepackage[linesnumbered,ruled,lined]{algorithm2e}
\usepackage{algpseudocode}
\usepackage{multirow} 
\usepackage{longtable}
\usepackage{threeparttable}
\usepackage{array}
\usepackage{cases}
\usepackage{color}
\usepackage[T1]{fontenc}
\usepackage[utf8]{inputenc}
\usepackage{authblk}
\usepackage{xfrac}
\usepackage{caption}
\usepackage{subcaption}

\newtheorem{prop}{Proposition}

\newtheorem{corollary}{Corollary}

\newtheorem{remark}{Remark}

\newcolumntype{C}[1]{>{\centering\arraybackslash}m{#1}}
\usepackage{tabularx}

\def\BibTeX{{\rm B\kern-.05em{\sc i\kern-.025em b}\kern-.08em T\kern-.1667em\lower.7ex\hbox{E}\kern-.125emX}}

\begin{document}

\title{Is NOMA Efficient in Multi-Antenna Networks? \\  A Critical Look at Next Generation Multiple Access Techniques}

\author{Bruno Clerckx, \textit{Senior Member, IEEE}, Yijie Mao, \textit{Member, IEEE},\\  Robert Schober, \textit{Fellow, IEEE}, Eduard Jorswieck, \textit{Fellow, IEEE}, David J. Love, \textit{Fellow, IEEE},\\ Jinhong Yuan, \textit{Fellow, IEEE}, Lajos Hanzo, \textit{Fellow, IEEE}, Geoffrey Ye Li, \textit{Fellow, IEEE},\\ Erik G. Larsson, \textit{Fellow, IEEE}, and Giuseppe Caire, \textit{Fellow, IEEE}
\thanks{This work has been partially supported by the EPSRC of the UK under grant EP/N015312/1, EP/R511547/1.}
\thanks{Bruno Clerckx, Yijie Mao and Geoffrey Ye Li are with the Department of Electrical and Electronic Engineering, Imperial College London, London SW7 2AZ, UK (email: \{b.clerckx,y.mao16,geoffrey.li\}@imperial.ac.uk). 

Robert Schober is with the Institute for Digital Communications, Friedrich-Alexander University Erlangen-Nürnberg (FAU), Erlangen, Germany (email: robert.schober@fau.de). 

Eduard Jorswieck is with the Institute for Communications Technology, Technische Universität Braunschweig, 38106 Brunswick, Germany (e-mail: e.jorswieck@tu-bs.de).

David J. Love is with the School of Electrical and Computer Engineering,
Purdue University, West Lafayette, IN, USA (email: djlove@purdue.edu).

Jinhong Yuan is with the School of Electrical Engineering and
Telecommunications, University of New South Wales, Sydney, NSW 2052,
Australia (e-mail: j.yuan@unsw.edu.au).

Lajos Hanzo is with the School of Electronics and Computer Science, University of Southampton, Southampton SO17 1BJ, U.K. (e-mail:
lh@ecs.soton.ac.uk).

Erik G. Larsson is with the Department of Electrical Engineering,
Linköping University, SE-581 83 Linköping, Sweden (e-mail:
erik.g.larsson@liu.se).

Giuseppe Caire is with the Communications and Information Theory Group, Faculty of Electrical Engineering and Computer Science, Technische Universität Berlin, 10587 Berlin, Germany (e-mail: caire@tu-berlin.de).}}

\IEEEtitleabstractindextext{\begin{abstract}In the past few years, a large body of literature has been created on downlink Non-Orthogonal Multiple Access (NOMA), employing superposition coding and Successive Interference Cancellation (SIC), in multi-antenna wireless networks. Furthermore, the benefits of NOMA over Orthogonal Multiple Access (OMA) have been highlighted. In this paper, we take a critical and fresh look at this downlink multiple access literature. Instead of contrasting NOMA with OMA, we contrast NOMA with two other baselines. The first is conventional Multi-User Linear Precoding (MU--LP), as used in Space-Division Multiple Access (SDMA) and multi-user Multiple-Input Multiple-Output (MIMO) in 4G and 5G. The second is Rate-Splitting Multiple Access (RSMA) based on multi-antenna Rate-Splitting (RS), which is also a non-orthogonal transmission strategy relying on SIC developed in the past few years in parallel and independently from  NOMA. We show that there is some confusion about the benefits of NOMA, and we dispel the associated \textit{misconceptions}. 
\textit{First}, we highlight why NOMA is inefficient in multi-antenna settings (compared to the baselines) based on basic multiplexing gain analysis. We stress that the issue lies in how the NOMA literature, originally developed for single-antenna setups, has been hastily applied to multi-antenna setups, resulting in a misuse of spatial dimensions and therefore loss in multiplexing gains and rate. 
\textit{Second}, we show that NOMA incurs a severe multiplexing gain loss despite an increased receiver complexity due to an inefficient use of SIC receivers.
\textit{Third}, we emphasize that much of the merits of NOMA are due to the constant comparison to OMA instead of comparing it to MU--LP and RS baselines. We then expose the pivotal \textit{design constraint} that multi-antenna NOMA requires one user to fully decode the messages of the other users. This design constraint is responsible for the multiplexing gain erosion, rate loss, and inefficient use of SIC receivers in multi-antenna settings. Our analysis and simulations confirm that NOMA should not be applied blindly to multi-antennas settings, highlight the scenarios where MU--LP outperforms NOMA and vice versa, and demonstrate the inefficiency, performance loss and complexity disadvantages of NOMA compared to RS. The first takeaway message is that, while NOMA is suited for single-antenna settings (as originally intended), it is not beneficial in most multi-antenna deployments. The second takeaway message is that other non-orthogonal transmission frameworks, such as RS, exist which fully exploit the multiplexing gain and the benefits of SIC to boost the rate in multi-antenna settings.\end{abstract}

}

\maketitle

\section{Introduction}
\IEEEPARstart{I}{n} contrast to Orthogonal Multiple Access (OMA) that
assigns users to orthogonal dimensions (e.g., Time-Division Multiple Access - TDMA, Frequency-Division Multiple Access - FDMA), (power-domain) Non-Orthogonal Multiple Access (NOMA)\footnote{Although there is a broad range of NOMA schemes in the power and code domains, in this treatise, we focus only on power-domain NOMA and simply use NOMA to represent power-domain NOMA.} superposes users in the same time-frequency resource and distinguishes them in the power domain \cite{Saito:2013,Dai:2015,Ding:2017,Liu:2017,Shin:2017b}. By doing so, NOMA has been promoted as a solution for 5G and beyond to deal with the vast throughput, access, and quality of service (QoS) requirements that are projected to grow exponentially for the foreseeable future.

\par In the downlink, NOMA refers to communication schemes where at least one user is forced to fully decode the message(s) of other co-scheduled user(s). This operation is commonly performed through the use of transmit-side superposition coding (SC) and receiver-side Successive Interference Cancellation (SIC) in downlink multi-user communications. Such techniques have been studied for years before being branded with the NOMA terminology. NOMA has indeed been known in the information theory and wireless communications literature for several decades, under the terminology of superposition coding with successive interference cancellation (denoted in short as SC–SIC), as the strategy that achieves (and has been used in achievability proofs for) the capacity region of the Single-Input Single-Output (SISO) (Gaussian) Broadcast Channel (BC) \cite{Cover:1972}. The superiority of NOMA over OMA was shown in the seminal paper by Cover in 1972. It is indeed well known that the capacity region of the SISO BC (achieved by NOMA) is larger than the rate region achieved by OMA (i.e. contains the achievable rate region of OMA as a subset) \cite{Cover:1972,Tse:2005,Goldsmith:2005}. The use of SIC receivers is a major difference between NOMA and OMA, although it should be mentioned that SIC has also been studied for a long time in the 3G and 4G research phases in the context of interference cancellation and receiver designs \cite{Li:2010}.

\par In today's wireless networks, access points commonly employ more than one antenna, which opens the door to multi-antenna processing. The key building block of the downlink of multi-antenna networks is the multi-antenna (Gaussian) BC. Contrary to the SISO BC that is degraded and where users can be ordered based on their channel strengths, the multi-antenna BC is nondegraded and users cannot be ordered based on their channel strengths \cite{Weingarten:2006,Tse:2005}. This is the reason why SC–SIC/NOMA is not capacity-achieving in this case, and Dirty Paper Coding (DPC) is the only known strategy that achieves the capacity region of the multi-antenna (Gaussian) BC with perfect Channel State Information at the Transmitter (CSIT) \cite{Weingarten:2006}. Due to the high computational burden of DPC, linear precoding is often considered the most attractive alternative to simplify the transmitter design \cite{Spencer:2004,Stojnic:2006,Dabbagh:2007,Dabbagh:2008,Clerckx:2013}. Interestingly, in a multi-antenna BC, Multi-User Linear Precoding (MU–LP) relying on treating the residual multi-user interference as noise, although suboptimal, is often very useful since the interference can be significantly reduced by spatial precoding. This is the reason why it has received significant attention in the past twenty years and it is the basic principle behind numerous 4G and 5G techniques such as Space-Division Multiple Access (SDMA) and multi-user (potentially massive) Multiple-Input Multiple-Output (MIMO) \cite{Clerckx:2013}.  
\par In view of the benefits of NOMA over OMA and multi-antenna over single-antenna, numerous attempts have been made in recent years to combine multi-antenna and NOMA schemes \cite{Saito:2013,Dai:2015,Ding:2017,Liu:2017,Shin:2017b,Liu:2018,Hanif:2016,Choi:2015,Sun:2015,Zhang:2016,Zeng:2017,Ding:2016a,Chen:2016a, Chen:2016b,Ding:2016,Choi:2017,Shin:2017,Nguyen:2017,Zeng:2017b,Cheng:2017,Zhu:2018,Chen:2017,Liu:2016,Alavi:2018,Jeong:2019,Zhang:2020,Chu:2020,Yalcin:2019,Liu:2020} (and references therein). Although there are a few contributions considering the comparison of NOMA with MU–LP schemes such as Zero-Forcing Beamforming (ZFBF) or DPC \cite{Chen:2016a, Chen:2016b, Nguyen:2017, Dai:2018}, much emphasis is put in the NOMA literature on comparing (single/multi-antenna) NOMA and OMA, and showing that NOMA outperforms OMA. But there is a lack of emphasis in the NOMA literature on contrasting multi-antenna NOMA to other multi-user multi-antenna baselines developed for the multi-antenna BC such as MU--LP (or other forms of multi-user MIMO techniques) and other forms of (power-domain) non-orthogonal transmission strategies such as Rate-Splitting Multiple Access (RSMA) based on multi-antenna Rate-Splitting (RS) \cite{Clerckx:2016}. 
RS designed for the multi-antenna BC also relies on SIC and has been developed in parallel and independently from NOMA \cite{Clerckx:2016,Yang:2013,Joudeh:2016a,Joudeh:2016b,Hao:2015,Dai:2016,Mao:2017}. Such a comparison is essential to assess the benefits and the efficiency of NOMA, since all these communication strategies can be viewed as different achievable schemes for the multi-antenna BC and all aim in their own way for the same objective, namely meet the throughput, reliability, QoS, and connectivity requirements of beyond-5G multi-antenna wireless networks.

\par In this paper, we take a critical look at multi-antenna NOMA for the downlink of communication systems and ask ourselves the important question ``\textit{Is multi-antenna NOMA an efficient strategy?}'' To answer this question, we go beyond the conventional NOMA vs. OMA comparison, and contrast multi-antenna NOMA with MU--LP and RS-based non-orthogonal transmission strategies. This allows us to  highlight some misconceptions and shortcomings of multi-antenna NOMA. Explicitly, we show that in most scenarios the short answer to that question is no, and demonstrate based on first principles and numerical performance evaluations why this is the case. Our discussions and results unveil the scenarios where MU--LP outperforms NOMA and vice versa, and demonstrate that multi-antenna NOMA is inefficient compared to RS. 

\par By contrasting multi-antenna NOMA to MU--LP and RS, we show that there is some confusion about multi-antenna NOMA and its merits, expose major misconceptions and reveal new insights. The contributions of this paper are summarized as follows.

\par \textit{First}, we analytically derive both the sum multiplexing gain as well as the max-min fair multiplexing gain of multi-antenna NOMA and compare them to those of MU--LP and RS. The scenarios considered are very general and include multi-antenna transmitter with single-antenna receivers, perfect and imperfect CSIT, in underloaded and overloaded regimes. On the one hand, multi-antenna NOMA can achieve gains, but can also incur losses compared to MU--LP. On the other hand, multi-antenna NOMA \textit{always} leads to a waste of multiplexing gain compared to RS. The multiplexing gain analysis provides a firm theoretical ground to infer that multi-antenna NOMA is not as efficient as RS in exploiting the spatial dimensions and the available CSIT. This analysis is instrumental to identify the scenarios where the multiplexing gain gaps among NOMA, MU--LP, and RS are the smallest/largest, therefore highlighting deployments that are suitable/unsuitable for the different multiple access strategies.   
\par \textit{Second}, we show that multi-antenna NOMA leads to a high receiver complexity due to the inefficient use of SIC. For instance, we show that the higher the number of SIC operations (and therefore the higher the receiver complexity) in multi-antenna NOMA, the lower the sum multiplexing gain (and therefore the lower the sum-rate at high SNR). Comparison with MU--LP and RS show that higher multiplexing gains can be achieved at a lower receiver complexity and a reduced number of SIC operations. 

\setlength\extrarowheight{3pt}
\begin{table*}[t!]
\centering
\caption{\label{tab:overview} Overview of the paper.}
\begin{tabular}{|l l|}
\hline
\multicolumn{2}{|c|}{\textbf{Section \ref{twouser}. Two-User  MISO  NOMA  with  Perfect  CSIT: The  Basic  Building  Block}} \\ 
\ref{twouserModel}. System Model                              & \ref{DoF_def}. Definition of Multiplexing Gain                                                 \\ 
\ref{discussion_twouser}. Discussions                               &                                                                                    \\ \hline \hline
\multicolumn{2}{|c|}{\textbf{Section \ref{kuser}. $K$-User  MISO  NOMA  with  Perfect  CSIT}}                                                                       \\ 
\ref{sec: MISO NOMA system}. MISO NOMA System Model                   & \ref{multiplexing_gain_kuser}. Multiplexing Gains                                                                                 \\ \hline \hline
\multicolumn{2}{|c|}{\textbf{Section \ref{Imperfect_CSIT_section}. $K$-User  MISO  NOMA  with  Imperfect  CSIT}}                                                                      \\ 
\ref{CSIT_Error_model_section}. CSIT Error Model                          & \ref{DoFimperfect}. Multiplexing Gains                                                                                  \\ \hline \hline
\multicolumn{2}{|c|}{\textbf{Section \ref{MULP_section}. Baseline  Scheme  I: Conventional  Multi-user  Linear  Precoding}}                                                  \\ 
\ref{sec: MULP system model}. MU–LP System Model                         &  \ref{DoF_MULP_perfect}. Multiplexing Gains with Perfect CSIT                       \\
\ref{DoF_MULP_imperfect}. Multiplexing Gains with Imperfect CSIT     &                                                                                                           \\ \hline \hline
\multicolumn{2}{|c|}{\textbf{Section \ref{RS_section}. Baseline  Scheme  II:  Rate-Splitting}}                                                                            \\ 
\ref{sec: RS}. Rate-Splitting System Model               & \ref{sec: multiplex gain}. Multiplexing Gains with Perfect CSIT                                                                \\ 
\ref{sec: multiplex gain imperfect CSIT}. Multiplexing Gains with Imperfect CSIT    &                                                                                                           \\ \hline \hline
\multicolumn{2}{|c|}{\textbf{Section \ref{sec: misconceptions NOMA}. Shortcomings   and   Misconceptions   of   Multi-Antenna  NOMA}}                                                  \\ 
\ref{NOMA_vs_MULP}. NOMA vs. Baseline I (MU–LP)              & \ref{NOMA_vs_RS}. NOMA vs. Baseline II (RS)                                                                          \\ 
\ref{misconceptions}. Misconceptions of Multi-Antenna NOMA     & \ref{example_sumdof}. Illustration  of  the  Misconceptions  with an Example   \\ 
\ref{shortcomings}. Shortcomings of Multi-Antenna NOMA       &                                                                                                           \\ \hline \hline
\multicolumn{2}{|c|}{\textbf{Section \ref{evaluations}.   Numerical  Results}}                                                                                           \\ 
\ref{evaluations_perfect}. Perfect CSIT                            & \ref{evaluations_imperfect}. Imperfect CSIT                                                                                    \\ 
\ref{evaluation_discussion}. Discussions                             &                                                                                                           \\ \hline \hline
\multicolumn{2}{|c|}{\textbf{Section \ref{conclusions}. Conclusions  and  Future  Works}}                                                                                  \\ \hline
\end{tabular}
\end{table*}

\par \textit{Third}, we show that most of the misconceptions behind NOMA are due to the prevalent comparison to OMA instead of comparing to MU--LP and RS. We show and explain that the misconceptions, the multiplexing gain reduction, and the inefficient use of SIC receivers in both underloaded and overloaded multi-antenna settings reyling on both perfect and imperfect CSIT originate from a limitation of the multi-antenna NOMA design philosophy, namely that one user is forced to fully decode the messages of the other users. Hence, while forcing a user to fully decode the messages of the  other users is an efficient approach in single-antenna degraded BC, it may not be an efficient approach in multi-antenna networks.  
\par \textit{Fourth}, we stress that an efficient design of non-orthogonal transmission and multiple access strategies ensures that the use of SIC  never leads to a performance loss but rather leads to a performance gain over MU--LP. We show that such non-orthogonal solutions based on RS exist and truly benefit from the multi-antenna multiplexing gain and from the use of SIC receivers in both underloaded and overloaded regimes relying on perfect and imperfect CSIT. In fact, multi-antenna RS completely resolves the design limitations of multi-antenna NOMA.
\par \textit{Fifth}, we depart from the multiplexing gain analysis and design the transmit precoders to maximize the sum-rate and max-min rate for multi-antenna NOMA, followed by numerically comparing the sum-rate and the max-min fair rate of NOMA to those of MU--LP and RS.
We show that the multiplexing gain analysis is accurate and instrumental to predict the rate performance of the multiple access strategies considered.

\par \textit{Sixth}, our numerical simulations confirm the inefficiency of multi-antenna NOMA in general settings. Multi-antenna NOMA is shown to lead to performance gains over MU--LP in some settings but also to losses in other settings despite the use of SIC receivers and a higher receiver complexity. Our results also highlight the significant benefits, performance-wise and receiver complexity-wise, of RSMA and multi-antenna RS over multi-antenna NOMA. It is indeed possible to achieve a significantly better performance than MU--LP and NOMA with just one layer of SIC by adopting RS so as to partially decode messages of other users (instead of fully decoding them as in NOMA).

\emph{Organization:} The remainder of this paper is organized as follows. Section \ref{twouser} introduces two-user Multiple-Input Single-Output (MISO) NOMA (with single-antenna receivers) as a basic building block (and toy example) for our subsequent studies, compares to MU--LP, and raises some questions about the efficiency of NOMA. Section \ref{kuser} studies the multiplexing gain of $K$-user MISO NOMA with perfect CSIT. Section \ref{Imperfect_CSIT_section}  extends the discussion to imperfect CSIT. Section \ref{MULP_section} and Section \ref{RS_section} study the multiplexing gains of the baseline schemes considered, namely MU--LP and RS, respectively. Section \ref{sec: misconceptions NOMA} compares the multiplexing gains of all multiple access schemes considered and exposes the misconceptions and shortcomings of multi-antenna NOMA. Section \ref{evaluations} provides simulation results. Section \ref{conclusions} concludes this paper. An overview of the paper is illustrated in Table \ref{tab:overview}.

\emph{Notation:} $|\cdot|$ refers to the absolute value of a scalar or to the cardinality of a set depending on the context. $\left\|\cdot\right\|$ refers to the $l_2$-norm of a vector. $\max\{a_1,...,a_n\}$ refers to the maximum value between $a_1$ to $a_n$. $\mathbf{a}^{H}$ denotes the Hermitian transpose of vector $\mathbf{a}$. $\textrm{Tr}(\mathbf{Q})$ refers to the trace of matrix $\mathbf{Q}$. $\mathbf{I}$ is the identity matrix. $P\nearrow$ means as $P$ grows large. $\mathcal{CN}(0,\sigma^2)$ denotes the circularly symmetric complex Gaussian distribution with zero mean and variance $\sigma^2$. $\sim$ stands for ``distributed as''. $O(\cdot)$ refers to the big O notation. $\mathbb{E}\big\{\cdot\big\}$ denotes statistical expectation. $A\cap B$ and $A\cup B$ refer to the intersection ($A$ and $B$ have to be satisfied) and the union ($A$ or $B$ to be satisfied) of two sets/events $A$ and $B$, respectively.

\section{Two-User MISO NOMA with Perfect CSIT: \\ The Basic Building Block}\label{twouser}
We commence by studying two-user MISO NOMA and show that, by comparing NOMA to MU--LP instead of to OMA, the potential merits of NOMA are less obvious. Limited to two single-antenna users with perfect CSIT, this system model illustrates the simplest though fundamental building block of multi-antenna NOMA.

\subsection{System Model}
\label{twouserModel}
\par We consider a downlink single-cell multi-user multi-antenna scenario with $K=2$ users, also known as two-user MISO BC, consisting of one transmitter with $M\geq 2$ antennas communicating with two single-antenna users. The transmitter aims to transmit simultaneously two messages $W_1$ and $W_2$ intended for user-1 and user-2, respectively. 

\begin{figure}
	\centering
	\includegraphics[width=0.8\columnwidth]{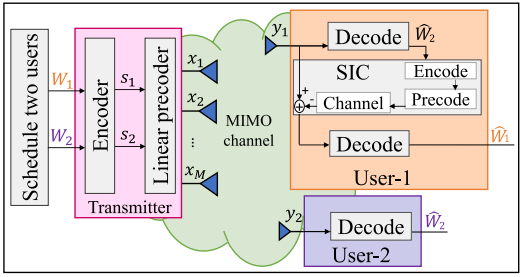}
	\caption{Two-user system architecture with NOMA (decoding order: user-2$\rightarrow$user-1). }
	\label{fig_sys_noma}
\end{figure}
\par The transmitter adopts the so-called multi-antenna NOMA or MISO NOMA strategy, illustrated in Fig. \ref{fig_sys_noma}, that encodes one of the two messages using a codebook shared by both users\footnote{This is not an issue in modern systems since, for example, in an LTE/5G NR system, all codebooks are shared since all users use the same family of modulation and coding schemes (MCS) specified in the standard.} so that it can be decoded and cancelled from the received signal at the co-scheduled user (following the same principle as superposition coding for the degraded BC). Consider $W_2$ is encoded into $s_2$ using the shared codebook and $W_1$ is encoded into $s_1$. The two streams are then linearly precoded by $M \times 1$ precoders\footnote{The precoders $\mathbf{p}_1$ and $\mathbf{p}_2$ can be any vectors that satisfy the power constraint, though the best choice of precoders would depend on the objective function.} $\mathbf{p}_1$ and $\mathbf{p}_2$ and superposed at the transmitter so that the transmit signal is given by 
\begin{equation}\label{NOMA_2user_system_model}
\mathbf{x}=\mathbf{p}_1 s_1+\mathbf{p}_2 s_2.
\end{equation}
Defining $\mathbf{s}=[s_1,s_2]^T$ and assuming that $\mathbb{E}[\mathbf{s}\mathbf{s}^H]=\mathbf{I}$, the average transmit (sum) power constraint is written as $P_{1}+P_{2}\leq P$ where $P_{k}=\left\|\mathbf{p}_{k}\right\|^2$ with $k=1,2$. 

\par The channel vector for user $k$ is denoted by $\mathbf{h}_k$, and the received signal at user $k$ can be written as $y_k=\mathbf{h}_k^H \mathbf{x}+n_k$, $k=1,2$, where $n_k\sim\mathcal{CN}(0,1)$ is Additive White Gaussian Noise (AWGN). We assume perfect CSIT and perfect Channel State Information at the Receivers (CSIR).

\par At both users, stream $s_2$ is decoded first into\footnote{Though not expressed explicitly, $\widehat{W}_{2}$ is receiver dependent since both receivers decode $s_2$ and the same estimate is not necessarily obtained at both receivers. Hence, more rigorously, we could have written $\widehat{W}_{2,k}$, $k=1,2$ to refer to the estimate at user-k. For simplicity of exposure, we have nevertheless opted to drop the index $k$.} $\widehat{W}_{2}$ by treating the interference from $s_1$ as noise. Using SIC at user-1, $\widehat{W}_{2}$ is re-encoded, precoded, and subtracted from the received signal, such that user-$1$ can decode its stream $s_1$ into $\widehat{W}_{1}$. Assuming proper Gaussian signalling and perfect SIC\footnote{Note there is no error in the SIC operation since the rates are achievable under Gaussian signalling and infinite block length.}, the achievable rates of the two streams with MISO NOMA are given by\footnote{Superscript (N) stands for NOMA. Similarly we will user (M) for MU--LP, (R) for Rate-Splitting, and $\star$ for the information theoretic optimum.} 
\begin{align}
R_{1}^{(\textnormal{N})}&=\log_2\left(1+\left|\mathbf{h}_1^H \mathbf{p}_{1}\right|^2\right),\label{R1_NOMA}\\
R_{2}^{(\textnormal{N})}&=\min\left(\log_2\left(1+A\right),\log_2\left(1+B\right)\right),\label{R2_NOMA} 
\end{align}
where 
\begin{equation}
\label{eq: B}
A=\frac{\left|\mathbf{h}_1^H \mathbf{p}_{2}\right|^2}{1+\left|\mathbf{h}_1^H \mathbf{p}_{1}\right|^2}, \hspace{1cm} B=\frac{\left|\mathbf{h}_2^H \mathbf{p}_{2}\right|^2}{1+\left|\mathbf{h}_2^H \mathbf{p}_{1}\right|^2}.
\end{equation}
In \eqref{R2_NOMA}, $\log_2\left(1+A\right)$ is the rate supportable by the channel of user-1 when user-1 decodes $s_2$ and treats its own stream $s_1$ as noise. Similarly, $\log_2\left(1+B\right)$ is the rate supportable by the channel of user-2 when user-2 decodes its own stream $s_2$ while treating stream $s_1$ of user-1 as noise.
The $\min$ in \eqref{R2_NOMA} is due to the fact that $s_2$, though carrying message $W_2$ intended to user-2, is decoded by both users and is therefore transmitted at a rate decodable by both users.
\par The most common performance metric of a multi-user system is the sum-rate. In this two-user MISO NOMA system model, the sum-rate is defined as $R_{\textnormal{s}}^{(\textnormal{N})}=R_{1}^{(\textnormal{N})}+R_{2}^{(\textnormal{N})}$ and can be upper bounded\footnote{This is an upper bound since when $A < B$, it is achieved with equality, and when $B < A$, $\log_2(1+B)<\log_2(1+A)$ and it is a strict upper bound.} as
\begin{align}
R_{\textnormal{s}}^{(\textnormal{N})}&\leq\log_2\left(1+\frac{\left|\mathbf{h}_1^H \mathbf{p}_{2}\right|^2}{1+\left|\mathbf{h}_1^H \mathbf{p}_{1}\right|^2}\right)+\log_2\left(1+\left|\mathbf{h}_1^H \mathbf{p}_{1}\right|^2\right),\nonumber\\
&=\log_2\left(1+\left|\mathbf{h}_1^H \mathbf{p}_{2}\right|^2+\left|\mathbf{h}_1^H \mathbf{p}_{1}\right|^2\right).\label{eq_MAC}
\end{align}
\par It is important to note that \eqref{eq_MAC} can be interpreted as the sum-rate of a two-user multiple access channel (MAC) with a single-antenna receiver. Indeed, user-1 acts as the receiver of a two-user MAC whose effective SISO channels for both links are given by $\mathbf{h}_1^H \mathbf{p}_{2}$ and $\mathbf{h}_1^H \mathbf{p}_{1}$, respectively. This observation will be revisited in the next few sections, and will be shown very helpful to explain the performance of multi-antenna NOMA.

\par A drawback of the sum-rate is that it does not capture the concept of rate fairness among the users. Another popular system performance metric is the Max-Min Fair (MMF) rate or symmetric rate defined as $R_{\textnormal{mmf}}^{(\textnormal{N})}=\min_{k=1,2}R_{k}^{(\textnormal{N})}$. MMF metric provides uniformly good quality of service since it aims for maximizing the minimum rate among all users. 
\par Throughout the manuscript, we will focus on the sum-rate and the MMF rate as two very different metrics to assess the system performance. We choose these two metrics as they are  commonly used in wireless networks, and in the NOMA literature in particular (see e.g., \cite{Hanif:2016,Zeng:2017,Zeng:2017b,Zhu:2018,Chen:2017} for the sum-rate and \cite{Liu:2016,Alavi:2018,Yalcin:2019,Timotheou:2015,Choi:2016} for the MMF rate). They are representative for two very different operational regimes, with the former focusing on high system throughput and the latter on user fairness.  

\par In the sequel, we introduce some useful definitions and then make some observations based on this two-user system model.

\subsection{Definition of Multiplexing Gain}\label{DoF_def}
\par Throughout the manuscript, we will often refer to the multiplexing gain to quantify how well a communication strategy can exploit the available spatial dimensions. We define the multiplexing gain, also referred to as Degrees-of-Freedom (DoF), of user-$k$ achieved with communication strategy\footnote{Throughout this paper, $j$ will be either N for NOMA, M for MU--LP, R for Rate-Splitting, or $\star$ for the information theoretic optimum, i.e., $j\in\{\textrm{N, M, R, }\star\}$.} $j$ as 
\begin{equation}
d_k^{(j)}=\lim_{P\rightarrow \infty}\frac{R_k^{(j)}(P)}{\log_2(P)},
\end{equation}
and the sum multiplexing gain as
\begin{equation}
d_{\textnormal{s}}^{(j)}=\lim_{P\rightarrow \infty}\frac{R_{\textnormal{s}}^{(j)}(P)}{\log_2(P)}=\sum_{k=1}^K d_k^{(j)},
\end{equation}
where $R_{\textnormal{s}}^{(j)}=\sum_{k=1}^K R_k^{(j)}$ is the sum-rate. We also define the MMF multiplexing gain as
\begin{equation}
d_{\textnormal{mmf}}^{(j)}=\lim_{P\rightarrow \infty}\frac{R_{\textnormal{mmf}}^{(j)}(P)}{\log_2(P)}=\min_{k=1,...,K} d_k^{(j)},
\end{equation}
where $R_{\textnormal{mmf}}^{(j)}=\min_{k=1,...,K} R_k^{(j)}$ is the MMF rate.
\par The multiplexing gain $d_k^{(j)}$ is a first-order approximation of the rate of user-$k$ at high Signal-to-Noise Ratio (SNR). $d_k^{(j)}$ can be viewed as the pre-log factor of the rate of user-$k$ at high SNR and be interpreted as the number or fraction of interference-free stream(s) that can be simultaneously communicated to user-$k$ by employing communication strategy $j$. The larger $d_k^{(j)}$, the faster the rate of user-$k$ increases with the SNR. Hence,  ideally a communication strategy should achieve the highest  multiplexing gain possible.
\par The sum multiplexing gain $d_{\textnormal{s}}^{(j)}$ is a first-order approximation of the sum-rate at high SNR and therefore the pre-log factor of the sum-rate and can be interpreted as the total number of interference-free data streams that can be simultaneously communicated to all $K$ users by employing communication strategy $j$. In other words, $R_{\textnormal{s}}^{(j)}$ scales as $d_{\textnormal{s}}^{(j)} \log_2(P)+\delta$ where $\delta$ is a term that scales slowly with SNR such that $\lim_{P\rightarrow \infty}\frac{\delta}{\log_2(P)}=0$ (e.g. $O(1)$, $O(\log_2(\log_2(P)))$ or $O(\sqrt{\log_2(P)})$), and the larger $d_{\textnormal{s}}^{(j)}$, the faster the sum-rate increases with the SNR. 
\par The MMF multiplexing gain $d_{\textnormal{mmf}}^{(j)}$, also referred to as symmetric multiplexing gain, corresponds to the maximum multiplexing gain that can be simultaneously achieved by all users, and reflects the pre-log factor of the MMF rate at high SNR. In other words, $R_{\textnormal{mmf}}^{(j)}$ scales as $d_{\textnormal{mmf}}^{(j)} \log_2(P)+\delta$, and the larger $d_{\textnormal{mmf}}^{(j)}$, the faster the MMF rate increases with the SNR. 

\begin{remark} Much of the analysis and discussion in this paper emphasizes the (sum and MMF) multiplexing gain as a metric to assess the capability of a strategy to exploit multiple antennas. As it becomes plausible from its definition, the multiplexing gain is an asymptotic metric valid in the limit of high SNR, and hence, does not precisely reflect  specific finite-SNR rates. Nevertheless, it provides firm theoretical grounds for performance comparisons and has been used in the MIMO literature for two decades \cite{Zheng:2003}. Furthermore, the multiplexing gain also impacts the performance at finite SNRs as shown in numerous papers \cite{Joudeh:2017,Joudeh:2016a,Joudeh:2016b} and in our simulation results in Section \ref{evaluations}. Moreover, it enables to gain deep insights into the performance limits and to guide the design of efficient communications strategies, as we will see throughout this paper.
\end{remark}

\subsection{Discussions}\label{discussion_twouser}
Equations \eqref{R1_NOMA} and \eqref{R2_NOMA}, respectively, suggest that $s_1$ is received interference-free at user-1, and that $s_2$ is always decoded in the presence of interference from $s_1$. This has as a consequence that MISO NOMA limits the sum multiplexing gain to 
$d_{\textnormal{s}}^{(\textnormal{N})}=d_1^{(\textnormal{N})}+d_2^{(\textnormal{N})}=1$, i.e., the same as OMA. Indeed, the sum-rate bound \eqref{eq_MAC} achieved by this two-user MISO NOMA strategy and user ordering user-2$\rightarrow$user-1 can be further upper bounded as
\begin{align}
R_{\textnormal{s}}^{(\textnormal{N})}\leq\log_2\left(1+\left\|\mathbf{h}_1\right\|^2 P\right),\label{eq_OMA_int_1}
\end{align}
where the equality in \eqref{eq_OMA_int} is achieved (i.e. upper bound is tight) by choosing $\mathbf{p}_{1}=\sqrt{P_1}\mathbf{h}_1/\left\|\mathbf{h}_1\right\|$ and $\mathbf{p}_{2}=\sqrt{P_2}\mathbf{h}_1/\left\|\mathbf{h}_1\right\|$ with $P_1+P_2= P$.
\par Had we considered the other decoding order where the shared codebook is used to encode $W_1$ and user-2 decodes $s_1$, the role of user-1 and user-2 in Fig. \ref{fig_sys_noma} would have been switched (user-1$\rightarrow$user-2) and we would have obtained 
\begin{align}
R_{\textnormal{s}}^{(\textnormal{N})}\leq\log_2\left(1+\left\|\mathbf{h}_2\right\|^2 P\right). \label{eq_OMA_int_2}
\end{align}
Hence, the sum-rate of MISO NOMA considering adaptive decoding order is upper bounded as
\begin{align}
R_{\textnormal{s}}^{(\textnormal{N})}&\leq\log_2\left(1+\max\{\left\|\mathbf{h}_{1}\right\|^2,\left\|\mathbf{h}_{2}\right\|^2\} P\right)\label{eq_OMA_int}
\end{align}
and the sum-rate is increased by letting the strong user $\arg \max_{k=1,2}\left\|\mathbf{h}_{k}\right\|$ decode the weak user $\arg \min_{k=1,2}\left\|\mathbf{h}_{k}\right\|$.

\par Considering the high SNR regime, \eqref{eq_OMA_int_1}, \eqref{eq_OMA_int_2}, \eqref{eq_OMA_int} all scale at most as $\log_2(P)$, i.e. 
\begin{align}
R_{\textnormal{s}}^{(\textnormal{N})}\stackrel{P\nearrow}{=} \log_2\left(P\right) + \delta,\label{eq_OMA}
\end{align}
which highlights that the sum multiplexing gain of two-user MISO NOMA (irrespectively of the decoding order) is (at most) one, i.e. $d_{\textnormal{s}}^{(\textnormal{N})}=1$. Moreover, \eqref{eq_OMA_int} reveals the stronger result that the sum-rate of MISO NOMA is actually no higher than that of OMA for any SNR! This fact is not surprising in the SISO case ($M=1$) since it is well known that to achieve the sum capacity of the SISO BC, one can simply transmit to the strongest user all the time (i.e., OMA) \cite{Knopp:1995}. The above result shows that this also holds for the two-user MISO NOMA basic building block.  
\par The sum multiplexing gain of one can be further split equally amongst the two users, which leads to an MMF multiplexing gain of the two-user MISO NOMA given by $d_{\textnormal{mmf}}^{(\textnormal{N})}=\frac{1}{2}$. This is achieved by scaling the power allocated to user-1  as $O(P^{1/2})$ and that to user-2 as $O(P)$. In other words, the MMF rate of this two-user MISO NOMA scales at most as $\frac{1}{2}\log_2(P)$ at high SNR. 
\begin{figure}
\centering
\includegraphics[width=0.7\columnwidth]{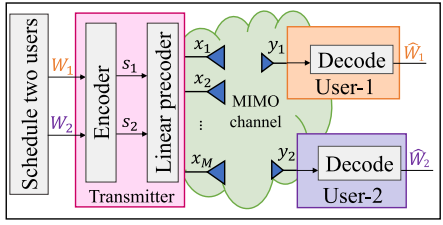}
\caption{Two-user system architecture with MU--LP/SDMA.}
\label{fig_sys_sdma}
\end{figure}
\par The above contrasts with the optimal sum multiplexing gain $d_{\textnormal{s}}^{(\star)}$ of the two-user MISO BC, that is equal to 2, i.e., two interference-free streams can be transmitted\footnote{This assumes that the two channel directions are not aligned, or in other words, that the rank of the matrix $\left[\begin{array}{cc}\mathbf{h}_1 & \mathbf{h}_2\end{array}\right]$ is equal to 2. Note that this condition is met in practice.}. This can be achieved by performing conventional MU--LP, illustrated in Fig. \ref{fig_sys_sdma}. Recall MU--LP system model where $W_1$ and $W_2$ are independently encoded into streams $s_1$ and $s_2$ and respectively precoded by $\mathbf{p}_1$ and $\mathbf{p}_2$ such that the transmit signal is given by
\begin{equation}\label{MULP_2user_system_model}
\mathbf{x}=\mathbf{p}_1 s_1+\mathbf{p}_2 s_2.
\end{equation}
At the receivers, $y_k=\mathbf{h}_k^H \mathbf{x}+n_k$, $k=1,2$, and $s_1$ and $s_2$ are respectively decoded by user-1 and user-2 by treating any residual interference as noise, leading to MU--LP rates
\begin{equation}
R_{1}^{(\textnormal{M})}=\log_2\left(1+C\right), R_{2}^{(\textnormal{M})}=\log_2\left(1+B\right),\label{R_MUMIMO}
\end{equation}
with 
\begin{equation}
C=\frac{\left|\mathbf{h}_1^H \mathbf{p}_{1}\right|^2}{1+\left|\mathbf{h}_1^H \mathbf{p}_{2}\right|^2},
\end{equation}
and $B$ as specified in  (\ref{eq: B}).
It is then indeed sufficient\footnote{More complicated precoders can be used to enhance the rate performance, but the sum and MMF multiplexing gains will not improve.} to transmit two streams using uniform power allocation and Zero-Forcing Beamforming (ZFBF), so that $\mathbf{h}_1^H \mathbf{p}_{2}=\mathbf{h}_2^H \mathbf{p}_{1}=0$, to reap the sum multiplexing gain $d_{\textnormal{s}}^{(\textnormal{M})}=d_{\textnormal{s}}^{(\star)}=2$ and the MMF multiplexing gain $d_{\textnormal{mmf}}^{(\textnormal{M})}=d_{\textnormal{mmf}}^{(\star)}=1$ (i.e., each user gets one full interference-free stream). Indeed, with MU--LP, the sum-rate scales as $2\log_2(P)$ and the MMF rate as $\log_2(P)$ at high SNR \cite{Caire:2003,Jindal:2005,Ding:2007}. Such sum-rate and MMF rate would always strictly outperform that of NOMA (and OMA) at high SNR. Since  both  OMA  and NOMA  achieve  only  half  the  (sum/MMF)  multiplexing  gain of MU–LP in the two-user MISO BC considered, it is not clear whether (and under what conditions) multi-antenna NOMA can outperform MU–LP and other forms of multi-user multi-antenna communication strategies, and if it does, whether multi-antenna NOMA is worth the associated increase in receiver complexity. The above discussion exposes some weaknesses of multi-antenna NOMA and highlights the uncertainty regarding the potential benefits of multi-antenna NOMA. Hence, in the following sections, we derive the multiplexing gains of generalized $K$-user multi-antenna NOMA, so as to better assess its potential. 

\begin{remark}
It appears from \eqref{NOMA_2user_system_model} and \eqref{MULP_2user_system_model} that the transmit signal vectors for 2-user MISO NOMA and 2-user MU--LP are the same, therefore giving the impression that NOMA is the same as MU--LP. This is obviously incorrect. Recall the major differences in the encoding and the decoding of NOMA and MU--LP:
\begin{itemize}
    \item \textit{Encoding}: In NOMA, $W_1$ is encoded into $s_1$ and $W_2$ is encoded into $s_2$ at a rate such that $s_2$ is decodable by both users, while $W_1$ and $W_2$ are independently encoded into streams $s_1$ and $s_2$ in MU--LP. 
    \item \textit{Decoding}: user-1 decodes $s_1$ and $s_2$ and user-2 decodes $s_2$ by treating $s_1$ as noise in NOMA while $s_1$ is decoded by user-1 by treating $s_2$ as noise and $s_2$ is decoded by user-2 by treating $s_1$ as noise in MU--LP. 
\end{itemize} 
Consequently the rate expressions \eqref{R1_NOMA}, \eqref{R2_NOMA} and \eqref{R_MUMIMO} are different, which therefore suggests that the best pair of precoders $\mathbf{p}_1$ and $\mathbf{p}_2$ that maximizes a given objective function (e.g. sum-rate, MMF rate, etc) would be different for NOMA and MU--LP. Choosing $\mathbf{p}_1$ and $\mathbf{p}_2$ according to ZFBF would commonly work reasonably well for MU--LP but would lead to $R_{2}^{(\textnormal{N})}=0$ in \eqref{R2_NOMA} for NOMA. Nevertheless, the above discussion on multiplexing gain loss of MISO NOMA always holds, even in the event where MISO NOMA is implemented with the best choice of precoders, since the above analysis for MISO NOMA is based on upperbound.
\end{remark}

\section{$K$-User MISO NOMA with Perfect CSIT}\label{kuser}
We now study $K$-user MISO NOMA relying on perfect CSIT and derive the sum and MMF multiplexing gains.

\subsection{MISO NOMA System Model}
\label{sec: MISO NOMA system}

\par
\begin{figure}
	\centering
	\includegraphics[width=0.85\columnwidth]{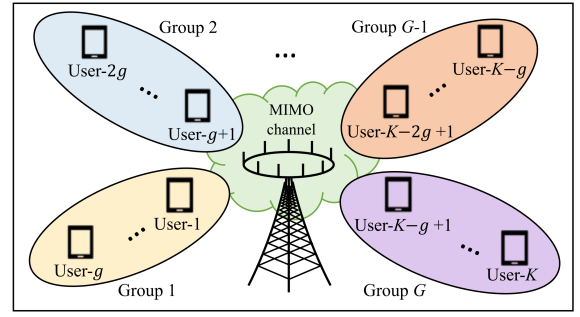}
	\caption{$K$-user system architecture with MISO NOMA  (containing $G$ user groups and $g$ users within each group).}
	\label{fig_sys_MISONOMA}
\end{figure}

 We consider a $K$-user MISO NOMA scenario where a single transmitter equipped with $M$ transmit antennas serves $K$ single-antenna users indexed by $\mathcal{K}=\{1,\cdots,K\}$. The $K$ users are grouped into $1\leq G<K$ groups\footnote{\label{footnote_NOMA_MULP} Note that $1\leq G<K$ is a widely considered option for MISO NOMA in which there exists (at least) one user decoding the  message of (at least) one another user in each  group. Importantly, $G=K$ corresponds to MU--LP as per Section \ref{MULP_section} and is not a MISO NOMA scheme since all $K$ messages are independently encoded into $K$ streams and residual interference is treated as noise at the receivers, i.e. there is no shared codebook and users therefore do not decode the messages of other users.} with groups indexed by $\mathcal{G}=\{1,\cdots,G\}$. There are $g$ users per group, i.e., we therefore assume for simplicity that $K=gG$. Users in group $i$ are  indexed by $\mathcal{K}_i=\{ig-g+1,\cdots,ig\}$. Hence, $\mathcal{K}=\bigcup_{i\in\mathcal{G}}\mathcal{K}_i$ and $|\mathcal{K}_i|=g$.
Without loss of generality, we assume that users 1, $g+1$, $2g+1$, ..., $K-g+1$ are the ``strong users''\footnote{``strong users" here refer to the users who decode the messages of other users in a group. Given the nondegraded nature of the multi-antenna BC, the strong users do not necessarily have to be the users with the largest channel vector norm. The multiplexing gain analysis is general and holds for any ordering. Nevertheless, following \cite{Sun:2015, Zhang:2016}, we consider in the simulation section the decoding order in each group to be the ascending order of users' channel strength such that ``strong users" refer to the users with the largest channel vector norm respectively in group 1 to $G$. } respectively in group 1 to $G$, and perform $g-1$ layers of SIC to fully decode the messages (and therefore remove interference) from the other $g-1$ users within the same group. Similarly, the second user in each group (i.e., $ig-g+2$ in group $i$) performs $g-2$ layers of SIC to fully decode messages from  other $g-2$ users within the same group, and so on.
The two most popular MISO NOMA strategies employ either $G=1$ \cite{Hanif:2016,Choi:2015,Sun:2015,Zhang:2016} or $G=K/2$ \cite{Ding:2016,Choi:2017,Shin:2017,Nguyen:2017,Zeng:2017b,Cheng:2017} but we here keep here the scenario general for any value of $1\leq G<K$. The general architecture of MISO NOMA is illustrated in Fig. \ref{fig_sys_MISONOMA}. The two-user building block in Section \ref{twouser} can be viewed as a particular setup with $K=2$ and $G=1$. 

\par At the transmitter, the messages $W_1$ to $W_K$ intended for user-1 to user-$K$, respectively, are encoded into $s_1$ to $s_K$. However, some of the messages in each group have to be encoded using codebooks shared by a subset of the users in that group so that they can be decoded and cancelled from the received signals at the co-scheduled users in that group. In particular, taking group 1 as an example, $W_2$ to $W_g$ are encoded using codebooks shared with user-1 such that user-1 can decode all of these $g-1$ messages. After encoding, the $K$ streams are linearly precoded by precoders\footnote{Further constraints can be imposed on the precoder design such that the same precoder is used for all users in the same group. This constraint would however further reduce the optimization space and therefore the rate performance.} $\mathbf{p}_1$ to $\mathbf{p}_K$, where $\mathbf{p}_k\in\mathbb{C}^M$ is the precoder of $s_k$, and superposed at the transmitter. The resulting transmit signal is 
\begin{equation}
\label{eq: tx signal NOMA}
\mathbf{x}=\sum_{k=1}^K\mathbf{p}_k s_k.
\end{equation}
Defining $\mathbf{s}=[s_1,...,s_K]^T$ and assuming that $\mathbb{E}[\mathbf{s}\mathbf{s}^H]=\mathbf{I}$, the average transmit power constraint is written as $\sum_{k=1}^K P_{k}\leq P$, where $P_{k}=\left\|\mathbf{p}_{k}\right\|^2$. 

\par At the receiver side, the signal received at user-$k$ is $y_k=\mathbf{h}_k^H \mathbf{x}+n_k$, $k\in\mathcal{K}$, where $\mathbf{h}_k$ is the channel vector\footnote{The rank of the matrix $\left[\begin{array}{ccc}\mathbf{h}_1 & \ldots & \mathbf{h}_K\end{array}\right]$ is assumed equal to $\min\{M,K\}$ for simplicity. Note that this condition is met in practice.} of each user-$k$  perfectly known at the transmitter and that user, and $n_k\sim\mathcal{CN}(0,1)$ is the AWGN. By employing SIC, user-$j$ in group $i$ (i.e., $j\in\mathcal{K}_i$)  decodes the messages of users-$\{k\mid k\geq j,k\in\mathcal{K}_i\}$ within the same user group in a descending order of the user index while treating the interference from users in different  groups  as noise. Under the assumption of  Gaussian signalling and perfect SIC, the rate at user-$j, j\in\mathcal{K}_i$, to decode the message of user-$k,   k\geq j,k\in\mathcal{K}_i$, is given by
\begin{equation}
\label{eq: rate K MIMONOMA}
R_{j,k}=\log_2\left(1+\frac{|\mathbf{h}_j^H\mathbf{p}_k|^2}{I_{j,k}^{(in)}+I_{j,k}^{(ou)}+1}\right),
\end{equation}
where 
\begin{equation}
\begin{aligned}
\label{eq: intefe K MIMONOMA}
I_{j,k}^{(in)}=\sum\limits_{m<k,m\in\mathcal{K}_i}|\mathbf{h}_j^H\mathbf{p}_m|^2,
I_{j,k}^{(ou)}=\sum\limits_{l\neq i, l\in\mathcal{G}}\sum\limits_{m\in\mathcal{K}_l}|\mathbf{h}_j^H\mathbf{p}_m|^2
\end{aligned}
\end{equation} 
are the intra-group interference and inter-group interference received at user-$k$, respectively. 
As the message of user-$k, k\in\mathcal{K}_i$, has to be decoded by users-$\{j|j\leq k,j\in\mathcal{K}_i\}$, to ensure  decodability, the rate of user-$k$ should not exceed
\begin{equation}
\label{eq: min rate K MIMONOMA}
R_k^{(\textnormal{N})}=\min_{j\leq k, j\in\mathcal{K}_i}R_{j,k}.
\end{equation}

\par In the next subsection, we study the sum multiplexing gain and the MMF multiplexing gain of $K$-user MISO NOMA.

\subsection{Multiplexing Gains}\label{multiplexing_gain_kuser}

\par The following proposition provides  the sum multiplexing gain of MISO NOMA for perfect CSIT.
\begin{prop}\label{Theorem_NOMA_DoF} The sum multiplexing gain of $K$-user MISO NOMA with $M$ transmit antennas, $G$ groups of $g=K/G$ users, and perfect CSIT is $d_{\textnormal{s}}^{(\textnormal{N})}=\min\left(M,G\right)$. 
\end{prop}
\par \textit{Proof:} The proof is obtained by showing that an upper bound on the sum multiplexing gain is achievable. The upper bound is obtained by applying the MAC argument (used in \eqref{eq_MAC}) to the strong user in each group and noticing that the sum-rate in groups 1 to $G$ is upper bounded as
\begin{align}
\sum_{k=1}^g R_k^{(\textnormal{N})} &\leq \log_2\left(1+\sum_{k=1}^g\left|\mathbf{h}_1^H \mathbf{p}_{k}\right|^2\right),\label{sumrate_1g}\\
\sum_{k=g+1}^{2g} R_k^{(\textnormal{N})} &\leq \log_2\left(1+\sum_{k=g+1}^{2g}\left|\mathbf{h}_{g+1}^H \mathbf{p}_{k}\right|^2\right),\label{sumrate_2g}\\
\vdots \nonumber\\
\sum_{k=K-g+1}^{K} R_k^{(\textnormal{N})} &\leq \log_2\left(1+\sum_{k=K-g+1}^{K}\left|\mathbf{h}_{K-g+1}^H \mathbf{p}_{k}\right|^2\right).\label{sumrate_3g}
\end{align}
Note that the left-hand sides of \eqref{sumrate_1g}, \eqref{sumrate_2g}, and \eqref{sumrate_3g} refer to the sum of the rates of the messages in group 1, 2, and $G$, respectively, but can also be viewed as the total rate to be decoded by user 1, $g+1$, and $K-g+1$ (since those users decode all the messages in their respective group). We now notice that the right-hand sides of \eqref{sumrate_1g}, \eqref{sumrate_2g}, and \eqref{sumrate_3g} scale as $\log_2(P)+ \delta$ for large $P$ (following the same argument as in the two-user case). This implies that each group $i$ achieves at most a (group) sum multiplexing gain $d_{\textnormal{s},i}^{(\textnormal{N})}=\sum_{k=ig-g+1}^{ig}d_k^{(N)}$ of 1, i.e., at most one interference-free stream can be transmitted to each group. Summing up all inequalities, we obtain in the limit of large $P$ that
\begin{equation}
R_{\textnormal{s}}^{(\textnormal{N})}=\sum_{k=1}^{K} R_k^{(\textnormal{N})} \stackrel{P\nearrow}{\leq} G \log_2(P) +\delta,
\end{equation}
which shows that $d_{\textnormal{s}}^{(\textnormal{N})}=\sum_{i=1}^{G}d_{\textnormal{s},i}^{(\textnormal{N})}\leq G$. Additionally, since $d_{\textnormal{s}}^{(\textnormal{N})}\leq d_{\textnormal{s}}^{\star}=\min\left(M,K\right)$, we have $d_{\textnormal{s}}^{(\textnormal{N})}\leq \min\left(M,G\right)$.
\par The achievability part shows that $d_{\textnormal{s}}^{(\textnormal{N})}\geq \min\left(M,G\right)$. To this end, it is indeed sufficient to perform ZFBF and transmit $\min\left(M,G\right)$ interference-free streams to $\min\left(M,G\right)$ of the $G$ ``strong users''. Combining the upper bound and achievability leads to the conclusion that $d_{\textnormal{s}}^{(\textnormal{N})}=\min\left(M,G\right)$.
$\hfill\Box$ 

\par The following result derives the MMF multiplexing gain of MISO NOMA with perfect CSIT.
\begin{prop}\label{Theorem_NOMA_MMF_DoF} The MMF multiplexing gain of $K$-user MISO NOMA with $M$ transmit antennas, $G$ groups of $g=K/G$ users and perfect CSIT is 
\begin{equation}
d_{\textnormal{mmf}}^{(\textnormal{N})}=\left\{\begin{array}{l}
\frac{1}{g}, \hspace{0.2cm} M\geq K-g+1 \\
0, \hspace{0.2cm} M < K-g+1. 
\end{array}
\right. 
\end{equation}
For $G=1$, i.e., $g=K$, $d_{\textnormal{mmf}}^{(\textnormal{N})}=\frac{1}{K}$.
\end{prop}
\par \textit{Proof:} Let us first consider $M\geq K-g+1$. The MMF multiplexing gain is always upperbounded by ignoring the inter-group interference, i.e., the $G$ groups are non-interfering. Following again the MAC argument, the sum multiplexing gain of one in each group can then be further split equally amongst the $g$ users, which leads to an upper bound on the MMF multiplexing gain of $\frac{1}{g}$. Achievability is simply obtained by designing the precoders using ZFBF to eliminate all inter-group interference, and allocating the power similarly to Subsection \ref{discussion_twouser}, i.e., consider group 1 for simplicity, and allocate the power to user $k=1,\ldots,g$ as $O(P^{k/g})$, which leads to an SINR for user-$k$ scaling as $O(P^{1/g})$ and to an achievable MMF multiplexing gain of $\frac{1}{g}$. For $G=1$, one can simply allocate the power to user $k=1,\ldots,K$ as $O(P^{k/K})$, which leads to an achievable MMF multiplexing gain of $\frac{1}{K}$. 
\par Let us now consider $M < K-g+1$. Take $M=K-g$ (any smaller $M$ cannot improve the multiplexing gain). Precoder $\mathbf{p}_k$ of any user-$k$ can be made orthogonal to the channel of $K-g-1$ co-scheduled users and will therefore cause interference to at least one user in another group. As a result, the MMF multiplexing gain collapses to 0. 
$\hfill\Box$

\begin{remark}\label{MMF_assumption} For the MMF multiplexing gain analysis, it should be noted that we consider one-shot transmission schemes with no time-sharing between
strategies. This is suitable for systems with rigid scheduling
and/or tight latency constraints, and also allows for simpler designs. This assumption is also commonly used in the NOMA literature \cite{Liu:2016,Alavi:2018,Yalcin:2019,Timotheou:2015,Choi:2016}.
\end{remark}

\section{$K$-user MISO NOMA with Imperfect CSIT}\label{Imperfect_CSIT_section}
We now go one step further and extend the multiplexing gain analysis to the imperfect CSIT setting. The results in this section therefore generalize the results in the previous section (with perfect CSIT being a particular case of imperfect CSIT). In this section, the achievable rates are defined in the ergodic sense in a standard Shannon theoretic fashion, and the corresponding sum and MMF mutiplexing gains are defined similarly to Subsection \ref{DoF_def} using ergodic rates. We first introduce the CSIT error model before deriving the multiplexing gains of MISO NOMA relying on imperfect CSIT.

\subsection{CSIT Error Model}\label{CSIT_Error_model_section}
For each user, the transmitter acquires an imperfect estimate of the channel vector $\mathbf{h}_k$, denoted as $\hat{\mathbf{h}}_k$. The CSIT imperfection is modelled by
\begin{equation}
    \mathbf{h}_k=\hat{\mathbf{h}}_k+\tilde{\mathbf{h}}_k,
\end{equation}
where $\tilde{\mathbf{h}}_k$ denotes the corresponding channel estimation error at the transmitter. For compactness, we define $\mathbf{H}=[\mathbf{h}_1 \ldots \mathbf{h}_K]$, $\hat{\mathbf{H}}=[\hat{\mathbf{h}}_1 \ldots \hat{\mathbf{h}}_K]$, and $\tilde{\mathbf{H}}=[\tilde{\mathbf{h}}_1 \ldots \tilde{\mathbf{h}}_K]$, which implies $\mathbf{H}=\hat{\mathbf{H}}+\tilde{\mathbf{H}}$. The joint fading process is characterized by the joint distribution $f_{\mathbf{H},\hat{\mathbf{H}}}\big(\mathbf{H},\hat{\mathbf{H}}\big)$ of $\{\mathbf{H},\hat{\mathbf{H}}\}$, assumed to be stationary and ergodic 
. The joint distribution $f_{\mathbf{H},\hat{\mathbf{H}}}\big(\mathbf{H},\hat{\mathbf{H}}\big)$ is continuous and known to the transmitter. The ergodic rates capture
the long-term performance over a long sequence of channel
uses $\{\mathbf{H},\hat{\mathbf{H}}\}$ spanning almost all possible joint channel states.

\par For each user-$k$, we define the average channel (power) gain as $\Gamma_k=\mathbb{E}\big\{\left\|\mathbf{h}_k\right\|^2\big\}$. Similarly, we define $\hat{\Gamma}_k=\mathbb{E}\big\{\big\|\hat{\mathbf{h}}_k\big\|^2\big\}$ and $\tilde{\Gamma}_k=\mathbb{E}\big\{\big\|\tilde{\mathbf{h}}_k\big\|^2\big\}$. For many CSIT acquisition mechanisms \cite{Kobayashi:2011}, $\hat{\mathbf{h}}_k$ and $\tilde{\mathbf{h}}_k$ are uncorrelated  according to the orthogonality principle \cite{Poor:2013}. By further assuming that $\hat{\mathbf{h}}_k$ and $\tilde{\mathbf{h}}_k$ have zero means, we have $\Gamma_k=\hat{\Gamma}_k+\tilde{\Gamma}_k$, based on which we can write $\hat{\Gamma}_k=(1-\sigma_{e,k}^2)\Gamma_k$ and $\tilde{\Gamma}_k=\sigma_{e,k}^2\Gamma_k$ for some $\sigma_{e,k}^2\in [0,1]$. Note that $\sigma_{e,k}^2$ is the normalized estimation error variance for user-$k$'s CSIT, e.g., $\sigma_{e,k}^2=1$ corresponds to no instantaneous CSIT, while $\sigma_{e,k}^2=0$ represents perfect instantaneous CSIT. 
\par For simplicity, we assume identical normalized CSIT error variances for all users, i.e., $\sigma_{e,k}^2=\sigma_{e}^2$ for all $k=1,\ldots,K$. To facilitate the multiplexing gain analysis, we assume that $\sigma_{e}^2$ scales with SNR as $\sigma_{e}^2=P^{-\alpha}$ for some CSIT quality parameter $\alpha\in[0,\infty)$ \cite{Jindal:2006,Yang:2013,Davoodi:2016,Joudeh:2016a,Ding:2007}. This is a convenient and tractable model extensively used in the information theoretic literature that allows us to assess the performance of the system in a wide range of CSIT quality conditions. Indeed, the larger $\alpha$, the faster the CSIT error decreases with the SNR. The two extreme cases, $\alpha=0$ and $\alpha=\infty$, correspond to no or constant CSIT (i.e. that does not scale or improve with SNR) and perfect CSIT, respectively. As far as the multiplexing gain analysis is concerned, however, we may truncate the CSIT quality parameter as $\alpha\in[0,1]$, where $\alpha=1$ amounts to perfect CSIT in the multiplexing gain sense. The regime $\alpha\in(0,1)$ corresponds to partial CSIT, resulting from imperfect CSI acquisition. The CSIT quality $\alpha$ can be interpreted in many different ways, but a plausible interpretation of $\alpha$ is related to the number of feedback bits, where $\alpha=0$ corresponds to a fixed number of feedback bits for all SNRs, $\alpha=\infty$ corresponds to an infinite number of feedback bits, and $0<\alpha<\infty$ reflects how quickly the number of feedback bits increases with the SNR. As a reference, a system like 4G and 5G use $\alpha=0$ when limited feedback (or codebook-based feedback) is used to report the CSI, since the number of feedback bits is constant and does not scale with SNR, e.g. 4 bits of CSI feedback in 4G LTE for $M=4$.

\subsection{Multiplexing Gains}
\label{DoFimperfect}
\par The following result quantifies the sum multiplexing gain of MISO NOMA for imperfect CSIT.
\begin{prop}\label{Theorem_NOMA_DoF_alpha} The sum multiplexing gain of  $K$-user MISO NOMA with $M$ transmit antennas, $G$ groups of $g=K/G$ users, and CSIT quality $0\leq\alpha \leq 1$ is $d_{\textnormal{s}}^{(\textnormal{N})}=\max\left(1,\min\left(M,G\right)\alpha\right)$. 
\end{prop}
\par \textit{Proof:} Similarly to the proof of Proposition \ref{Theorem_NOMA_DoF}, let us look at the $G$ strong users, since they are the ones who have to decode all messages. We recall that $d_{\textnormal{s},i}^{(\textnormal{N})}=\sum_{k=ig-g+1}^{ig}d_k^{(N)}$ reflects the multiplexing gain of the total rate to be decoded by the strong user $ig-g+1$ in group $i$ as a consequence of the fact that this user decodes all $g$ messages in group $i$. Making use of the results of MU--LP in the $G$-user MISO BC with imperfect CSIT \cite{Joudeh:2016a}\footnote{See also Proposition \ref{prop_MULP_alpha}.}, we obtain $d_{\textnormal{s}}^{(\textnormal{N})}=\sum_{i=1}^G d_{\textnormal{s},i}^{(\textnormal{N})}=\sum_{k=1}^K d_k^{(N)}\leq \max\left(1,\min\left(M,G\right)\alpha\right)$. 
\par The achievability part shows that $d_{\textnormal{s}}^{(\textnormal{N})}\geq \max\left(1,\min\left(M,G\right)\alpha\right)$. It is indeed sufficient to perform ZFBF and transmit $\min\left(M,G\right)$ streams, each at a power level of $P^{\alpha}/\min\left(M,G\right)$, to $\min\left(M,G\right)$ of the $G$ ``strong users''. If $\min\left(M,G\right)\alpha<1$, one can simply transmit a single stream (i.e., perform OMA) and reap a sum multiplexing gain of 1. Combining the upper bound and achievability leads to the conclusion that we have $d_{\textnormal{s}}^{(\textnormal{N})}=\max\left(1,\min\left(M,G\right)\alpha\right)$.
$\hfill\Box$

\par For $\alpha=1$ (perfect CSIT from a multiplexing gain perspective), Proposition \ref{Theorem_NOMA_DoF_alpha} boils down to the perfect CSIT result in Proposition \ref{Theorem_NOMA_DoF}. 


\par The following proposition provides the MMF multiplexing gain of MISO NOMA with imperfect CSIT.
\begin{prop}\label{Theorem_NOMA_MMF_DoF_alpha} The MMF multiplexing gain of  $K$-user MISO NOMA with $M$ transmit antennas, $G$ groups of $g=K/G$ users, and CSIT quality $0\leq\alpha \leq 1$ is 
\begin{equation}
d_{\textnormal{mmf}}^{(\textnormal{N})}=\left\{\begin{array}{l}
\frac{\alpha}{g}, \hspace{0.3cm} G>1 \:\textrm{and}\: M\geq K-g+1 \\
0, \hspace{0.4cm} G>1 \:\textrm{and}\: M < K-g+1\\
\frac{1}{K}, \hspace{0.23cm} G=1.
\end{array}
\right. 
\end{equation}

\end{prop}
\par The proof is relegated to Appendix \ref{app: prop4}.

\par It is interesting to note that the sensitivity of the multiplexing gain of MISO NOMA to the CSIT quality $\alpha$ is different for $G>1$ and  $G=1$. Indeed the sum and MMF multiplexing gains of MISO NOMA with $G>1$ decay as $\alpha$ decreases, while the multiplexing gains of MISO NOMA with $G=1$ are not affected by $\alpha$. This can be interpreted in two different ways. On the one hand, this suggests that MISO NOMA $G=1$ is inherently robust to CSIT imperfections since the multiplexing gains are not affected by $\alpha<1$. On the other hand, this also reveals that MISO NOMA with $G=1$ is unable to exploit the presence of CSIT since its multiplexing gains are the same as in the absence of CSIT ($\alpha=0$).

\section{Baseline Scheme I: \\ Conventional Multi-user Linear Precoding}\label{MULP_section}
The first baseline to assess the performance of multi-antenna NOMA is conventional Multi-User Linear Precoding. In the sequel, we recall the multiplexing gains achieved by MU--LP.
\subsection{MU--LP System Model}
\label{sec: MULP system model}
\begin{figure}
	\centering
	\includegraphics[width=0.7\columnwidth]{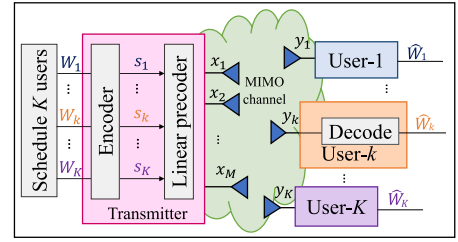}
	\caption{$K$-user system architecture with MU--LP. Receiver architecture is illustrated for user-$k$ though the same applies to other users, i.e. all $K$ users are equipped with a decoder that maps the received signal into an estimated message by treating residual interference as noise.}
	\label{fig_sys_MULP}
\end{figure}

Following Subsection \ref{sec: MISO NOMA system}, we consider a $K$-user MISO BC with one transmitter equipped with $M$ transmit antennas and $K$ single-antenna users. As per Fig. \ref{fig_sys_MULP}, the messages $W_1,\ldots,W_K$ respectively for user-1 to user-$K$ are independently encoded into $s_1$ to $s_K$, which are then mapped to the transmit antennas through the precoders $\mathbf{p}_1,\ldots,\mathbf{p}_K$. The resulting transmit signal is 
$
\mathbf{x}=\sum_{k=1}^K\mathbf{p}_k s_k.
$

The signal received at user-$k$ is $y_k=\mathbf{h}_k^H \mathbf{x}+n_k$ with $n_k\sim\mathcal{CN}(0,1)$. Each user-$k$ directly  decodes the intended message $W_k$ by treating the interference from other users  as noise. Under the assumption of  Gaussian signalling, the rate of user-$k$ for $k\in\mathcal{K}$ is given by
\begin{equation}
R_{k}^{(\textnormal{M})}=\log_2\left(1+\frac{\left|\mathbf{h}_k^H \mathbf{p}_{k}\right|^2}{1+\sum_{q\neq k}\left|\mathbf{h}_k^H \mathbf{p}_{q}\right|^2}\right).
\end{equation}
The sum-rate of MU--LP is therefore  $R_{\mathrm{s}}^{(\textnormal{M})}=\sum_{k=1}^K R_k^{(\textnormal{M})}$, and the MMF rate of MU--LP is given as $R_{\mathrm{mmf}}^{(\textnormal{M})}=\min_{k=1,\ldots,K}R_{k}^{(\textnormal{M})}$.

\subsection{Multiplexing Gains with Perfect CSIT}
\label{DoF_MULP_perfect}
\par We recall the sum multiplexing gain and the MMF multiplexing gain of MU--LP with perfect CSIT from \cite{Jindal:2005} and \cite{Joudeh:2017}, respectively.
\begin{prop}\label{Theorem_MULP_DoF} The sum multiplexing gain of $K$-user MU--LP with $M$ transmit antennas and perfect CSIT is $d_{\textnormal{s}}^{(\textnormal{M})}=\min\left(M,K\right)$. 
\end{prop}
\par This result\footnote{It is implicitly assumed here that the coherence block is much larger than $\min(M,K)$ such that the resource needed to estimate the channel vanishes.} is simply achieved by choosing the MU--LP precoders based on ZFBF and transmitting $\min\left(M,K\right)$ interference-free streams. Note that $\min\left(M,K\right)$ is also the optimal\footnote{This is easily proved by showing that an upper bound on the sum multiplexing gain is equal to $\min\left(M,K\right)$, which is the same as the lower bound achieved by MU--LP. The upper bound is obtained by noticing that enabling full cooperation among receivers does not decrease the sum multiplexing gain and leads to an effective point-to-point MIMO channel with $M$ transmit and $K$ receive antennas, which has a sum multiplexing gain of $\min\left(M,K\right)$.} sum multiplexing gain of the $K$-user MISO BC\footnote{More generally, in MIMO BC, $d_{\textnormal{s}}^{(\textnormal{M})}=d_{\textnormal{s}}^{(\star)}=\min\left(M,KN\right)$ \cite{Jindal:2005}.}  \cite{Jindal:2005}. In other words, $d_{\textnormal{s}}^{(\textnormal{M})}=d_{\textnormal{s}}^{(\star)}=\min\left(M,K\right)$.  

\begin{prop}\label{Theorem_MULP_MMF_DoF} The MMF multiplexing gain of the $K$-user MU--LP with $M$ transmit antennas and perfect CSIT is  
\begin{equation}
d_{\textnormal{mmf}}^{(\textnormal{M})}=\left\{\begin{array}{l}
1, \hspace{0.2cm} M\geq K \\
0, \hspace{0.2cm} M < K. 
\end{array}
\right. 
\end{equation}
\end{prop}
\par When $M\geq K$, ZFBF can be used to fully eliminate interference. On the other hand, for $M < K$ interference cannot be eliminated anymore and $d_{\textnormal{mmf}}^{(\textnormal{M})}$ collapses, therefore leading to a rate saturation at high SNR. 

\subsection{Multiplexing Gains with Imperfect CSIT}
\label{DoF_MULP_imperfect}
\par We use the CSIT error model introduced in Subsection \ref{CSIT_Error_model_section}. We recall the sum multiplexing gain and the MMF multiplexing gain of MU--LP with imperfect CSIT from \cite{Joudeh:2016a} and \cite{Joudeh:2016b,Yin:2020}, respectively.

\begin{prop}\label{prop_MULP_alpha} The sum multiplexing gain of the $K$-user MU--LP with $M$ transmit antennas and CSIT quality $0\leq\alpha \leq 1$ is $d_{\textnormal{s}}^{(\textnormal{M})}=\max\left(1,\min\left(M,K\right)\alpha\right)$. 
\end{prop}
\par This result is simply achieved by choosing the MU--LP precoders based on ZFBF and transmitting $\min\left(M,K\right)$ streams, each with power level $P^{\alpha}/\min\left(M,K\right)$. This enables each stream to reap a multiplexing gain of $\alpha$ and therefore a sum multiplexing gain of $\min\left(M,K\right)\alpha$. If $\min\left(M,K\right)\alpha<1$, one can simply transmit a single stream (i.e., perform OMA) and reap a sum multiplexing gain of 1.
\par Comparing Propositions \ref{Theorem_MULP_DoF} and \ref{prop_MULP_alpha}, we note that imperfect CSIT leads to a reduction of the sum multiplexing gain. For $\alpha=1$ (perfect CSIT in a multiplexing gain sense), Proposition \ref{prop_MULP_alpha} matches Proposition \ref{Theorem_MULP_DoF}. Importantly, in contrast to the $K$-user MISO BC with perfect CSIT setting where MU--LP achieves the information theoretic optimal sum multiplexing gain $d_{\textnormal{s}}^{(\textnormal{M})}=d_{\textnormal{s}}^{(\star)}$, in the imperfect CSIT setting, MU--LP does not achieve the information theoretic optimal sum multiplexing gain \cite{Davoodi:2016,Joudeh:2016a}.

\begin{prop}\label{Theorem_MULP_MMF_DoF_alpha} The MMF multiplexing gain of the $K$-user MU--LP with $M$ transmit antennas and CSIT quality $0\leq\alpha \leq 1$ is  
\begin{equation}
d_{\textnormal{mmf}}^{(\textnormal{M})}=\left\{\begin{array}{l}
\alpha, \hspace{0.2cm} M\geq K \\
0, \hspace{0.2cm} M < K. 
\end{array}
\right. 
\end{equation}
\end{prop}
\par This is achieved by performing ZFBF when $M \geq K$. When $M<K$, rate saturation occurs (similarly to the perfect CSIT setting).

\section{Baseline Scheme II: Rate-Splitting}\label{RS_section}
The second baseline to assess multi-antenna NOMA performance is multi-antenna Rate-Splitting (RS) and Rate-Splitting Multiple Access (RSMA)  for the multi-antenna BC \cite{Clerckx:2016,Yang:2013,Joudeh:2016a,Joudeh:2016b,Hao:2015,Dai:2016,Mao:2017}. This literature leverages and extends the concept of RS, originally developed in \cite{Han:1981} for the two-user single-antenna interference channel, to design multi-antenna non-orthogonal transmission strategies for the multi-antenna BC.

\subsection{Rate-Splitting System Model}
\label{sec: RS}

\begin{figure}
	\centering
	\includegraphics[width=0.9\columnwidth]{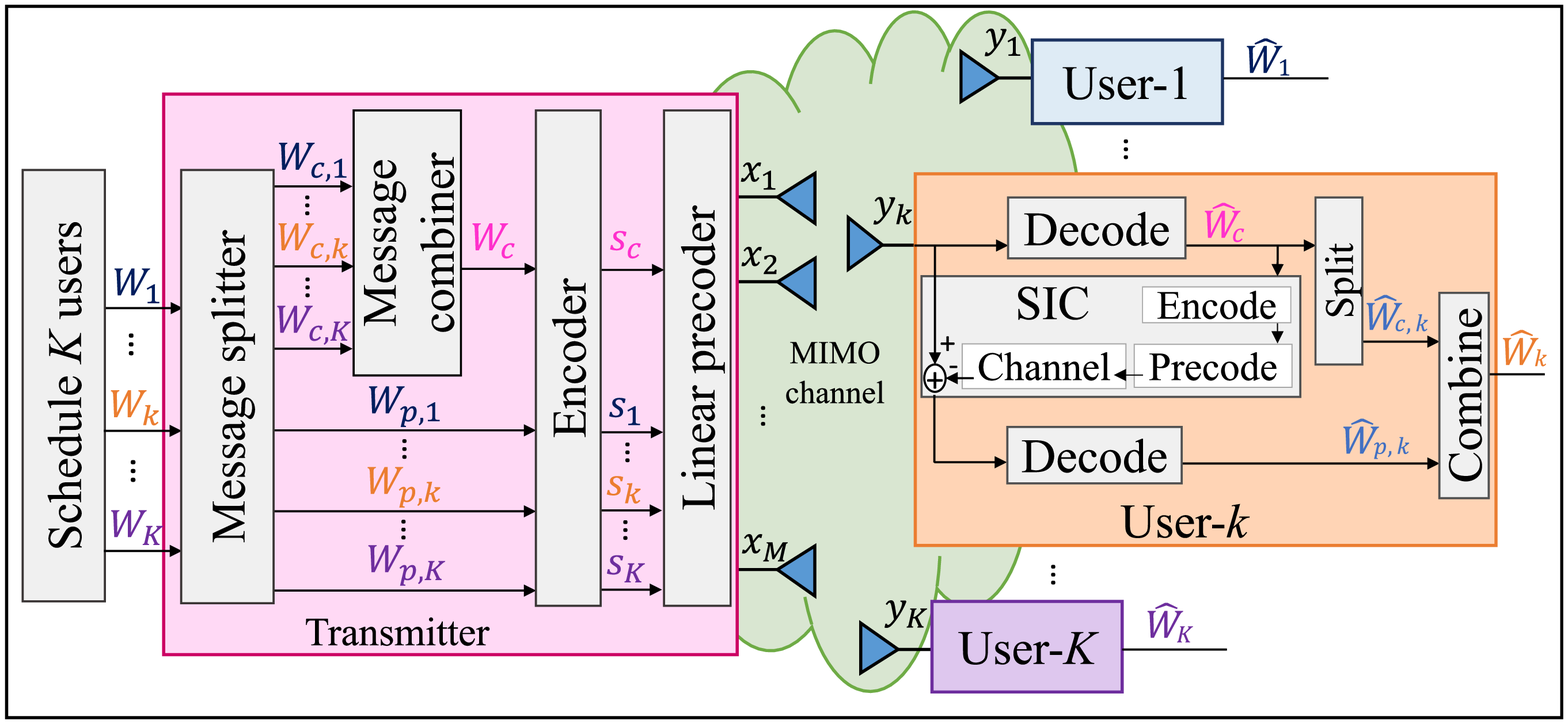}
	\caption{$K$-user system architecture with 1-layer rate-splitting. Receiver architecture is illustrated for user-$k$ though the same applies to other users.}
	\label{fig_sys_rsma}
\end{figure}

\par We consider again a MISO BC consisting of one transmitter with $M$ antennas and $K$ single-antenna users. As per Fig. \ref{fig_sys_rsma}, the architecture relies on rate-splitting of messages $W_1$ to $W_K$ intended for user-1 to user-$K$, respectively. To that end,  message $W_k$ of user-$k$ is split into a common part $W_{\mathrm{c},k}$ and a private part $W_{\mathrm{p},k}$. The common parts $W_{\mathrm{c},1},\ldots, W_{\mathrm{c},K}$ of all users are combined into the common message $W_{\mathrm{c}}$, which is encoded into the common stream $s_{\mathrm{c}}$ using a codebook shared by all users. Hence, $s_{\mathrm{c}}$ is a common stream required to be decoded by all users, and contains parts of  messages $W_1$ to $W_K$ intended for user-1 to user-$K$, respectively. The private parts $W_{\mathrm{p},1},\ldots,W_{\mathrm{p},K}$, respectively containing the remaining parts of  messages $W_1$ to $W_K$, are independently encoded into the private stream $s_1$ for user-1 to $s_K$ for user-$K$. Out of the $K$ messages, $K+1$ streams $s_{\mathrm{c}}, s_1,\ldots, s_K$ are therefore created. The streams are linearly precoded such that the transmit signal is given by
\begin{equation}
\mathbf{x}=\mathbf{p}_{\mathrm{c}} s_{\mathrm{c}}+\sum_{k=1}^K\mathbf{p}_k s_k.
\end{equation}
Defining $\mathbf{s}=[s_{\mathrm{c}},s_1,\ldots,s_K]^T$ and assuming that $\mathbb{E}[\mathbf{s}\mathbf{s}^H]=\mathbf{I}$, the average transmit power constraint is written as $P_{\mathrm{c}}+\sum_{k=1}^K P_{k}\leq P$, where $P_{\mathrm{c}}=\left\|\mathbf{p}_{\mathrm{c}}\right\|^2$ and $P_{k}=\left\|\mathbf{p}_{k}\right\|^2$.

\par At each user-$k$, the common stream $s_{\mathrm{c}}$ is first decoded into $\widehat{W}_{\mathrm{c}}$ by treating the interference from the private streams as noise. Using SIC, $\widehat{W}_{\mathrm{c}}$ is re-encoded, precoded, and subtracted from the received signal, such that user-$k$ can decode its private stream $s_k$ into $\widehat{W}_{\mathrm{p},k}$ by treating the remaining interference from the other private stream as noise. User-$k$ reconstructs the original message by extracting $\widehat{W}_{\mathrm{c},k}$ from $\widehat{W}_{\mathrm{c}}$, and combining $\widehat{W}_{\mathrm{c},k}$ with $\widehat{W}_{\mathrm{p},k}$ into $\widehat{W}_{k}$. Assuming proper Gaussian signalling, the rate of the common stream is given by
\vspace{-0.2cm}
\begin{equation}
\label{eq: SINR common}
R_{\mathrm{c}}=\min_{k=1,...,K} \log_2\left(1+\frac{\left|\mathbf{h}_k^H \mathbf{p}_{\mathrm{c}}\right|^2}{1+\sum_{q=1}^K\left|\mathbf{h}_k^H \mathbf{p}_{q}\right|^2}\right).
\end{equation}
Assuming perfect SIC, the rates of the private streams are obtained as
\begin{equation}
\label{eq: SINR private}
R_{k}=\log_2\left(1+\frac{\left|\mathbf{h}_k^H \mathbf{p}_{k}\right|^2}{1+\sum_{q\neq k}\left|\mathbf{h}_k^H \mathbf{p}_{q}\right|^2}\right).
\end{equation}
The rate of user-$k$ is given by $R_k+R_{\mathrm{c},k}$ where $R_{\mathrm{c},k}$ is the rate of the common part of the $k$th user’s message, i.e., $W_{\mathrm{c},k}$, and it satisfies $\sum_{k=1}^K R_{\mathrm{c},k}=R_{\mathrm{c}}$. The sum-rate is therefore simply written as $R_{\mathrm{s}}^{(\textnormal{R})}=\sum_{k=1}^K (R_k+R_{\mathrm{c},k})=R_{\mathrm{c}}+\sum_{k=1}^K R_{k}$, and the MMF rate is written as $R_{\mathrm{mmf}}^{(\textnormal{R})}=\min_{k=1,\ldots,K}R_{k}+R_{\mathrm{c},k}$.
\par The above RS architecture is called 1-layer RS since it only relies on a single common stream and a single layer of SIC at each user as illustrated in Fig. \ref{fig_sys_rsma}.

\subsection{Multiplexing Gains with Perfect CSIT}
\label{sec: multiplex gain}

We here summarize the sum and MMF multiplexing gains achieved by 1-layer RS with perfect CSIT.
\begin{prop}\label{Theorem_RS_DoF} The sum multiplexing gain of the $K$-user 1-layer RS with $M$ transmit antennas and perfect CSIT is $d_{\textnormal{s}}^{(\textnormal{R})}=\min\left(M,K\right)$. 
\end{prop}
\par \textit{Proof:} Since MU--LP is a subscheme of 1-layer  RS\footnote{By allocating no power to the common stream, 1-layer RS boils down to MU-LP.}, it is sufficient\footnote{More complicated precoders for both the common and private streams can be used to enhance the rate performance, but the multiplexing gain will not improve.} to design the private precoders using ZFBF and allocate zero power to the common stream at high SNR. Note that $d_{\textnormal{s}}^{(\textnormal{R})}=d_{\textnormal{s}}^{(\textnormal{M})}=d_{\textnormal{s}}^{(\star)}=\min\left(M,K\right)$. 
$\hfill\Box$

\begin{prop}\label{Theorem_RS_MMF_DoF} The MMF multiplexing gain of the $K$-user 1-layer RS with $M$ transmit antennas and perfect CSIT is  
\begin{equation}
d_{\textnormal{mmf}}^{(\textnormal{R})}=\left\{\begin{array}{l}
1, \hspace{1.2cm} M\geq K \\
\frac{1}{1+K-M}, \hspace{0.2cm} M < K.\\
\end{array}
\right. 
\end{equation}
\end{prop}

\par The MMF multiplexing gain of 1-layer RS was derived and proved in \cite{Joudeh:2017}\footnote{The MMF multiplexing gain derived in \cite{Joudeh:2017} considers a more  complex scenario involving the simultaneous transmission of distinct messages to multiple multicast groups (each message is intended for a group of users), known as multigroup multicasting. By considering the special case where there is a single user per group, we obtain the MMF multiplexing gain of 1-layer RS in this section.}, under the same assumption as in Remark \ref{MMF_assumption}. Readers are referred to \cite{Joudeh:2017} for more details of the proof of Proposition \ref{Theorem_RS_MMF_DoF}.


\subsection{Multiplexing Gains with Imperfect CSIT}
\label{sec: multiplex gain imperfect CSIT}
 Again, we use the CSIT error model introduced in Subsection \ref{CSIT_Error_model_section}. We recall the sum multiplexing gain of RS with imperfect CSIT from \cite{Joudeh:2016a}.
\begin{prop}\label{Theorem_RS_DoF_alpha} The sum multiplexing gain of $K$-user 1-layer RS with $M$ transmit antennas and CSIT quality $0\leq\alpha \leq 1$ is $d_{\textnormal{s}}^{(\textnormal{R})}=1+\left(\min\left(M,K\right)-1\right)\alpha$. 
\end{prop}

\par  Achievability of $d_{\textnormal{s}}^{(\textnormal{R})}$ in Proposition \ref{Theorem_RS_DoF_alpha} is obtained by  using random precoding to design $\mathbf{p}_c$ with power level $P_c=O(P)$, transmitting $\min(M,K)$ private streams and using ZFBF to design the precoders  of those $\min(M,K)$ private streams, each with power level  $P_k=O(P^{\alpha})$.
From the SINR expressions at the right-hand side of (\ref{eq: SINR common}), it follows that the received SINR of the common stream at each user scales as $O(P^{1-\alpha})$, leading to the multiplexing gain of  $1-\alpha$  achieved by the common stream $s_c$.
By performing ZFBF, the transmitter transmits $\min\left(M,K\right)$ interference-free private streams. The received SINR of each private stream  scales as  $O(P^{\alpha})$ leading to  multiplexing gain $\alpha$. Hence, we obtain the sum multiplexing gain of $1+(\min\left(M,K\right)-1)\alpha$.

Importantly, for the underloaded regime $M \geq K$, 1-layer RS achieves the information theoretic optimal sum multiplexing gain $d_{\textnormal{s}}^{(\textnormal{M})}=d_{\textnormal{s}}^{(\star)}$ in the imperfect CSIT setting \cite{Davoodi:2016,Joudeh:2016a}. Hence, 1-layer RS attains the optimal sum multiplexing gain in both perfect CSIT and imperfect CSIT (underloaded regime). Actually, for $M \geq K$, 1-layer RS is optimal, achieving the maximum multiplexing gain region of the underloaded $K$-user MISO BC\footnote{The optimality of RS is not limited to MISO BC but also extends to MIMO BC. Indeed, a more complicated form of RS is multiplexing gain region-optimal for the two-user MIMO BC with imperfect CSIT in the general case of an asymmetric number of receive antennas \cite{Hao:2017,Davoodi:2020}.} with imperfect CSIT \cite{Piovano:2017,Joudeh:2019}. 
 
This optimality of RS (including 1-layer RS), shown through multiplexing gain analysis, is very significant since it implies that one cannot find any other scheme achieving a better multiplexing gain region in multi-antenna BC. As a consequence of this optimality, MU--LP and multi-antenna NOMA will always incur a multiplexing gain loss or at best will achieve the same multiplexing gain as RS for both perfect and imperfect CSIT.

\begin{prop}\label{Theorem_RS_MMF_DoF_alpha} The MMF multiplexing gain of $K$-user 1-layer RS with $M$ transmit antennas and CSIT quality $0\leq\alpha \leq 1$ is  
\begin{equation}\label{eq_MMF_RS}
d_{\textnormal{mmf}}^{(\textnormal{R})}=\left\{\begin{array}{l}
\frac{1+(K-1)\alpha}{K}, \hspace{0.2cm} M\geq K \\
\frac{1+(M-1)\alpha}{K}, \hspace{0.2cm} M < K \:\textnormal{and}\: \alpha\leq\frac{1}{1+K-M}\\
\frac{1}{1+K-M}, \hspace{0.4cm} M < K \:\textnormal{and}\: \alpha > \frac{1}{1+K-M}.\\
\end{array}
\right. 
\end{equation}
\end{prop}
\par The MMF multiplexing gain of 1-layer RS with imperfect CSIT was derived in \cite{Yin:2020} (by considering the specific case where there is a single user per group), under the same assumption as in Remark \ref{MMF_assumption}. Readers are referred to \cite{Yin:2020} for more details of the proof of Proposition \ref{Theorem_RS_MMF_DoF_alpha}.

\par This highlights that when $M<K$, the CSIT quality $\alpha$ can be reduced to $\frac{1}{1+K-M}$ without impacting the MMF multiplexing gain of 1-layer RS.

Following our discussion of Proposition \ref{Theorem_RS_DoF_alpha}, we know that when $M\geq K$, the respective multiplexing gains of the common and each private streams are $1-\alpha$ and $\alpha$. The MMF multiplexing gain when $M\geq K$ is achieved by evenly sharing the common stream among users, which  is the sum of evenly allocated multiplexing gain of the common stream $\frac{1-\alpha}{K}$
and the multiplexing gain of one private stream $\alpha$, yielding $\frac{1+(K-1)\alpha}{K}$.

When $M<K$, the achievability is obtained by partitioning users into two
subsets $\mathcal{K}_1$ and $\mathcal{K}_2$ with set size of $|\mathcal{K}_1|=M$ and $|\mathcal{K}_2|=K-M$. Users in $\mathcal{K}_1$ are served via  the common and private streams while users in $\mathcal{K}_2$ are served using the common stream only. Random precoding and ZFBF are respectively used for the common stream and the private streams with power allocation $P_c=O(P)$ and $P_k=O(P^{\beta}), \forall k\in\mathcal{K}_1$. It may be readily shown that the respective multiplexing gains of the common stream and each private stream are given by $1-\beta$ and $\min\{\alpha,\beta\}$, respectively. 
By further introducing a fraction $z\in[0,1]$ to specify the fraction of the rate of the common stream allocated to the users in the two subsets, we obtain that the respective sum multiplexing gains of the common stream for the users in $\mathcal{K}_1$ and $\mathcal{K}_2$ are $z(1-\beta)$ and $(1-z)(1-\beta)$, respectively.
By equally dividing the multiplexing gain of the common stream between the users in the two subsets,  the multiplexing gain of each user in $\mathcal{K}_2$ is $d_{k,2}=\frac{(1-z)(1-\beta)}{K-M}$, and 
the multiplexing gain of each user in $\mathcal{K}_1$ is $d_{k,1}=\min\{\alpha,\beta\}+\frac{z(1-\beta)}{M}$.  The MMF multiplexing gain among the users is $\max_{z}\min\{d_{k,1}, d_{k,2}\}$.  
When $\beta=\alpha$, the optimal rate allocation factor  $z^{\star}$ is obtained when $\frac{(1-z)(1-\alpha)}{K-M}= \frac{z(1-\alpha)}{M}+\alpha$. We have $z^{\star}=\frac{(1-\alpha-\alpha K+\alpha M)M}{(1-\alpha)K}$ and the optimal MMF multiplexing gain is $\frac{1+(M-1)\alpha }{K}$. As $z^{\star}\in[0,1]$, we have $1-\alpha-\alpha K+\alpha M\geq 0$. Hence, when $\alpha \leq \frac{1}{1+K-M}$, $d_{\textnormal{mmf}}^{(\textnormal{R})}=\frac{1+(M-1)\alpha }{K}$.
When $ \beta<  \alpha$ and $z=0$, the optimal power allocation $\beta^{\star}$ is obtained when $\frac{1-\beta}{K-M}=\beta$.
We have $\beta^{\star}=\frac{1}{1+K-M}$ and the optimal MMF multiplexing gain is $\frac{1}{1+K-M}$.
Hence, when $\alpha > \frac{1}{1+K-M}$, $d_{\textnormal{mmf}}^{(\textnormal{R})}=\frac{1}{1+K-M}$.

\par For $\alpha=1$, the results in Propositions \ref{Theorem_RS_DoF_alpha} and \ref{Theorem_RS_MMF_DoF_alpha} boil down to the perfect CSIT results in Propositions \ref{Theorem_RS_DoF} and \ref{Theorem_RS_MMF_DoF}, respectively.

\section{Shortcomings and Misconceptions of Multi-Antenna NOMA}\label{sec: misconceptions NOMA}

In this section, we first compare the multiplexing gains of multi-antenna NOMA to those of the MU--LP and 1-layer RS baselines. The sum and MMF multiplexing gains of multi-antenna NOMA, MU--LP, and 1-layer RS for both perfect and imperfect CSIT are summarized in Table \ref{tab: DoF compare}. The objective of this section is to identify under which conditions NOMA provides performance gains/losses over the two baselines. We then use these comparisons to reveal several misconceptions and shortcomings of multi-antenna NOMA. 

\begin{table*}
		\centering
		\caption{Comparison of sum  and MMF multiplexing gains of different strategies with  perfect and imperfect CSIT}
		\label{tab: DoF compare}
		\addtolength\tabcolsep{-2pt}
		\begin{tabular}{|C{2.1cm}|C{2.3cm}|C{3.2cm}|C{5.6cm}|}
        \hline   \vspace{0.2cm}
\textbf{Strategy}    \vspace{0.2cm}                 & \textbf{Sum/MMF Multiplexing Gain}                                               & \textbf{Perfect CSIT}                                                                                                                                                  & \textbf{Imperfect CSIT}                                                                                                                                                                                                                                                                                              \\ \hline 
\multirow{2}{*}{ \textbf{MISO NOMA}} & \vspace{0.1cm}$d_{\textnormal{s}}^{(\textnormal{N})}$  \vspace{0.1cm}          & $\min\left(M,G\right)$               & $\max\left(1,\min\left(M,G\right)\alpha\right)$              \\ \cline{2-4} 
&
\textbf{$d_{\textnormal{mmf}}^{(\textnormal{N})}$}
& \vspace{0.1cm} \begin{tabular}[c]{@{}l@{}}
$ \left\{\begin{matrix}
\frac{1}{g}, &M\geq K-g+1 \\ 
\\
 0,& M < K-g+1
\end{matrix}\right.$ \end{tabular} \vspace{0.1cm}& \vspace{0.1cm}
\begin{tabular}[c]{@{}l@{}}$\left\{\begin{matrix}
\frac{\alpha}{g},  & G>1 \:\textnormal{and}\: M\geq K-g+1 \\ 
 0,&G>1 \:\textnormal{and}\: M < K-g+1\\
 \frac{1}{K}, & \hspace{-2.55cm} G=1 
\end{matrix}\right.$\end{tabular}                                \vspace{0.1cm}                                                                                                          \\ \hline
\multirow{2}{*}{\textbf{MU--LP}}    & \vspace{0.1cm}\textbf{$d_{\textnormal{s}}^{(\textnormal{M})}$} \vspace{0.1cm}  
& $\min\left(M,K\right)$                       & $\max\left(1,\min\left(M,K\right)\alpha\right)$                \\ \cline{2-4} 
& \textbf{$d_{\textnormal{mmf}}^{(\textnormal{M})}$} &
  \vspace{0.1cm} \begin{tabular}[c]{@{}l@{}}
      $ \left\{\begin{matrix}
1, &M\geq K \\ 
\\
 0,& M < K
\end{matrix}\right.$ \end{tabular} \vspace{0.1cm}               & 
  \vspace{0.1cm} \begin{tabular}[c]{@{}l@{}}
      $ \left\{\begin{matrix}
\alpha, &M\geq K \\ 
\\
 0,& M < K
\end{matrix}\right.$ \end{tabular}     \vspace{0.1cm}                          \\ \hline
\multirow{2}{*}{\textbf{ 1-layer RS}}        & \vspace{0.1cm}\textbf{$d_{\textnormal{s}}^{(\textnormal{R})}$} \vspace{0.1cm} 
& $\min\left(M,K\right)$                       & $1+\left(\min\left(M,K\right)-1\right)\alpha$                  \\ \cline{2-4} 
& \textbf{$d_{\textnormal{mmf}}^{(\textnormal{R})}$} & 
  \vspace{0.1cm} \begin{tabular}[c]{@{}l@{}}
      $ \left\{\begin{matrix}
1, &M\geq K \\ 
\\
\frac{1}{1+K-M},& M < K
\end{matrix}\right.$ \end{tabular} \vspace{0.1cm}           
& 
  \vspace{0.1cm} \begin{tabular}[c]{@{}l@{}}
      $ \left\{\begin{matrix}
 \frac{1+(K-1)\alpha}{K},&\hspace{-2.2cm}M\geq K \\ 
\frac{1+(M-1)\alpha}{K}, & M < K \:\textnormal{and}\: \alpha\leq\frac{1}{1+K-M}\\ 
 \frac{1}{1+K-M}, &  M < K \:\textnormal{and}\: \alpha > \frac{1}{1+K-M}
\end{matrix}\right.$ \end{tabular}
\vspace{0.1cm}
  \\ \hline
\end{tabular}
\end{table*}

\subsection{NOMA vs. Baseline I (MU--LP)}\label{NOMA_vs_MULP} 
We show in the following corollaries that MISO NOMA can achieve a performance gain over MU--LP but it may also incur a performance loss, depending on the values of $M$, $K$, $G$, and $\alpha$.

\par The performance (expressed in terms of multiplexing gain) gain/loss of multi-antenna NOMA vs. MU--LP is obtained by comparing Propositions \ref{Theorem_NOMA_DoF_alpha} and \ref{prop_MULP_alpha} (for sum multiplexing gain), and Propositions \ref{Theorem_NOMA_MMF_DoF_alpha} and \ref{Theorem_MULP_MMF_DoF_alpha} (for MMF multiplexing gain), which are summarized in Corollary \ref{corollary_sumdof_imperfectcsit}, and \ref{corollary_mmfdof_imperfectcsit_G1} ($G=1$), and \ref{corollary_mmfdof_imperfectcsit_Glarger1} ($G>1$), respectively. For the MMF multiplexing gain with imperfect CSIT, we consider $G=1$ and $G>1$ in two different corollaries.

\begin{corollary}\label{corollary_sumdof_imperfectcsit} 
The sum multiplexing gain comparison between MISO NOMA  and  MU--LP is summarized in \eqref{NOMA_vs_MULP_eq}. MISO NOMA never achieves a sum multiplexing gain higher than MU--LP.
\begin{table*}
\begin{equation}
d_{\textnormal{s}}^{(\textnormal{N})}-d_{\textnormal{s}}^{(\textnormal{M})}\left\{\begin{matrix}
<0, & \textrm{if } ([\min\left(M,G\right)\alpha < 1]\cap[\min\left(M,K\right)\alpha > 1])\cup([M>G]\cap[\min\left(M,G\right)\alpha \geq 1]) \\ 
=0, & \textrm{if }  (\min\left(M,K\right)\alpha\leq 1)\cup([\min\left(M,G\right)\alpha \geq 1]\cap[M\leq G]). 
\end{matrix}\right.    \label{NOMA_vs_MULP_eq}
\end{equation}\hrulefill
\end{table*}
\end{corollary}


\par Corollary \ref{corollary_sumdof_imperfectcsit} shows that MISO NOMA can achieve a lower or the same sum multiplexing gain compared to MU--LP, but cannot outperform MU--LP.
\par If $\alpha=1$ (perfect CSIT), Corollary \ref{corollary_sumdof_imperfectcsit} boils down to $d_{\textnormal{s}}^{(\textnormal{N})}<d_{\textnormal{s}}^{(\textnormal{M})}$ whenever $M>G$, and $d_{\textnormal{s}}^{(\textnormal{N})}=d_{\textnormal{s}}^{(\textnormal{M})}$ whenever $M\leq G$. This is instrumental as it says that the slope of the sum-rate of MISO NOMA at high SNR will be strictly lower than that of MU--LP (i.e., the sum-rate of MISO NOMA will grow more slowly than that of MU--LP) whenever the number of transmit antennas is larger than the number of groups, and hence in this case, MU--LP is guaranteed to outperform MISO NOMA at high SNR. Consequently, in the massive MIMO regime where $M$ grows large, MISO NOMA would achieve a sum multiplexing gain strictly lower than MU--LP (and the role of NOMA in massive MIMO is therefore questionable as highlighted in \cite{Senel:2019}). If $G=1$ as in e.g., \cite{Hanif:2016, Choi:2015, Sun:2015, Zhang:2016}, MISO NOMA always incurs a sum multiplexing gain loss compared to MU--LP irrespective of $M$ (except in single-antenna systems when $M=1$). In other words, from a sum multiplexing gain perspective, one cannot find any multi-antenna configuration at the transmitter, i.e., any value of $M$, that would motivate the use MISO NOMA with $G=1$ compared to MU--LP. If $G=K/2$ as in \cite{Ding:2016,Choi:2017,Shin:2017,Nguyen:2017,Zeng:2017b}, MISO NOMA incurs a sum multiplexing gain loss compared to MU--LP whenever $M>K/2$. In other words, from a sum multiplexing gain perspective, the only multi-antenna deployments for which MISO NOMA with $G=K/2$ would not incur a multiplexing gain loss (but no improvement either) over MU--LP is when $M\leq K/2$. 
\par If $\alpha<1$ (imperfect CSIT), a sum multiplexing gain loss of MISO NOMA over MU--LP occurs in two different scenarios: 1) medium CSIT quality setting with $\frac{1}{\min\left(M,K\right)}<\alpha < \frac{1}{\min\left(M,G\right)}$ or 2) sufficiently large number of antennas and high CSIT quality with $M>G$ and $\alpha \geq \frac{1}{\min\left(M,G\right)}$. In other scenarios where the CSIT quality is poor $\alpha\leq \frac{1}{\min\left(M,K\right)}$ or the CSIT quality is good $\alpha \geq \frac{1}{\min\left(M,G\right)}$ but the number of transmit antennas is low $M\leq G$, MISO NOMA and MU--LP achieve the same sum multiplexing gains.



\begin{corollary}\label{corollary_mmfdof_imperfectcsit_G1} 
The MMF multiplexing gain comparison between MISO NOMA with $G=1$ and  MU--LP is summarized as follows
\begin{equation}
d_{\textnormal{mmf}}^{(\textnormal{N})}-d_{\textnormal{mmf}}^{(\textnormal{M})}\left\{\begin{matrix}
<0, & \textrm{if }(M \geq K)\cap(\alpha>\frac{1}{K}) \\ 
=0, & \textrm{if }(M\geq K)\cap(\alpha=\frac{1}{K})\\ 
>0, & \textrm{if }(M<K)\cup((M \geq K)\cap(\alpha<\frac{1}{K})).
\end{matrix}\right.    
\end{equation}
\end{corollary}

\begin{corollary}\label{corollary_mmfdof_imperfectcsit_Glarger1} 
The MMF multiplexing gain comparison between MISO NOMA with $G>1$ and  MU--LP is summarized as follows
\begin{equation}
d_{\textnormal{mmf}}^{(\textnormal{N})}-d_{\textnormal{mmf}}^{(\textnormal{M})}\left\{\begin{matrix}
<0, & \textrm{if } M \geq K \\ 
=0, & \textrm{if } M < K-g+1 \\ 
>0, & \textrm{if } K > M\geq K-g+1.
\end{matrix}\right.    
\end{equation}
\end{corollary}

\par Corollaries \ref{corollary_mmfdof_imperfectcsit_G1} and \ref{corollary_mmfdof_imperfectcsit_Glarger1} show that MISO NOMA can achieve either a higher or a lower MMF multiplexing gain compared to MU--LP, depending on the values of $M$, $G$, $K$, and $\alpha$. 
\par If $\alpha=1$ (perfect CSIT), with $G=1$ as in e.g., \cite{Hanif:2016, Choi:2015, Sun:2015, Zhang:2016}, 
$d_{\textnormal{mmf}}^{(\textnormal{N})}>d_{\textnormal{mmf}}^{(\textnormal{M})}$ whenever $M<K$, and incurs an MMF multiplexing loss otherwise ($M\geq K$). With $G=K/2$ as in \cite{Ding:2016,Choi:2017,Shin:2017,Nguyen:2017,Zeng:2017b},
$d_{\textnormal{mmf}}^{(\textnormal{N})}<d_{\textnormal{mmf}}^{(\textnormal{M})}$ whenever $M\geq K$, and $d_{\textnormal{mmf}}^{(\textnormal{N})}>d_{\textnormal{mmf}}^{(\textnormal{M})}$ whenever $K>M\geq K-1$, and $d_{\textnormal{mmf}}^{(\textnormal{N})}=d_{\textnormal{mmf}}^{(\textnormal{M})}$ whenever $M<K-1$.
In other words, from an MMF multiplexing gain perspective, the multi-antenna deployments for which MISO NOMA with $G=1$ and $G=K/2$ can outperform or achieve the same performance as MU--LP  when $M<K$.

\par If $\alpha<1$ (imperfect CSIT), we note from Corollary \ref{corollary_mmfdof_imperfectcsit_Glarger1}, that for $G>1$,  CSIT quality $\alpha$ does not affect the operational regimes where MISO NOMA outperforms/incurs a loss compared to MU--LP. This is different from $G=1$ where the condition for $d_{\textnormal{mmf}}^{(\textnormal{N})}<d_{\textnormal{mmf}}^{(\textnormal{M})}$ is a function of $\alpha$ in Corollary \ref{corollary_mmfdof_imperfectcsit_G1}. MISO NOMA incurs an MMF multiplexing loss whenever the number of antenna and the CSIT quality are sufficiently large, i.e., $M \geq K$ and $\alpha>\frac{1}{K}$.

\subsection{NOMA vs. Baseline II (RS)} \label{NOMA_vs_RS}

\par We show in the following corollaries that, for all $M$, $K$, $\alpha$, 1-layer RS (that relies on a single SIC at each user) achieves the same or higher (sum and MMF) multiplexing gains than the best of the MISO NOMA schemes (i.e., whatever $G$ and the number of SICs). In other words, 1-layer RS outperforms (multiplexing gain-wise) MISO NOMA and simultaneously requires fewer SICs (only one) than MISO NOMA. Hence, employing MISO NOMA over 1-layer RS can only cause a multiplexing gain loss and/or a complexity increase at the receiver.




 The performance loss of MISO NOMA vs. RS is obtained by comparing Propositions \ref{Theorem_NOMA_DoF_alpha} and \ref{Theorem_RS_DoF_alpha} (for the sum multiplexing gain), and Propositions \ref{Theorem_NOMA_MMF_DoF_alpha} and \ref{Theorem_RS_MMF_DoF_alpha} (for the MMF multiplexing gain), and is summarized in Corollaries \ref{corollary_sumdof_imperfectcsit_RS}, and \ref{corollary_mmfdof_imperfectcsit_RS_G1} ($G=1$), and \ref{corollary_mmfdof_imperfectcsit_RS_Glarger1} ($G>1$), respectively.

\begin{corollary}\label{corollary_sumdof_imperfectcsit_RS} 
The sum multiplexing gain comparison between MISO NOMA  and 1-layer RS  is summarized as follows
\begin{equation}
d_{\textnormal{s}}^{(\textnormal{N})}-d_{\textnormal{s}}^{(\textnormal{R})}\left\{\begin{matrix}
<0, & \textrm{if } (0<\alpha<1)\cup([\alpha>0]\cap[M>G]) \\ 
=0, & \textrm{if } (\alpha=0)\cup([\alpha=1]\cap[M\leq G]).
\end{matrix}\right.    
\end{equation}
MISO NOMA never achieves a sum multiplexing gain higher than 1-layer RS.  
\end{corollary}
\par If $\alpha=1$ (perfect CSIT), Corollary \ref{corollary_sumdof_imperfectcsit_RS} boils down to  $d_{\textnormal{s}}^{(\textnormal{N})}<d_{\textnormal{s}}^{(\textnormal{R})}$, whenever $M>G$, and  $d_{\textnormal{s}}^{(\textnormal{N})}=d_{\textnormal{s}}^{(\textnormal{R})}$ whenever $M\leq G$.

\begin{corollary}\label{corollary_mmfdof_imperfectcsit_RS_G1} 
The MMF multiplexing gain comparison between MISO NOMA with $G=1$  and 1-layer RS is summarized as follows
\begin{equation}
d_{\textnormal{mmf}}^{(\textnormal{N})}-d_{\textnormal{mmf}}^{(\textnormal{R})}\left\{\begin{matrix}
<0, & \textrm{if } (\alpha>0)\cap(M>1) \\ 
=0, & \textrm{if } (\alpha=0)\cup(M = 1).
\end{matrix}\right.    
\end{equation}
MISO NOMA with $G=1$ never achieves an MMF multiplexing gain higher than 1-layer RS. 
\end{corollary}

\begin{corollary}\label{corollary_mmfdof_imperfectcsit_RS_Glarger1} 
The MMF multiplexing gain comparison between MISO NOMA with $G>1$  and 1-layer RS is summarized in \eqref{NOMA_vs_RS_eq}. MISO NOMA with $G>1$ never achieves an MMF multiplexing gain larger than 1-layer RS.
\begin{table*}
\begin{equation}\label{NOMA_vs_RS_eq}
d_{\textnormal{mmf}}^{(\textnormal{N})}-d_{\textnormal{mmf}}^{(\textnormal{R})}\left\{\begin{matrix}
<0, & \textrm{if } (M \neq K-g+1)\cup([]M = K-g+1]\cap[\alpha<1]) \\ 
=0, & \textrm{if } (M = K-g+1)\cap(\alpha=1).
\end{matrix}\right.    
\end{equation}\hrulefill
\end{table*}
\end{corollary}
\par If $\alpha=1$ (perfect CSIT), Corollaries \ref{corollary_mmfdof_imperfectcsit_RS_G1} and \ref{corollary_mmfdof_imperfectcsit_RS_Glarger1} simply boil down to $d_{\textnormal{mmf}}^{(\textnormal{N})}<d_{\textnormal{mmf}}^{(\textnormal{R})}$, whenever $M \neq K-g+1$, and $d_{\textnormal{mmf}}^{(\textnormal{N})}=d_{\textnormal{mmf}}^{(\textnormal{R})}$, whenever $M = K-g+1$. 
\par Recalling again from \cite{Piovano:2017,Joudeh:2019,Hao:2017,Davoodi:2020} that RS achieves the optimal multiplexing gain region in multi-antenna BC with imperfect CSIT, and multi-antenna NOMA (and MU--LP/MU-MIMO) will therefore always incur a multiplexing gain loss compared to RS.

\subsection{Misconceptions of Multi-Antenna NOMA}
\label{misconceptions}
\par The comparisons with the MU--LP and 1-layer RS baselines reveal that depending on the particular setting NOMA may incur a multiplexing gain loss at the additional expense of an increased receiver complexity, as detailed in the following.

\par \textit{First}, NOMA is an inefficient strategy to exploit the spatial dimensions. This issue could already be observed from the two-user MISO case with perfect CSIT, where NOMA limits the sum multiplexing gain to one, same as OMA, which is only half of the sum multiplexing gain obtained with MU--LP. Moreover, even when considering a fair metric such as MMF, NOMA limits the MMF multiplexing gain to $\frac{1}{2}$, which is again only half of the MMF multiplexing gain obtained by MU--LP. 
\par In the general $K$-user case, it is clear from Corollaries \ref{corollary_sumdof_imperfectcsit} and \ref{corollary_sumdof_imperfectcsit_RS} that NOMA incurs a loss in sum multiplexing gain in most scenarios, and the best NOMA can achieve is the same sum multiplexing gain as the baselines in some specific configurations. NOMA with $G=1$ achieves $d_{\textnormal{s}}^{(\textnormal{N})}=1$ irrespectively of the number of transmit antennas $M$, i.e., it achieves the same sum multiplexing gain as OMA and the same as a single-antenna transmitter (hence, wasting the transmit antenna array). NOMA with $G=K/2$ achieves $d_{\textnormal{s}}^{(\textnormal{N})}=\min\left(M,K/2\right)$ with $\alpha=1$. On the other hand, MU--LP and 1-layer RS achieve the full sum multiplexing gain $d_{\textnormal{s}}^{(\textnormal{M})}=\min\left(M,K\right)$ with $\alpha=1$.

\par Considering the MMF multiplexing gain of the general $K$-user case, the situation appears to be better for NOMA. Assuming $\alpha=1$, from Corollaries \ref{corollary_mmfdof_imperfectcsit_G1} and \ref{corollary_mmfdof_imperfectcsit_Glarger1}, we observe that NOMA incurs a loss compared to MU--LP in the underloaded regime $M \geq K$ but outperforms MU--LP in the overloaded regime. In particular, NOMA with $G=1$ achieves a higher MMF multiplexing gain than NOMA wtih $G=K/2$ and MU--LP whenever $M<K-1$. Hence, though the receiver complexity increases of NOMA does not pay off in the underloaded regime, it appears to pay off in the overloaded regime (since $G=1$ with more SICs outperforms $G=K/2$ with fewer SICs). Nevertheless, the MMF multiplexing gain of NOMA with $G=1$ is independent of $M$, suggesting again that the spatial dimensions are not properly exploited. This can indeed be seen from Corollary \ref{corollary_mmfdof_imperfectcsit_RS_G1} where NOMA is consistently outperformed by 1-layer RS, i.e., the increase in MMF multiplexing gain attained by NOMA ($G=1$) over MU--LP is actually marginal in light of the complexity increase, and is much lower than what can be achieved by 1-layer RS with just a single SIC operation. In other words, while NOMA has some merits over MU--LP in the overloaded regime, NOMA makes an inefficient use of the multiple antennas, and fails to boost the MMF multiplexing gain compared to the 1-layer RS baseline. 

\par We note that the above observations hold for both the perfect and imperfect CSIT settings. Nevertheless, it is interesting to stress that the sensitivity to the CSIT quality $\alpha$ differs largely between MU--LP, NOMA with $G>1$, NOMA with $G=1$, and 1-layer RS. Indeed the sum and MMF multiplexing gains of MU--LP, NOMA with $G>1$, and 1-layer RS decay as $\alpha$ decreases, while the multiplexing gains of NOMA with $G=1$ are not affected by $\alpha$. This can be interpreted in two different ways. On the one hand, this implies that NOMA with $G=1$ is inherently robust to CSIT imperfections since the multiplexing gains are unchanged. On the other hand, this means that NOMA with $G=1$ is unable to exploit the available CSIT since the resulting multiplexing gain is the same as in the absence of CSIT ($\alpha=0$). One can indeed see from the above Propositions and Corollaries that the sum and MMF multiplexing gains for 1-layer RS with imperfect CSIT are clearly larger than those of MU--LP and NOMA. In other words, NOMA and MU--LP are inefficient in fully exploiting the available CSIT in multi-antenna settings.

\par We conclude from the theoretical results and above discussions that NOMA fails to efficiently exploit the multiplexing gain of the multi-antenna BC and is an inefficient strategy to exploit the spatial dimensions and the available CSIT, especially compared to the 1-layer RS baseline.  
\textit{The first misconception behind NOMA is to believe that because NOMA is capacity achieving in the single-antenna BC, NOMA is an efficient strategy for multi-antenna settings. As a consequence, the single-antenna NOMA principle has been applied to multi-antenna settings without recognizing that such a strategy would waste the primary benefit of using multiple antennas, namely the capability of transmitting multiple interference-free streams.} In contrast to NOMA, other non-orthogonal transmission strategies such as 1-layer RS do not lead to any sum multiplexing gain loss. On the contrary, 1-layer RS achieves the information theoretic optimal sum multiplexing gain in both perfect and imperfect CSIT scenarios (and therefore has the capability of transmitting the optimal number of interference-free streams). 1-layer RS also achieves higher MMF multiplexing gains than NOMA and MU--LP. 

\par \textit{Second}, the multiplexing gain loss of NOMA is encountered despite the increased receiver complexity. In the two-user MISO BC with perfect CSIT, MU--LP does not require any SIC receiver to achieve the optimal sum multiplexing gain of two (assuming $M>1$) and a MMF multiplexing gain of one, while NOMA requires one SIC and only provides half the (sum and MMF) multiplexing gains of MU--LP. 
This is surprising since one would expect a performance gain from an increased architecture complexity. Here instead, NOMA brings together a complexity increase at the receivers and a (sum and MMF) multiplexing gain loss compared to MU--LP, therefore highlighting that the SIC receiver is inefficiently exploited.
\par This inefficient use of SIC in NOMA also persists in the general $K$-user scenario. Recall that NOMA with $G$ groups requires $g-1$ layers of SIC at the receivers. Among the two popular NOMA architectures $G=1$ and $G=K/2$, the former requires an even higher number of SIC layers than the latter (namely $K-1$ for $G=1$ and 1 for $G=K/2$) and has an even lower sum multiplexing gain ($d_{\textnormal{s}}^{(\textnormal{N})}=1$ for $G=1$ and $d_{\textnormal{s}}^{(\textnormal{N})}=\min\left(M,K/2\right)$ for $G=K/2$ with $\alpha=1$). On the other hand, MU--LP achieves the full sum multiplexing $d_{\textnormal{s}}^{(\textnormal{M})}=\min\left(M,K\right)$ with $\alpha=1$ without any need for SIC. This highlights the inefficient (and detrimental) use of SIC receivers in NOMA: the higher the number of SICs, the lower the sum multiplexing gain! 
\par Comparing to the 1-layer RS baseline further highlights the inefficient use of SIC in NOMA. We note that 1-layer RS causes a complexity increase at the receivers (due to the one SIC needed) but also an increase in the (sum and MMF) multiplexing gains compared to MU--LP (i.e., it is easy to see from Propositions \ref{prop_MULP_alpha}, \ref{Theorem_MULP_MMF_DoF_alpha}, \ref{Theorem_RS_DoF_alpha}, and \ref{Theorem_RS_MMF_DoF_alpha} that the sum and MMF multiplexing gains with RS are always either identical to or higher than those with MU--LP). Hence, in contrast to NOMA, the SIC in 1-layer RS is beneficial since it boosts the (sum and MMF) multiplexing gains and therefore introduces a performance gain compared to (or at least maintains the same performance as) MU--LP. Actually, 1-layer RS achieves the information theoretic optimal sum multiplexing gain for imperfect CSIT, and does so with a single SIC per user. This shows that to achieve the information theoretic optimality, it is sufficient to use a single SIC per user\footnote{Actually, though the analysis  here is limited to 1-layer RS, all RS schemes (from 1-layer to generalized RS) in \cite{Mao:2017} guarantee the optimal sum multiplexing gain and a higher MMF multiplexing gain than MU--LP and NOMA, and provide an improved rate performance as the number of SIC increases \cite{Mao:2017,Dai:2016,Mao:2019}.}.
This is in contrast to NOMA whose sum multiplexing gain is far from optimal and for which the sum multiplexing gain decreases as the number of SICs increases. The inefficient use of SIC in NOMA is also obvious from the MMF multiplexing gain. Indeed, from Proposition \ref{Theorem_NOMA_MMF_DoF} and \ref{Theorem_RS_MMF_DoF} and Corollary \ref{corollary_mmfdof_imperfectcsit_RS_G1}, the single SIC in 1-layer RS achieves a much larger MMF multiplexing gain than the $K-1$ layers of SIC needed for NOMA with $G=1$. This again illustrates how inefficient the use of SIC in NOMA often is. It also shows that there exist non-orthogonal (RS-based) transmission strategies with better performance and lower receiver complexity requiring just a single SIC per user. 
\par We  conclude from the theoretical analysis and above discussion that NOMA often does not make efficient use of the SIC receivers compared to the considered baselines. \textit{The second misconception regarding multi-antenna NOMA is to believe that adopting SIC receivers always boosts the rate since the interference is fully cancelled at the receiver.} Considering the two-user toy example, and comparing \eqref{R1_NOMA} and \eqref{R_MUMIMO}, the interference power term $\left|\mathbf{h}_1^H \mathbf{p}_{2}\right|^2$ appearing in the SINR of user-1 in the MU--LP rate has indeed disappeared in NOMA thanks to the SIC receiver, such that $R_{\textnormal{M},1}\leq R_{\textnormal{N},1}$. However, this comes at the cost of a reduced rate for user-2  since $R_{\textnormal{N},2}=\min\left(\log_2\left(1+A\right),R_{\textnormal{M},2}\right)\leq R_{\textnormal{M},2}$. In other words, for a given pair of precoders $\mathbf{p}_1$ and $\mathbf{p}_2$, NOMA increases the rate (or maintains the same rate) of user-1 but decreases the rate (or maintains the same rate) of user-2  compared to MU--LP. 

\par \textit{Third}, reflecting on the above two misconceptions, the NOMA design philosophy does not leverage the extensive research in multi-user MIMO, which has been fundamental to 4G and 5G in achieving the optimal sum multiplexing gain of multi-antenna BC with perfect CSIT and low-complexity transmitter and receiver architectures. 
\textit{The third misconception behind multi-antenna NOMA is to believe that, since NOMA is routinely compared to OMA in SISO BC, it is also sufficient to compare NOMA to OMA in multi-antenna settings to demonstrate its merits}.  In fact, the Corollaries in Sections \ref{NOMA_vs_MULP} and \ref{NOMA_vs_RS} show that NOMA is far from being an efficient strategy if NOMA is compared to alternative baselines. Unfortunately, simply comparing with OMA has led the NOMA literature to the misleading conclusion that multi-antenna NOMA is an efficient strategy. It should therefore be stressed that comparing NOMA to OMA does not demonstrate the merits of NOMA in multi-antenna settings and most importantly, the baseline for any multi-antenna NOMA design, optimization, and evaluation should be MU--LP and RS, not simply OMA\footnote{Recall also 4G and 5G are both based on MU--LP, and not simply on OMA.}! In contrast to MISO NOMA, the gain of 1-layer RS over MU--LP is guaranteed, i.e., the rate of 1-layer RS is equal to or higher than that of MU--LP, since MU--LP is a particular instance of RS when no power is allocated to the common stream. 

\par \textit{Fourth}, the SISO BC is naturally overloaded (more users than the number of transmit antennas, namely one), and NOMA was therefore concluded to be suitable for overloaded scenarios. \textit{The fourth misconception behind multi-antenna NOMA is to believe that MISO NOMA is an efficient strategy for  overloaded regimes, namely whenever $K > M$.} The Corollaries in Subsections \ref{NOMA_vs_MULP} and \ref{NOMA_vs_RS} nevertheless expose that this is incorrect. It is clear that NOMA incurs a sum multiplexing gain erosion compared to MU--LP and 1-layer RS whenever $M>G$. Such a loss can occur also in the overloaded regime, namely whenever we have $K > M > G$. Moreover, NOMA incurs an MMF multiplexing gain loss compared to 1-layer RS whenever $M\neq K-g+1$. Here again, such a loss occurs also in the overloaded regime. In contrast to NOMA (and MU--LP), 1-layer RS is an efficient strategy for both the underloaded and overloaded regimes. Though NOMA with $G=1$ was shown in Proposition \ref{Theorem_NOMA_MMF_DoF} to achieve a non-vanishing MMF multiplexing gain of $1/K$ in the overloaded regime, this MMF multiplexing gain is considerably smaller than that of 1-layer RS, therefore highlighting the inefficiency of NOMA in the overloaded regime. In particular, we note that the MMF multiplexing gain of 1-layer RS increases with $M$ in contrast to that of NOMA with $G=1$ which is constant regardless of $M$.

\begin{table}
\caption{Sum multiplexing gain with $K=6$ - perfect CSIT.}
\centering
\begin{tabular}{|p{0.6cm}|p{0.8cm}|p{1.45cm}|p{1.45cm}|p{1.6cm}|}
\hline $M$ & regime & $d_{\textnormal{s}}^{(\textnormal{N})}$\! ($G\!=\!1$)	& $d_{\textnormal{s}}^{(\textnormal{N})}$\! ($G\!=\!3$) &	$d_{\textnormal{s}}^{(\textnormal{M})}, d_{\textnormal{s}}^{(\star)}, d_{\textnormal{s}}^{(\textnormal{R})}$ \\
\hline
\hline 1 & O	& 1 & 1 & 1 \\
\hline 2 & O & \textcolor{red}{1} & 2 & 2 \\
\hline 3 & O	& \textcolor{red}{1} & 3	& 3 \\
\hline 4 & O	& \textcolor{red}{1} & \textcolor{red}{3}& 4 \\
\hline 5 & O & \textcolor{red}{1} & \textcolor{red}{3}& 5 \\
\hline $\geq 6$ & U	& \textcolor{red}{1} & \textcolor{red}{3}& 6 \\
\hline
\end{tabular}\\
\footnotesize{\vspace{1.0mm}O: Overloaded ($K>M$), U: Underloaded ($K\leq M$)}
\label{multiplexing_example}
\end{table}

\begin{table}
\caption{MMF multiplexing gain with $K\!=\!6$ \!-\! perfect CSIT.}
\centering
\begin{tabular}{|p{0.6cm}|p{0.8cm}|p{1.5cm}|p{1.5cm}|p{0.7cm}|p{0.7cm}|}
\hline $M$ & regime & $d_{\textnormal{mmf}}^{(\textnormal{N})}$\! ($G\!=\!1$)	& $d_{\textnormal{mmf}}^{(\textnormal{N})}$\! ($G\!=\!3$) &	$d_{\textnormal{mmf}}^{(\textnormal{M})}$ & $d_{\textnormal{mmf}}^{(\textnormal{R})}$ \\
\hline
\hline 1 & O	& \sfrac{1}{6} & \textcolor{red}{0} & \textcolor{red}{0} & \sfrac{1}{6} \\
\hline 2 & O & \textcolor{red}{\sfrac{1}{6}} & \textcolor{red}{0} & \textcolor{red}{0} & \sfrac{1}{5} \\
\hline 3 & O	& \textcolor{red}{\sfrac{1}{6}} & \textcolor{red}{0}	& \textcolor{red}{0} & \sfrac{1}{4} \\
\hline 4 & O	& \textcolor{red}{\sfrac{1}{6}} & \textcolor{red}{0} & \textcolor{red}{0} & \sfrac{1}{3} \\
\hline 5 & O & \textcolor{red}{\sfrac{1}{6}} & \sfrac{1}{2} & \textcolor{red}{0} & \sfrac{1}{2} \\
\hline $\geq 6$ & U	& \textcolor{red}{\sfrac{1}{6}} & \textcolor{red}{\sfrac{1}{2}} & 1 & 1 \\
\hline
\end{tabular}\\
\footnotesize{\vspace{1.0mm}O: Overloaded ($K>M$), U: Underloaded ($K\leq M$)}
\label{mmf_multiplexing_example}
\end{table}
\subsection{Illustration of the Misconceptions with an Example}\label{example_sumdof}
To illustrate the above discussion and make the statements more explicit based on numbers, we consider a MISO BC with $K=6$, and compare in Table \ref{multiplexing_example} the sum multiplexing gains $d_{\textnormal{s}}^{(\textnormal{N})}$ of NOMA with $G=1$ and $G=3$ and the sum multiplexing gain of MU--LP $d_{\textnormal{s}}^{(\textnormal{M})}$ and 1-layer RS $d_{\textnormal{s}}^{(\textnormal{R})}$ (recall that $d_{\textnormal{s}}^{(\textnormal{M})}=d_{\textnormal{s}}^{(\textnormal{R})}=d_{\textnormal{s}}^{(\star)}$) as a function of $M$ for perfect CSIT. We observe that NOMA incurs a sum multiplexing gain reduction (highlighted in red in Table \ref{multiplexing_example}) in the underloaded regime but also in the overloaded regime depending on the values of $M$ and $G$. Specifically, in this example with $K=6$, $G=1$ incurs a sum multiplexing erosion compared to MU--LP and 1-layer RS whenever $M\geq 2$ and $G=3$ whenever $M\geq 4$. This shows that in an overloaded regime associated with $M<K$, although $M$ is the limiting factor of the sum multiplexing gain in MU--LP and 1-layer RS, $\min\left(M,G\right)$ is the limiting factor in NOMA. Morever, Table \ref{multiplexing_example} clearly illustrates that the higher the number of SICs in NOMA, the lower the sum multiplexing gain. NOMA with $G=1$ requires 5 layers of SIC to achieve a multiplexing gain $d_{\textnormal{s}}^{(\textnormal{N})}=1$, NOMA with $G=3$ requires 1 layer of SIC and achieves at most $d_{\textnormal{s}}^{(\textnormal{N})}=3$. On the other hand, MU--LP does not require any SIC and achieves the optimal sum multiplexing gain $d_{\textnormal{s}}^{(\star)}$ (that can be as high as 6). 1-layer RS achieves the same (and optimal) sum multiplexing gain as MU--LP. 
\par Table \ref{mmf_multiplexing_example} highlights the MMF multiplexing gains of NOMA, MU--LP, and 1-layer RS for $K=6$ with perfect CSIT and stresses the significant benefit of 1-layer RS over NOMA and MU--LP. The entries highlighted in red relate to configurations for which 1-layer RS provides a multiplexing gain strictly higher than that of NOMA and MU--LP. Recall that 1-layer RS provides those multiplexing gains over NOMA and MU--LP with a single SIC per user! 

\begin{figure}[!t]
	\centering
	\includegraphics[width=0.9\columnwidth]{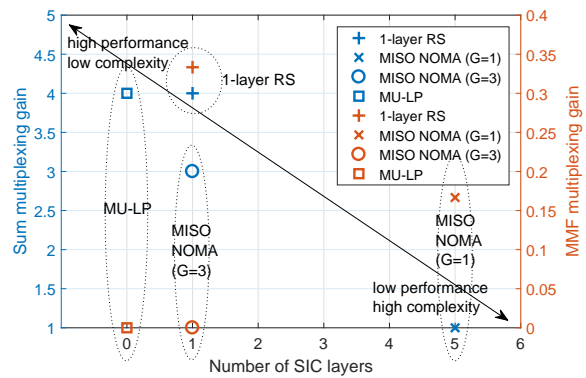}
	\caption{Multiplexing gains with single-antenna receivers and perfect CSIT vs. number of SIC layers for $M=4$, $K=6$.}
	\label{DoF_vs_SIC}
\end{figure}

\begin{figure}[!t]
	\centering
	\includegraphics[width=0.9\columnwidth]{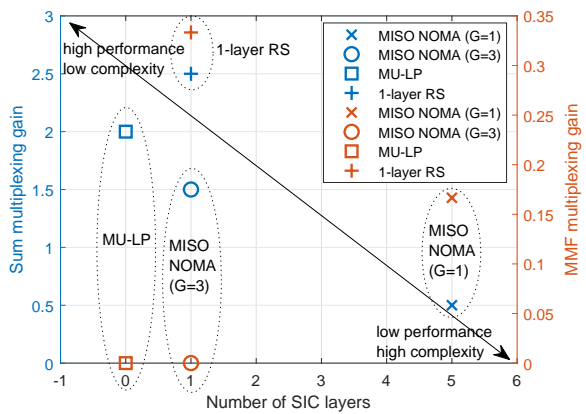}
	\caption{Multiplexing gains with single-antenna receivers and imperfect CSIT vs. number of SIC layers for $M=4$, $K=6$, $\alpha=0.5$.}
	\label{DoF_vs_SIC_imperfect}
\end{figure}

\par In Fig. \ref{DoF_vs_SIC}, we further illustrate the tradeoff between the multiplexing gains and the number of SIC layers for $M=4$, $K=6$ and perfect CSIT. We observe that 1-layer RS enables higher performance and lower receiver complexity compared to NOMA, stressing that the non-orthogonal transmission enabled by RS is much more efficient than NOMA. We see that  NOMA with different $G$ is suited for very different settings in this $M=4$, $K=6$ configuration, namely NOMA with $G=3$ performs better in terms of sum multiplexing gain, whereas NOMA with $G=1$ achieves a higher  MMF multiplexing gain. The baseline 1-layer RS achieves a higher performance for both metrics and entails a lower receiver complexity. 
\par Though the above example was provided for perfect CSIT ($\alpha=1$), it is easy to calculate from the above propositions the multiplexing gains for the imperfect CSIT setting for a given CSIT quality $\alpha$. For imperfect CSIT, the strict superiority of 1-layer RS over MU--LP and NOMA will become much more apparent, as illustrated in Fig. \ref{DoF_vs_SIC_imperfect} for $\alpha=0.5$.

\subsection{Shortcomings of Multi-Antenna NOMA}
\label{shortcomings}
\par The previous subsections have highlighted that comparing multi-antenna NOMA to MU--LP and 1-layer RS, instead of OMA, provides a completely different picture of the actual merits of multi-antenna NOMA. In view of the previous results highlighting the waste of multiplexing gain and the inefficient use of the SIC receivers by multi-antenna NOMA, we can ask ourselves multiple questions, which help to pinpoint the shortcomings and limitations of the multi-antenna NOMA design philosophy.
\par The \textit{first question} is ``\textit{What prevents multi-antenna NOMA from reaping the multiplexing gain of the system?}'' The answer lies in \eqref{eq_MAC}, and similarly in \eqref{sumrate_1g}, \eqref{sumrate_2g}, and \eqref{sumrate_3g}. Equation \eqref{eq_MAC} can be interpreted as the sum-rate of a two-user MAC with a single antenna receiver. Indeed, in \eqref{eq_MAC}, user-1 acts as the receiver of a two-user MAC whose effective SISO channels of both links are given by $\mathbf{h}_1^H \mathbf{p}_{2}$ and $\mathbf{h}_1^H \mathbf{p}_{1}$. Similarly, in \eqref{sumrate_1g}, user-1 acts as the receiver of a $g$-user MAC whose effective SISO channels of the $g$ links are given by $\mathbf{h}_1^H \mathbf{p}_{k}$ for $k=1,\ldots,g$. Such a MAC is well known to have a sum multiplexing gain of one \cite{Clerckx:2013,Tse:2005}. The multiplexing gain \textit{losses} compared to the MU--LP and 1-layer RS baselines therefore \textit{come from forcing one user to fully decode all streams in a group}, i.e., its intended stream and the co-scheduled streams in the group. This is radically different from MU--LP where streams are encoded independently and each receiver decodes its intended stream treating any residual interference as noise. By contrast, in 1-layer RS, no user is forced to fully decode the co-scheduled streams since all private streams are encoded independently and each receiver decodes its intended private stream treating any residual interference from other private streams as noise.

\par The \textit{second question} is ``\textit{Does an increase in the number of SICs always come with a reduction in the sum multiplexing gain?}'' The answer is clearly no. This anomaly is deeply rooted in the way MISO NOMA was developed by applying single-antenna NOMA principle to multi-antenna settings. The proof of Proposition \ref{Theorem_NOMA_DoF} indeed tells us that \textit{the fundamental principle of NOMA consisting in forcing one user in each group to fully decode the messages of $g-1$ co-scheduled users is an inefficient design in multi-antenna settings} that leads to a sum multiplexing gain reduction in each group. 

\par The \textit{third question} is ``\textit{Are non-orthogonal transmission strategies inefficient for multi-antenna settings?}'' The answer is no. As we have seen, there exist frameworks of non-orthogonal transmission strategies also relying on SIC, such as RS, that do not incur the limitations of multi-antenna NOMA and make efficient use of the non-orthogonality and SIC receivers in multi-antenna settings. The key for the design of such non-orthogonal strategies is not to fall into the trap of blindly  applying the SISO NOMA principle to multi-antenna settings, and therefore constraining the strategy to always fully decode the message of other users. Non-orthogonal transmission strategies and multiple access need to be re-thought for multi-antenna settings and one such strategy is based on the multi-antenna Rate-Splitting (RS) and Rate-Splitting Multiple Access (RSMA) literature for multi-antenna BC . 

\par
\begin{figure}[!t]
	\centering
	\includegraphics[width=0.8\columnwidth]{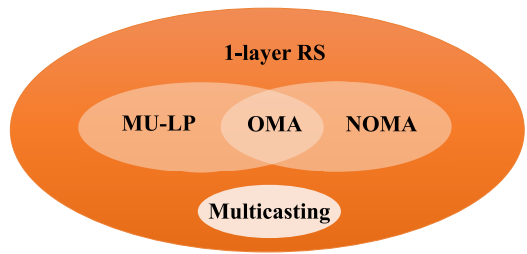}
	\caption{The relationship between existing strategies and 1-layer RS in two-user case. Each set  illustrates the optimization space of the corresponding communication strategy. The optimization space of 1-layer RS is larger such that MU--LP, NOMA and multicasting are just subsets.}
	\label{fig_sys_relation1}
\end{figure}

\begin{figure}[!t]
	\centering
	\includegraphics[width=0.8\columnwidth]{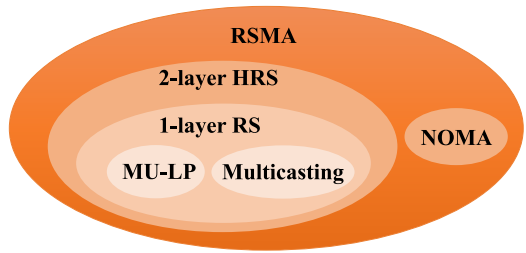}
	\caption{The relationship among existing strategies and the $K$-user RSMA framework.}
	\label{fig_sys_relation2}
\end{figure}

 The \textit{fourth question} is ``\textit{Since NOMA and RS both rely on SIC, is there any relationship between NOMA and RS?}'' The answer is yes in a two-user setting, but not necessarily in the general $K$-user case. In the two-user case,  1-layer RS is a superset of MU--LP, NOMA and multicasting, i.e., MU--LP, NOMA and multicasting are particular instances of 1-layer RS, as shown in \cite{Clerckx:2020} and in Table \ref{fig_mapping} and Fig. \ref{fig_sys_relation1}. Indeed, MU--LP is obtained as a special case from 1-layer RS by allocating no power to the common stream ($P_{\mathrm{c}}\!=\!0$) such that $W_k$ is encoded directly into $s_k$. No interference is decoded at the receiver using the common message, and the interference between $s_1$ and $s_2$ is fully treated as noise. NOMA is obtained by encoding $W_2$ entirely into $s_{\mathrm{c}}$ (i.e., $W_{\mathrm{c}}\!=\!W_2$) and $W_1$ into $s_{1}$, and turning off $s_2$ ($P_{2}\!=\!0$)\footnote{To better relate to the system model in Section \ref{twouser}, note that NOMA also has a common message/stream, though commonly not denoted using such terminology. Indeed, the stream of the weakest user, namely $s_2$ in Section \ref{twouser}, is a common stream since it is decoded by both users. $s_2$ in Section \ref{twouser} carries information, namely $W_2$, intended for user-2 but is decoded by both user-1 and 2.}. In this way, user-1 fully decodes the interference created by the message of user-2. OMA is a sub-strategy of MU--LP and NOMA, which is encountered when only user-1 (with the stronger channel gain) is scheduled ($P_{\mathrm{c}}\!=\!0, P_2\!=\!0$). Multicasting is obtained when both $W_1$ and $W_2$ are entirely encoded into $s_{\mathrm{c}}$. In the $K$-user case, 1-layer RS is a superset of MU--LP since by turning off (i.e., allocating no power to) the common stream,  1-layer RS boils down to MU--LP. On the other hand, 1-layer RS is \textit{not} a superset of NOMA. 1-layer RS and NOMA are particular instances/schemes of the RSMA framework based on the generalized RS relying on multiple layers of SIC at each receiver \cite{Mao:2017,Dai:2016,Mao:2019,Mao:2020_TCOM}\footnote{2-layer hierarchical RS (HRS) in  Fig. \ref{fig_sys_relation2} is proposed in \cite{Dai:2016} for massive MIMO. Besides one common message decoded by all users as in 1-layer RS, 2-layer HRS relies on multiple group-specific common messages being decoded by different groups of users to further manage inter-user interference. RSMA is a generalized framework that embraces both 1-layer RS and 2-layer HRS as subschemes \cite{Mao:2017}.}, as illustrated in Fig. \ref{fig_sys_relation2}. 

\begin{table}
	\caption{Messages-to-streams\! mapping\! in\! two-user\! MISO\! BC.}
	\centering
\begin{tabular}{|p{.2\textwidth}|p{.6\textwidth}|p{2.3cm}|p{2.5cm}|}
\hline
\multicolumn{1}{|c|}{}             & \multicolumn{1}{c|}{$s_1$}     & \multicolumn{1}{c|}{$s_2$}     & \multicolumn{1}{c|}{$s_c$}             \\ \hline
\multicolumn{1}{|c|}{ MU--LP}         & \multicolumn{1}{c|}{\color{blue} $\,\,\quad \quad W_1\quad\quad\,\,$}     & \multicolumn{1}{c|}{\color{blue} $W_2$}     & \multicolumn{1}{c|}{\color{red} --}                \\ \hline
\multicolumn{1}{|c|}{ NOMA}         & \multicolumn{1}{c|}{\color{blue} $W_1$}     & \multicolumn{1}{c|}{\color{blue} --}        & \multicolumn{1}{c|}{\color{red} $W_2$}             \\ \hline
\multicolumn{1}{|c|}{ OMA}          & \multicolumn{1}{c|}{\color{blue} $W_1$}     & \multicolumn{1}{c|}{\color{blue} --}        & \multicolumn{1}{c|}{\color{red} --}                \\ \hline
\multicolumn{1}{|c|}{Multicasting} & \multicolumn{1}{c|}{\color{blue} --}        & \multicolumn{1}{c|}{\color{blue} --}        & \multicolumn{1}{c|}{\color{red} $W_1,W_2$}         \\ \hline
\multicolumn{1}{|c|}{RS}           & \multicolumn{1}{c|}{\color{blue} $W_{\mathrm{p},1}$} & \multicolumn{1}{c|}{\color{blue} $W_{\mathrm{p},2}$} & \multicolumn{1}{c|}{\color{red} $W_{\mathrm{c},1},W_{\mathrm{c},2}$} \\ \hline
 \multicolumn{1}{c}{}  & \multicolumn{2}{c}{\color{blue} decoded by its intended user and}            &  \multicolumn{1}{c}{\color{red} decoded by}                         \\
 \multicolumn{1}{c}{}  &\multicolumn{2}{c}{\color{blue} treated as noise by the other user}          & \multicolumn{1}{c}{ \color{red} both users   }                         
\end{tabular}
	\label{fig_mapping}
\end{table}



\par The \textit{fifth question} is ``\textit{How does 1-layer RS achieve simultaneously higher multiplexing gains and a lower receiver complexity than NOMA?}'' In view of the previous sections, the key is to build non-orthogonal transmission strategies upon MU--LP (and therefore SDMA/multi-user MIMO) such that the performance benefits (including sum multiplexing gain) of MU--LP are guaranteed but extra performance (e.g., in MMF multiplexing gain) is observed by the use of SIC receivers. Indeed, a performance gain over MU--LP should be expected from a more complex receiver architecture in multi-antenna BC. To do so, one should enable the flexibility at the transmitter to encode messages such that parts of them can be decoded by all users using SIC while the remaining parts are decoded by their intended receivers and treated as noise by non-intended receivers. Hence, we provide the flexibility to partially decode interference and partially treat the remaining interference as noise. This contrasts with MU--LP where interference is always treated as noise, and with NOMA where interference is fully decoded. This flexibility is achieved by extending the concept of RS, originally developed in \cite{Han:1981} for the two-user single-antenna interference channel, to the multi-antenna BC. To manage multi-user interference by partially decoding the interference and treating the remaining interference as noise, RS facilitates a complete message-to-streams mapping flexibility for each user to have part of its message transmitted in the common stream and the remaining part in one of the $K$ private streams. By adjusting the power levels of the common and private streams, one can adjust the amount of interference caused to the private streams such that its level is weak enough to be treated as noise. This contrasts with MU--LP where the communication strategy is fundamentally constrained such that the messages are  mapped to private streams only (i.e., there is no common stream, and multi-user interference between private streams is treated as noise even when its level is not weak enough to be treated as noise), and with NOMA where the constraint is that the entire message of one of the users is mapped onto a common stream (e.g., $W_2$ mapped to $s_2$ decoded by both user-1 and 2 in Section \ref{twouser}). These constraints imposed by MU--LP and NOMA are well illustrated by the message-to-stream mapping in Table \ref{fig_mapping}.

\par A consequence of the above flexibility is that by decreasing the amount of power allocated to the common stream, $K$-user 1-layer RS progressively converges to $K$-user MU--LP and in the limit where no power is allocated to the common stream, $K$-user 1-layer RS swiftly boils down to $K$-user MU--LP. Hence, 1-layer RS really builds upon MU--LP and MU--LP is a subscheme of 1-layer RS, which provides guarantee to 1-layer RS that its rate and multiplexing gains are always teh same or better than those of MU--LP. This is completely different from MISO NOMA. MISO NOMA does not build upon MU--LP. With $G$ groups, $K$-user MISO NOMA can boil down to $G$-user MU--LP by turning off the power to the weaker users in each group, but $K$-user MISO NOMA can mathematically never boil down to $K$-user MU--LP (recall footnote \ref{footnote_NOMA_MULP}). The rate/multiplexing gains of $K$-user MISO NOMA can therefore be worse than that of $K$-user MU--LP. 

\par Another interpretation arises by noting that MU--LP (and other form of multi-user MIMO), as one extreme, can be viewed as a full transmit-side interference management strategy. On the other extreme, NOMA can be seen as a full receiver-side interference cancellation strategy. In between stands RS that can be viewed as a smart combination of transmit-side and receive-side interference management/cancellation strategies where the contribution of the common stream is adjusted according to the level of interference that can be canceled by the receiver.
\par Consequently, RS is an enabler of a general class of communication strategies and can cover a wider set of communication strategies than SDMA and NOMA, which leads to significant multiplexing gain and complexity reduction benefits.

\par The \textit{sixth question} is ``\textit{Can we use other types of receivers than SIC for NOMA and RS and would the multiplexing gains be improved?}'' We can indeed use other types of receivers but the multiplexing gains will not improve. Instead of using stream-by-stream SIC, we can use any other joint (Maximum Likelihood) decoder. Hence a strong user in NOMA could use a joint decoder to decode its intended stream jointly with all other streams intended to its co-scheduled users in the same group. The multiplexing gains would not improve since the strong user would still act as the receiver of an effective MAC (as discussed in relationship with \eqref{eq_MAC}, \eqref{sumrate_1g}, \eqref{sumrate_2g}, and \eqref{sumrate_3g} and the first question) which limits the multiplexing gains. Similarly, in 1-layer RS, each user could use a joint decoder to decode its private stream jointly with the common stream and the multiplexing gains would not improve (recall that 1-layer RS already achieves the information theoretic optimal multiplexing gain region, hence any other scheme, receiver or multi-layer RS would not increase the multiplexing gains any further).

\section{Numerical Results}\label{evaluations}
Through numerical evaluation, we illustrate the misconceptions and the shortcomings of MISO NOMA. Moreover, we show that, by adopting 1-layer RS, the optimal sum multiplexing gain of the MISO BC is guaranteed in both underloaded and overloaded deployments for both perfect and imperfect CSIT scenarios. Furthermore, results also demonstrate that the MMF multiplexing gain (and MMF rate) is significantly enhanced when using 1-layer RS compared to MU--LP and MISO NOMA, and the complexity of the receivers is reduced compared to MISO NOMA. In other words, our evaluations show that 1-layer RS makes a more efficient use of the spatial dimensions (multiplexing gains) and of the SIC receivers than MISO NOMA, and it is more robust to CSIT inaccuracy. 

\par The following two precoder optimization problems are solved in the simulation for the $K$-user MISO NOMA system model specified in Section \ref{sec: MISO NOMA system}. 
The first problem is maximizing the sum-rate of MISO NOMA subject to the transmit power constraint, which is given by
\begin{subequations}
	\label{eq: SR prob}
	\begin{align}
	\max_{\mathbf{{P}}}\quad &\sum_{k\in\mathcal{K}}R_{k}^{(\textnormal{N})}\\
	\mbox{s.t.}\quad
	&	\text{tr}(\mathbf{P}\mathbf{P}^{H})\leq P
	\end{align}
\end{subequations}
where $R_k^{(\textnormal{N})}$ is the rate of user-$k$ in the MISO NOMA system as specified in (\ref{eq: rate K MIMONOMA})--(\ref{eq: min rate K MIMONOMA}). 
The second problem is maximizing the minimum rate subject to the transmit power constraint, which is formulated as
\begin{subequations}
	\label{eq: maxmin prob}
	\begin{align}
	\max_{\mathbf{{P}}}\quad &\min_{k\in\mathcal{K}}R_{k}^{(\textnormal{N})}\\
	\mbox{s.t.}\quad
	&	\text{tr}(\mathbf{P}\mathbf{P}^{H})\leq P.
	\end{align}
\end{subequations}
The Weighted Minimum Mean Square Error (WMMSE) optimization framework proposed in \cite{Christensen:2008} (originally developed for MU--LP) is extended to solve both problems (\ref{eq: SR prob}) and (\ref{eq: maxmin prob}). The details of the algorithm are specified in Appendix \ref{sec: precOpt}. 
The optimization problems  requiring interior-point methods are solved using the CVX toolbox \cite{cvx:2008}. 

We will assume $K=6$ in the simulations, so as to be able to relate the numerical results to the theoretical results of Table \ref{tab: DoF compare}. The channel $\mathbf{h}_k$ of user-$k$  has i.i.d. complex Gaussian entries drawn from the distribution $\mathcal{CN}(0,\sigma_k^2)$. The presented results are averaged over 100 channel realizations. 

The  following five strategies are compared and analyzed for both perfect and imperfect CSIT:
\begin{itemize}
	\item  \textbf{MISO NOMA ($G=3$)}: MISO NOMA ($G=3$) is the MISO NOMA strategy  specified in Section \ref{sec: MISO NOMA system} when  $G=3$. Each user  requires $\frac{K}{3}-1=1$ layer of SIC (since each user is possible to be selected as the ``strong user" in the corresponding user group).  Ideally, the  sum-rate (or max-min) rate is maximized by solving (\ref{eq: SR prob}) (or (\ref{eq: maxmin prob})) for all possible user grouping methods and decoding orders within each group. 
	Due to the high computational complexity of jointly optimizing the precoders, grouping, and decoding order, we assume that the user grouping is fixed\footnote{For a given $K$ and $G$ (with $g=\frac{K}{G}$), there are in total $\frac{1}{G!}\prod _{i=0}^{G-1}\binom{K-ig}{g}$ user grouping methods. When $K=6$, the number of grouping methods for MISO NOMA ($G=3$) is 15. To  optimize the user grouping (for a fixed decoding order), the optimization problem (\ref{eq: SR prob}) (or (\ref{eq: maxmin prob})) has to be solved 15 times. The computational complexity is 15-fold increase compared with MU--LP/1-layer RS/OMA. To consider the complexity fairness among all the studied strategies, we fix the grouping method to be user-1 and user-2 in Group 1, user-3 and user-4 in Group 2 while user-5 and user-6 in Group 3. Recall however that the multiplexing gain analysis is general and holds for any decoding order and any grouping method.} while the decoding order in each group $i$ is the ascending order of  users' channel strength  $\|\mathbf{h}_k\|, \forall k\in\mathcal{K}_i$ in the following results.  To keep aligned with the system model in Section \ref{sec: MISO NOMA system}, user indices are updated within each group such that $\|\mathbf{h}_k\|\leq \|\mathbf{h}_j\|, \forall k<j$ and $ k,j\in\mathcal{K}_i$. When the CSIT is imperfect, the decoding order follows the same method but based on $\|\hat{\mathbf{h}}_k\|, \forall k\in\mathcal{K}_i$.

	\item  \textbf{MISO NOMA ($G=1$)}: MISO NOMA ($G=1$) is the MISO NOMA strategy  in Section \ref{sec: MISO NOMA system} when  $G=1$. Each user requires $K-1=5$ layers of SIC (since each user can potentially be selected as the ``strong user").  There is no user grouping optimization issue at the transmitter since all users are assumed to be in the same user group. However, the decoding order at users should be jointly optimized with precoders in order to maximize sum-rate (or max-min rate), which however, is computationally prohibitive.  Following the literature of single-cell MISO NOMA \cite{Sun:2015, Zhang:2016}, we assume that the decoding order is the ascending order of  users' channel strength $\|\mathbf{h}_k\|, \forall k\in\mathcal{K}$.  User indices are updated such that $\|\mathbf{h}_k\|\leq \|\mathbf{h}_j\|, \forall k<j$ and $ k,j\in\mathcal{K}$. Similarly,  the decoding order follows the same method but based on $\|\hat{\mathbf{h}}_k\|, \forall k\in\mathcal{K}$ when the CSIT is imperfect.

	\item  \textbf{MU--LP}: MU--LP is the baseline strategy studied in Section \ref{MULP_section}. Each user directly decodes the intended stream by fully treating the interference as noise. 
	The  WMMSE algorithm specified in Appendix \ref{sec: precOpt} can be applied and extended to solve the corresponding sum-rate and max-min problems of MU--LP \cite{Joudeh:2017,Christensen:2008}.
	The transmitter and receiver complexity of MU--LP is low since there is no SIC is deployed at each user and  no  user grouping and decoding order  optimization issue at the transmitter.
	
	\item  \textbf{Orthogonal Multiple Access (OMA)}: This is the single-user transmission where only the user with the highest channel strength is served.
	
	\item  \textbf{1-layer RS}: 1-layer RS is the RS strategy we specified in Section \ref{RS_section}. The corresponding sum-rate and max-min rate maximization problems are solved by using the WMMSE algorithm proposed in \cite{Joudeh:2016a,Joudeh:2017}. Compared with MISO NOMA schemes, the transmitter and receiver complexities of 1-layer RS are much reduced. Similarly to MU--LP, no user grouping and decoding order optimization is needed. Each user only requires a single layer of SIC.
\end{itemize}

\subsection{Perfect CSIT}
\label{evaluations_perfect}
Following \cite{Joudeh:2016a}, the initialization of the precoding matrix $\mathbf{P}$ in Algorithm \ref{AO algorithm} is designed by using Maximum Ratio Transmission (MRT) combined with Singular Value Decomposition (SVD). Specifically, the precoder for the message to be decoded by a group of users is designed based on the SVD of the channel matrix formed by the channel vectors of the corresponding users  while the precoder for the message to be decoded by a single user is designed based on MRT. For example, when considering MISO NOMA ($G=3$), the  message for user-$k, k\in\mathcal{K}_i$, is decoded by users-$\{j\mid j\leq k,j\in\mathcal{K}_i\}$. The precoders are initialized as
 $\mathbf{p}_k=\sqrt{p_k}\widehat{\mathbf{p}}_k$, where $\widehat{\mathbf{p}}_k$ is the largest left singular vector of the channel estimate ${\mathbf{H}}_k$ formed by channels $\{{\mathbf{h}}_j\mid j\leq k,j\in\mathcal{K}_i\}$. The precoder $\mathbf{p}_k$ of the  stream  to be decoded at last in each group  is initialized as $\mathbf{p}_k=\sqrt{p_k}\frac{{\mathbf{h}}_k}{||{\mathbf{h}}_k||}$, where $p_k$ is the power allocated to the corresponding precoder  $\mathbf{p}_k$ and it satisfies that $\sum_{k=1}^Kp_k=P_t$.

\begin{figure}
	\centering
	\begin{subfigure}[b]{0.38\columnwidth}
		\centering
		\includegraphics[width=\textwidth]{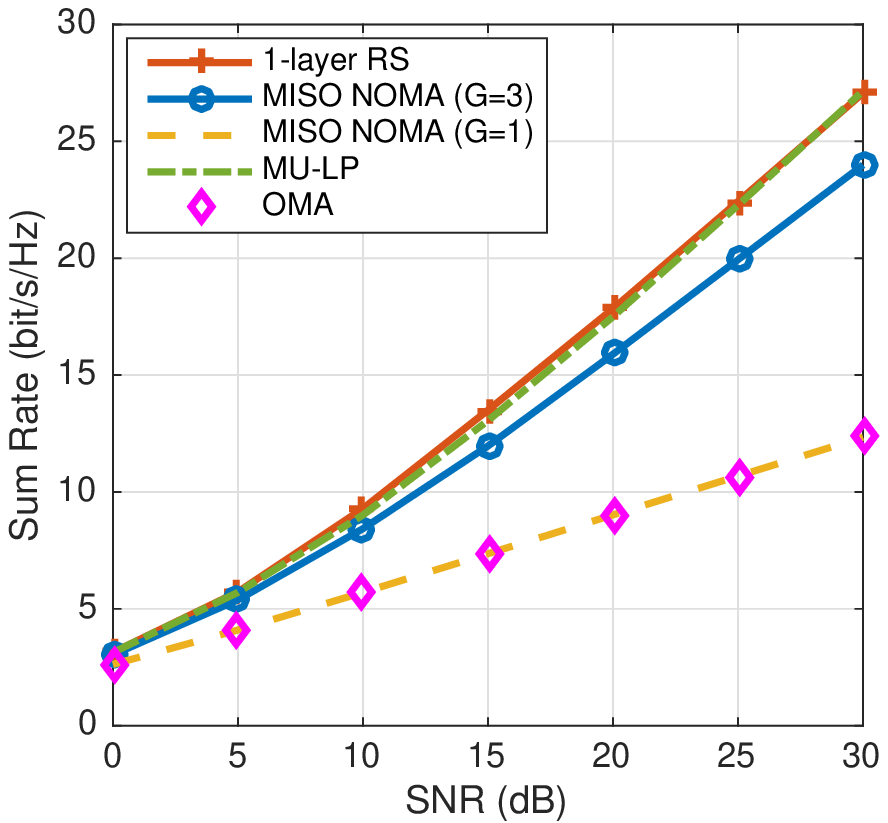}%
		\caption{$M=3$, $\sigma_k^2=1$ }
	\end{subfigure}%
	~
	\begin{subfigure}[b]{0.38\columnwidth}
		\centering
		\includegraphics[width=\textwidth]{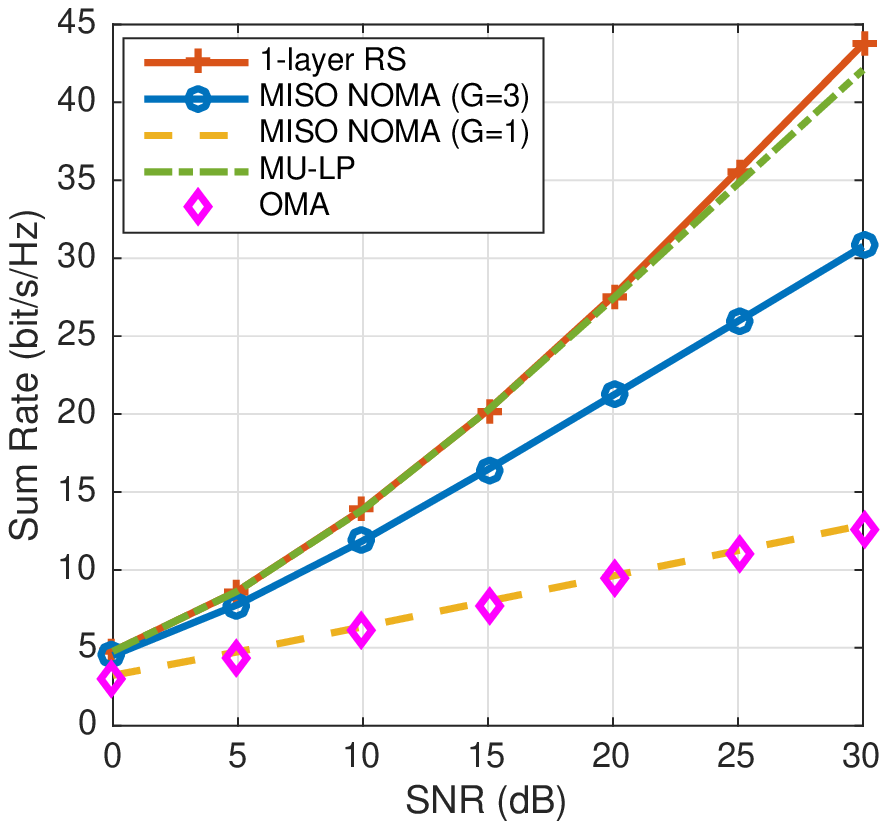}%
		\caption{$M=6, \sigma_k^2=1$}
	\end{subfigure}%
	~\\
	\begin{subfigure}[b]{0.38\columnwidth}
		\centering
		\includegraphics[width=\textwidth]{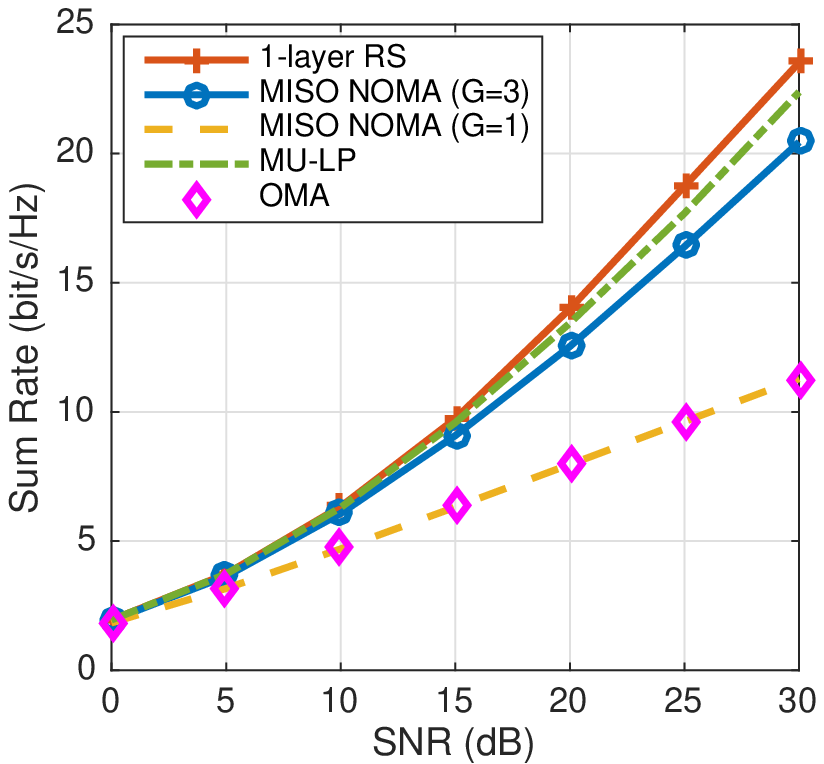}%
		\caption{$M=3, \sigma_k^2\in[0.1,1]$ }
	\end{subfigure}%
	~
	\begin{subfigure}[b]{0.38\columnwidth}
		\centering
		\includegraphics[width=\textwidth]{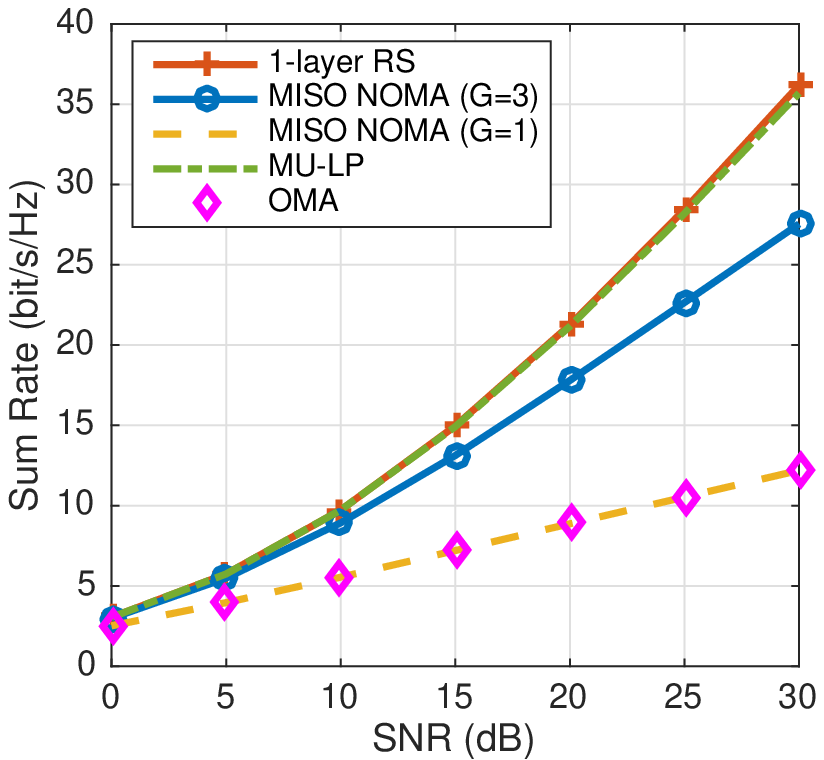}%
		\caption{$M=6, \sigma_k^2\in[0.1,1]$}
	\end{subfigure}%
	
	\caption{Sum-rate vs. SNR comparison of different strategies,  $K=6$.}
	\label{fig: ESR vs SNR}
\end{figure}

\par Fig. \ref{fig: ESR vs SNR} illustrates the sum-rate vs. SNR comparison of the five strategies considered when there are $K=6$ users and the number of transmit antennas is $M=3$ and $M=6$. In Fig. \ref{fig: ESR vs SNR}(a) and Fig. \ref{fig: ESR vs SNR}(b), all users have equal channel variances, i.e.,  $\sigma_k^2=1,\forall k\in \mathcal{K}$ while the users' channel variances are randomly generated from $[0.1, 1]$ in Fig. \ref{fig: ESR vs SNR}(c) and Fig. \ref{fig: ESR vs SNR}(d), i.e.,  $\sigma_k^2\in[0.1, 1],\forall k\in \mathcal{K}$. In other words, the average channel strength disparities among users are randomly generated between 0 and 10 dB\footnote{As a reference, at a carrier frequency of 2 GHz, the typical macro cell propagation model of \cite{ETSI} states that the path loss [dB] is equal to $128.1+37.6 \log_{10}(R)$ where $R$ is the transmitter-receiver distance in km. Considering a macro cell deployment with an inter-site distance of 750m \cite{ETSI}, a 0 to 10 dB channel gain disparity implies that users are distributed between e.g. 160m to 300m or between 200m and 375m from their serving base station, i.e. a user located at 300m (375m) will experience 10dB extra path loss compared to a user at 160m (200m).} in Fig. \ref{fig: ESR vs SNR}(c) and Fig. \ref{fig: ESR vs SNR}(d). In the high SNR regime of each subfigure, the multiplexing gains of all  strategies are found to match the theoretical sum multiplexing gains  specified in Table \ref{multiplexing_example}. Specifically,  when $M=3, K=6$, the sum multiplexing gains of 1-layer RS, MU--LP,  and MISO NOMA ($G=3$) in Fig. \ref{fig: ESR vs SNR}(a) and Fig. \ref{fig: ESR vs SNR}(c) approach  $d_{\textnormal{s}}^{(\star)}=3$ (which is optimal).   In Fig. \ref{fig: ESR vs SNR}(c) and Fig. \ref{fig: ESR vs SNR}(d) where $M=K=6$, the sum multiplexing gains of 1-layer RS and MU--LP  are  $d_{\textnormal{s}}^{(\star)}=6$.  The sum multiplexing gain of MISO NOMA ($G=3$) remains 3. The sum multiplexing gains of MISO NOMA ($G=1$)  and OMA are limited to 1 in all subfigures of  Fig. \ref{fig: ESR vs SNR}. Therefore, MISO NOMA has a reduced sum multiplexing gain, inefficeintly makes use of the available multiple antennas and incurs a significant rate loss, especially at medium and high SNRs. It is not an efficient strategy for multi-antenna settings. The first misconception behind multi-antenna NOMA is confirmed.

\par As pointed out earlier in this section, the complexity of MISO NOMA at both the transmitter and the receiver is the highest among all strategies studied in this work. At the transmitter, the scheduling complexity is high since the  user grouping and decoding order are required to be jointly optimized with the precoders. At the receivers, each user requires multiple layers of  SIC and the number of SIC layers at each user  increases with the number of users $K$ in the system. In addition to such a high complexity, as evident from Fig. \ref{fig: ESR vs SNR}, the  sum-rate performance of MISO NOMA is worse than that of MU--LP\footnote{Though multiplexing gains analysis hold for any antenna configurations, simulations are here conducted for small MIMO systems. For larger antenna regime, the same observation can be obtained and NOMA has an even less role to play as shown in \cite{Senel:2019} for the massive MIMO.} which exhibits a much lower complexity at the transmitter and each receiver. Adopting SIC receivers does not always boost the rate performance. On the contrary, an inefficient and inappropriate use of SIC as in MISO NOMA can make the rate performance worse than simply not using  SIC (as in MU--LP). This illustrates the second misconception behind multi-antenna NOMA. 

\par We also observe from  Fig. \ref{fig: ESR vs SNR} that the sum-rate performance of OMA  and MISO NOMA ($G=1$) is the worst, which is also reflected in their sum multiplexing gains. Hence, comparing MISO NOMA with OMA is not sufficient in multi-antenna settings. Both MU--LP and 1-layer RS should  be considered as the baselines for all MISO NOMA schemes. This verifies the third misconception behind multi-antenna NOMA.

\begin{figure}
	\centering
	\begin{subfigure}[b]{0.38\columnwidth}
		\centering
		\includegraphics[width=\textwidth]{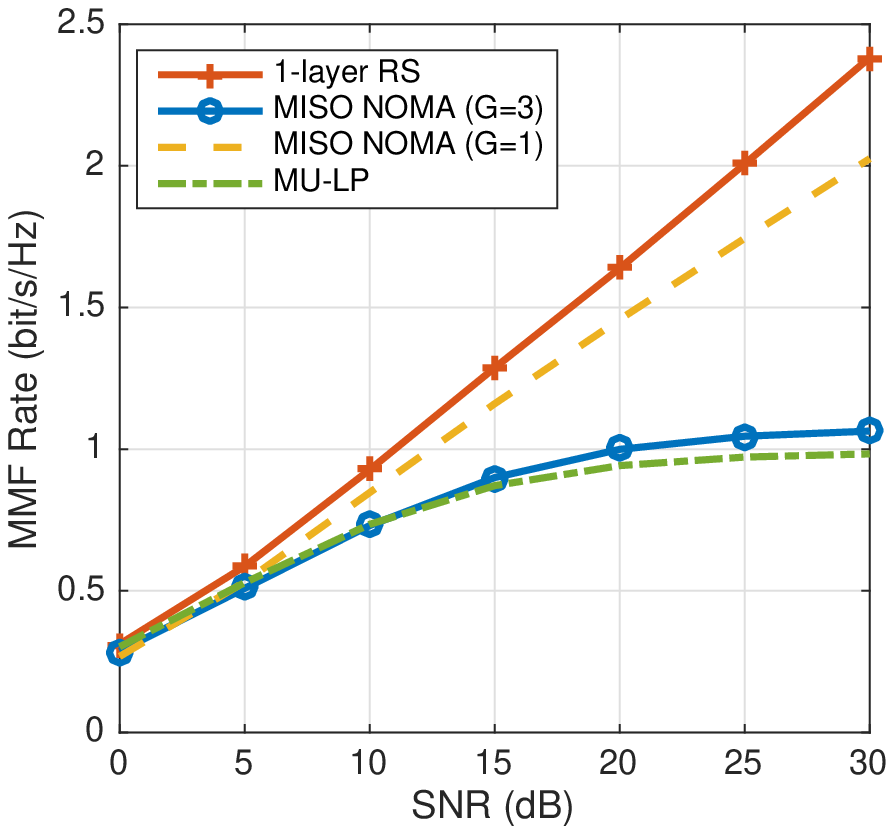}%
		\caption{$M=3$ }
	\end{subfigure}%
	~
	\begin{subfigure}[b]{0.38\columnwidth}
		\centering
		\includegraphics[width=\textwidth]{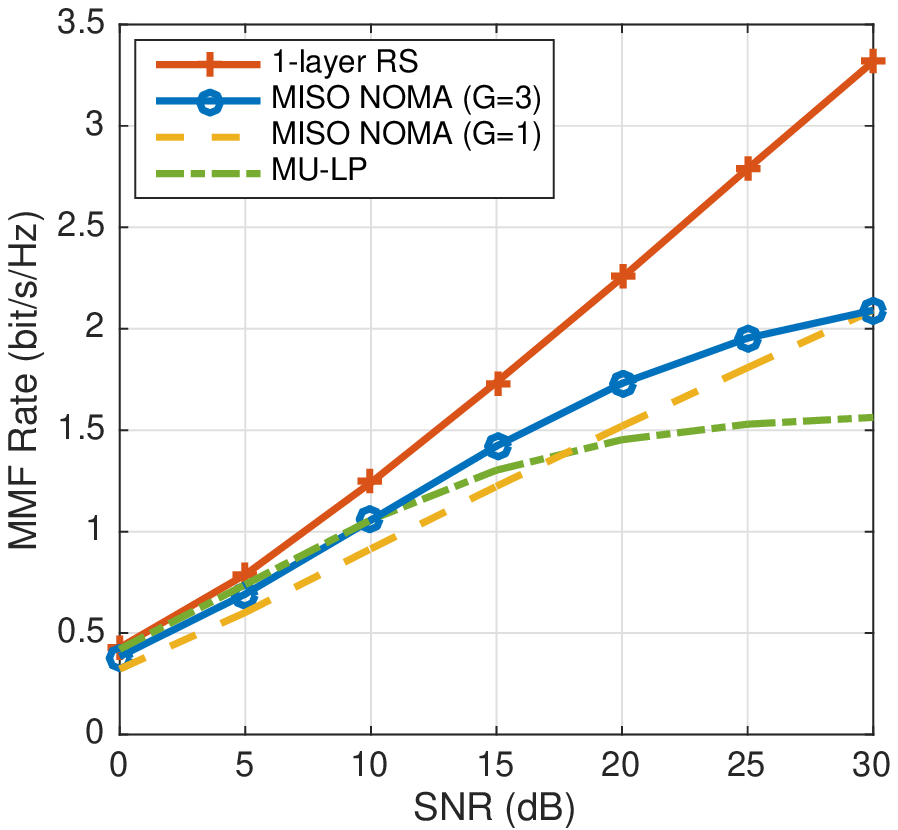}%
		\caption{$M=4$}
	\end{subfigure}%
	~\\
	\begin{subfigure}[b]{0.38\columnwidth}
		\centering
		\includegraphics[width=\textwidth]{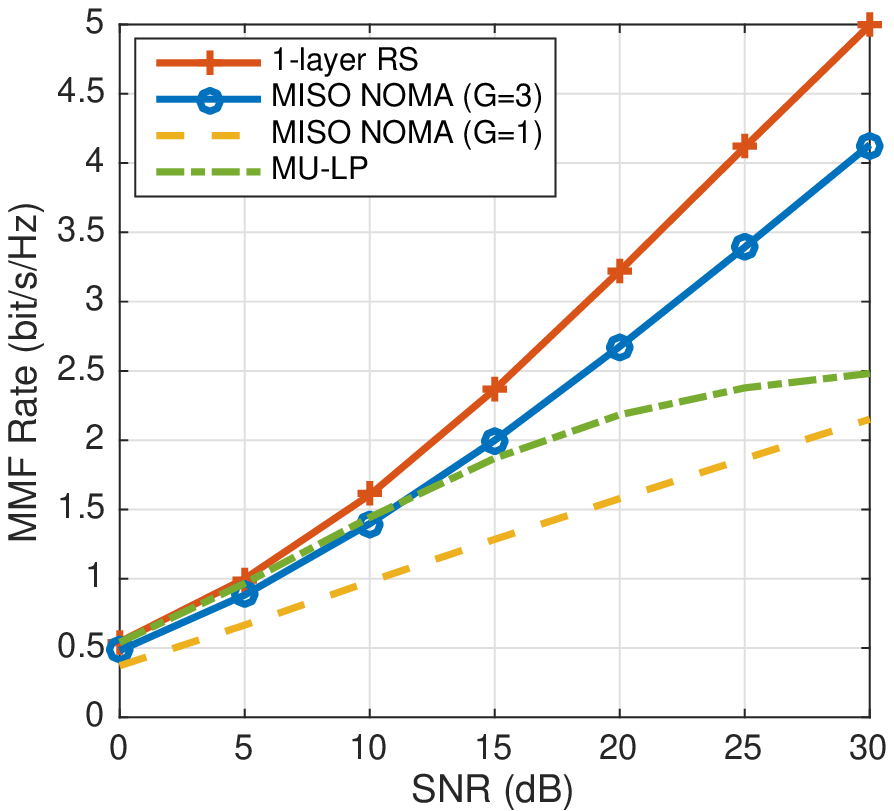}%
		\caption{$M=5$}
	\end{subfigure}%
	~
	\begin{subfigure}[b]{0.375\columnwidth}
		\centering
		\includegraphics[width=\textwidth]{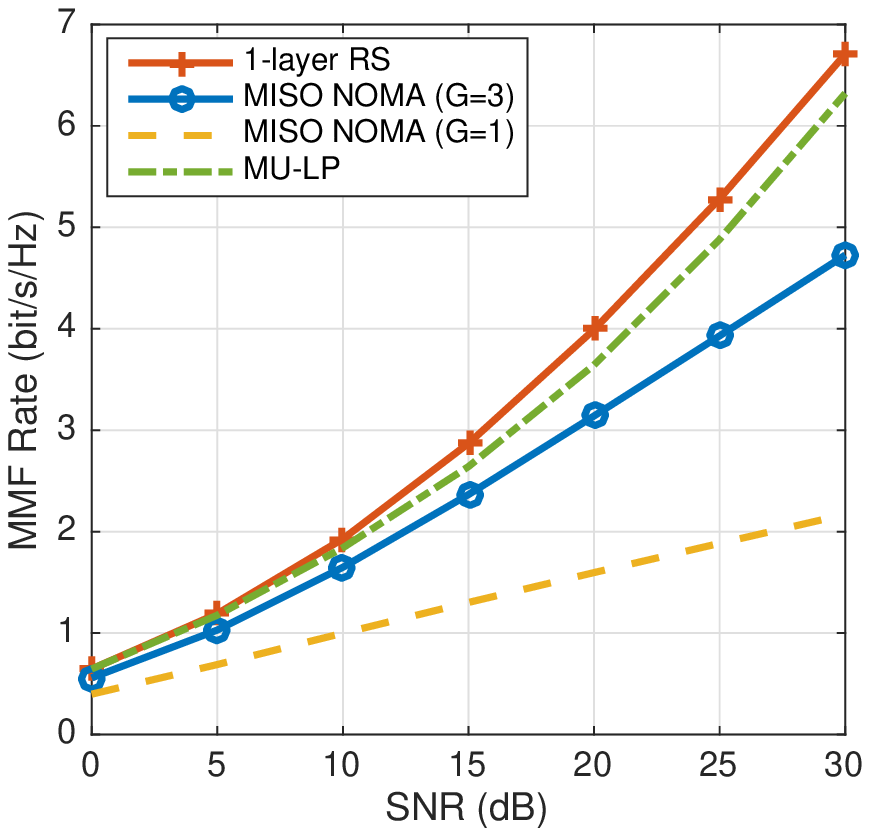}%
		\caption{$M=6$}
	\end{subfigure}%
	
	\caption{Max-min rate vs. SNR comparison of different strategies,  $K=6$, $\sigma_k^2=1, \forall k\in\mathcal{K}$.}
	\label{fig: maxMin vs SNR equalGain}
\end{figure}

\begin{figure}
	\centering
	\begin{subfigure}[b]{0.38\columnwidth}
		\centering
		\includegraphics[width=\textwidth]{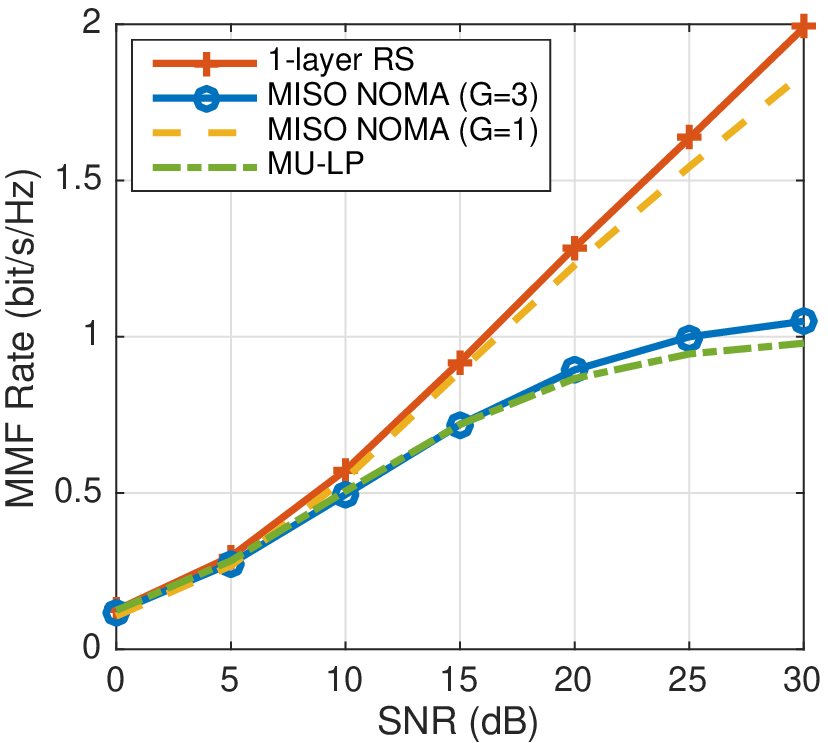}%
		\caption{$M=3$ }
	\end{subfigure}%
	~
	\begin{subfigure}[b]{0.38\columnwidth}
		\centering
		\includegraphics[width=\textwidth]{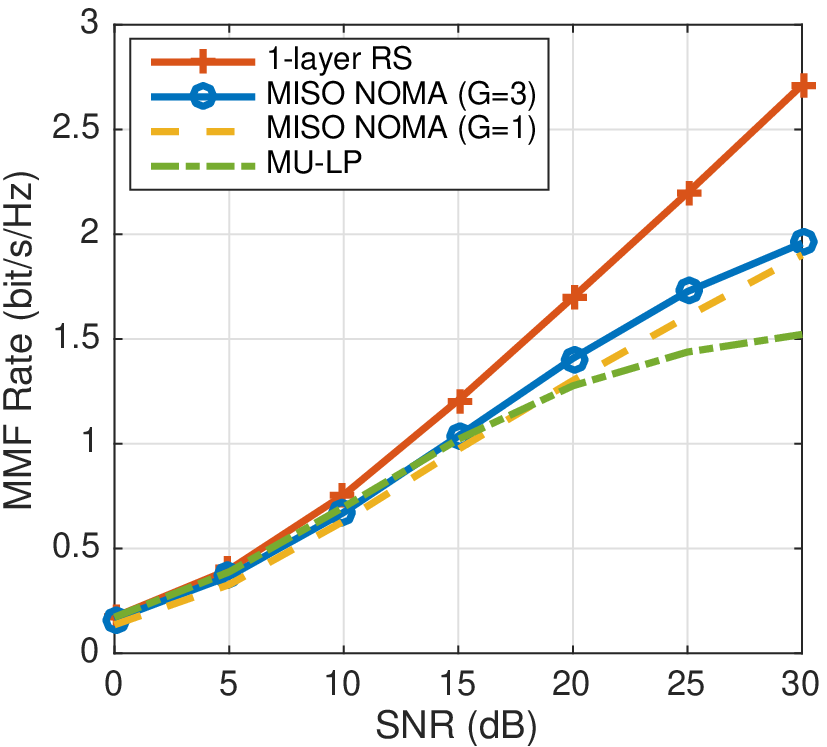}%
		\caption{$M=4$}
	\end{subfigure}%
	~\\
	\begin{subfigure}[b]{0.38\columnwidth}
		\centering
		\includegraphics[width=\textwidth]{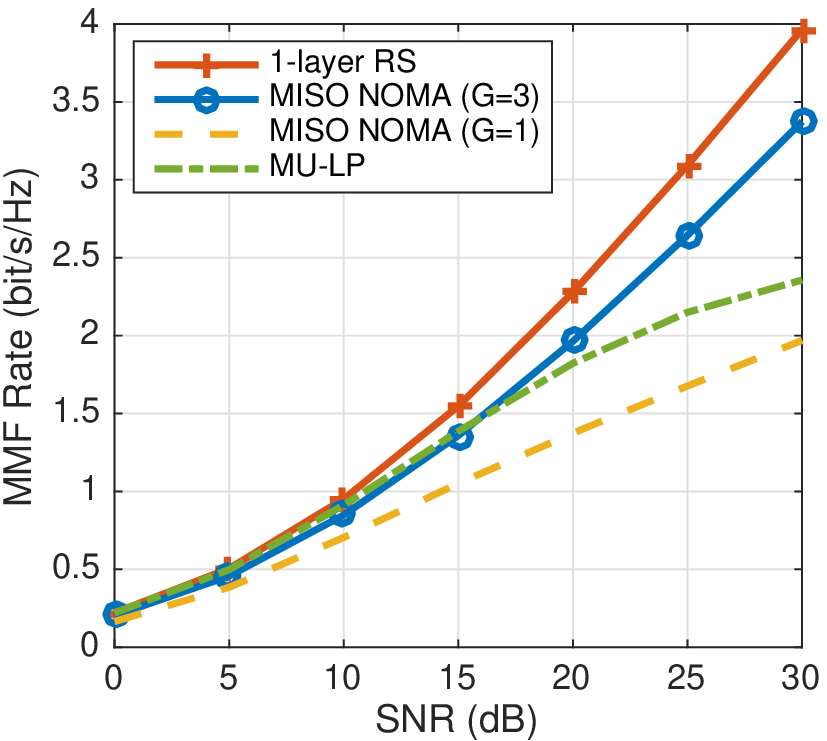}%
		\caption{$M=5$}
	\end{subfigure}%
	~
	\begin{subfigure}[b]{0.375\columnwidth}
		\centering
		\includegraphics[width=\textwidth]{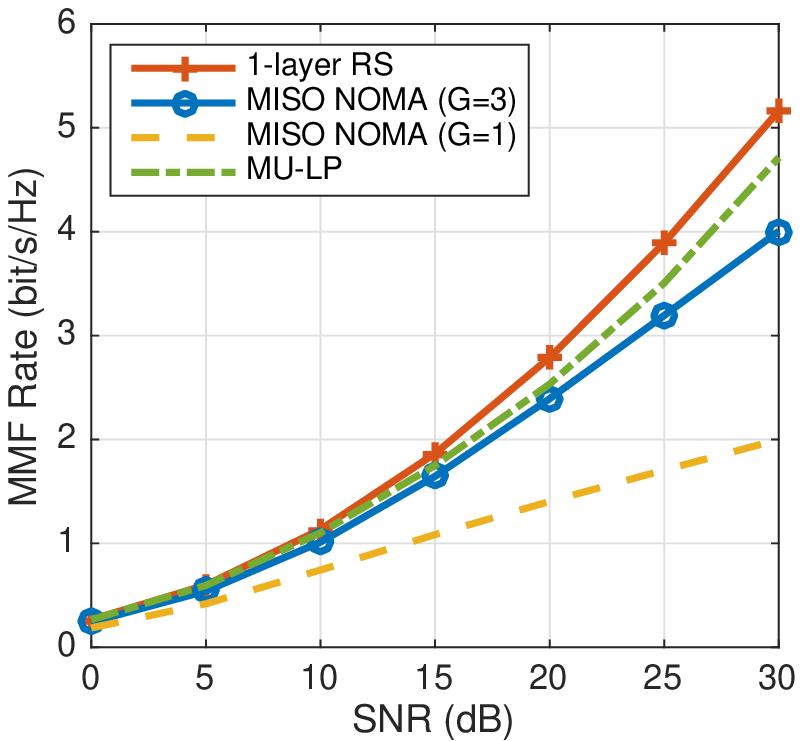}%
		\caption{$M=6$}
	\end{subfigure}%
	
	\caption{Max-min rate vs. SNR comparison of different strategies,  $K=6$, $\sigma_k^2\in[0.1, 1], \forall k\in\mathcal{K}$.}
	\label{fig: maxMin vs SNR diffGain}
\end{figure}

\par In Fig. \ref{fig: maxMin vs SNR equalGain} and  Fig. \ref{fig: maxMin vs SNR diffGain}, we focus on the MMF rate performance  when there are $K=6$ users and the number of transmit antennas is varied from $M=3$ to $M=6$. All users have equal channel variances in Fig. \ref{fig: maxMin vs SNR equalGain}  while  the users' channel variances are  randomly generated from $[0.1,1]$ in Fig. \ref{fig: maxMin vs SNR diffGain}. The MMF multiplexing gains of all the strategies in both Fig. \ref{fig: maxMin vs SNR equalGain} and  Fig. \ref{fig: maxMin vs SNR diffGain} match the corresponding theoretical MMF multiplexing gain results specified in Table \ref{mmf_multiplexing_example}. In the overloaded regime when $M=3/4/5$,  the corresponding MMF multiplexing gains of MISO NOMA ($G=3$) and MISO NOMA ($G=1$) are $d_{\textnormal{mmf}}^{(\textnormal{N,G=3})}=0/0/\frac{1}{2}$, and  $d_{\textnormal{mmf}}^{(\textnormal{N,G=1})}=\frac{1}{6}/\frac{1}{6}/\frac{1}{6}$, respectively. In contrast, the MMF multiplexing gain of 1-layer RS is $d_{\textnormal{mmf}}^{(\textnormal{R})}=\frac{1}{4}/\frac{1}{3}/\frac{1}{2}$ when $M=3/4/5$, which is significantly higher. The low MMF multiplexing gains  of the MISO NOMA strategy translates into a poor MMF rate performance as illustrated in Fig. \ref{fig: maxMin vs SNR equalGain} and  Fig. \ref{fig: maxMin vs SNR diffGain}. Though MISO NOMA has been promoted as a strategy to enhance user fairness and to deal with overloaded regimes, its MMF rate in the overloaded regime is actually worse than that of 1-layer RS. MISO NOMA is not an efficient strategy for  overloaded regimes. This underscores the validity of the fourth misconception behind multi-antenna NOMA.

\subsection{Imperfect CSIT}
\label{evaluations_imperfect}
\begin{figure}
	\centering
	\begin{subfigure}[b]{0.38\columnwidth}
		\centering
		\includegraphics[width=\textwidth]{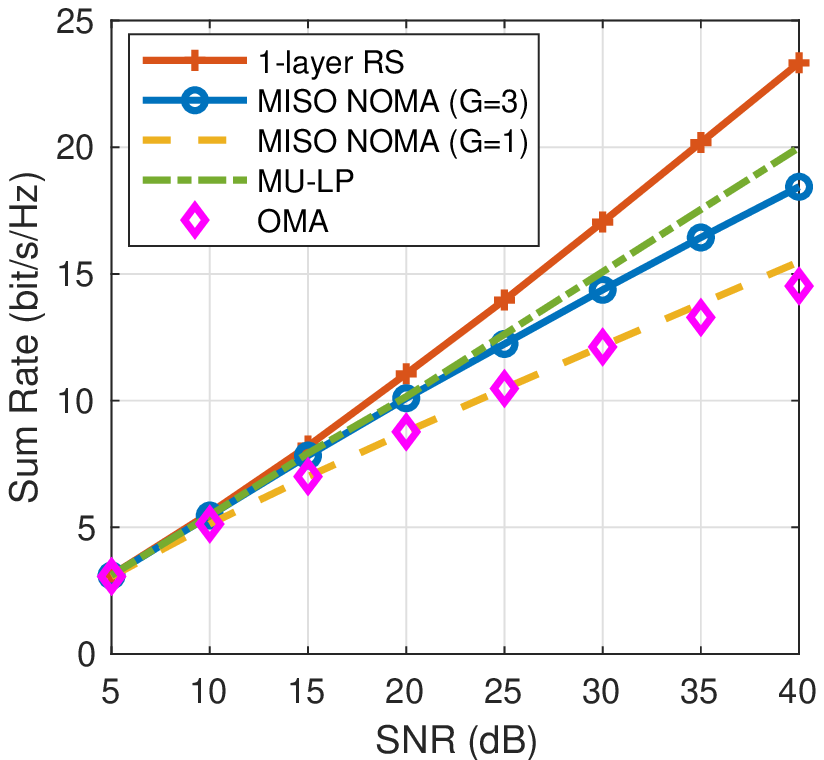}%
		\caption{$M=3$, $\sigma_k^2=1$ }
	\end{subfigure}%
	~
	\begin{subfigure}[b]{0.38\columnwidth}
		\centering
		\includegraphics[width=\textwidth]{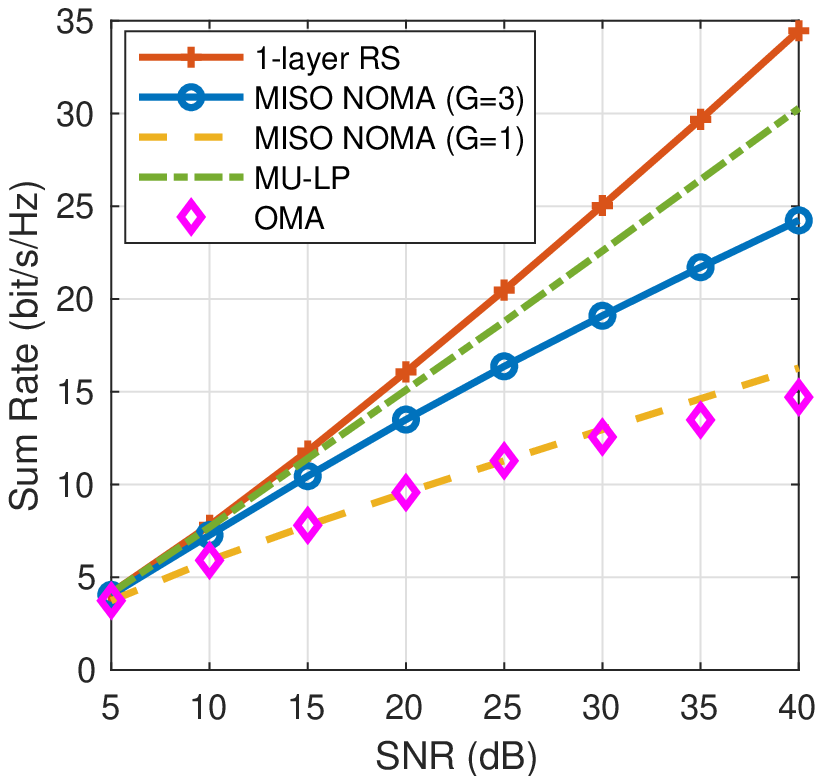}%
		\caption{$M=6, \sigma_k^2=1$}
	\end{subfigure}%
	~\\
	\begin{subfigure}[b]{0.38\columnwidth}
		\centering
		\includegraphics[width=\textwidth]{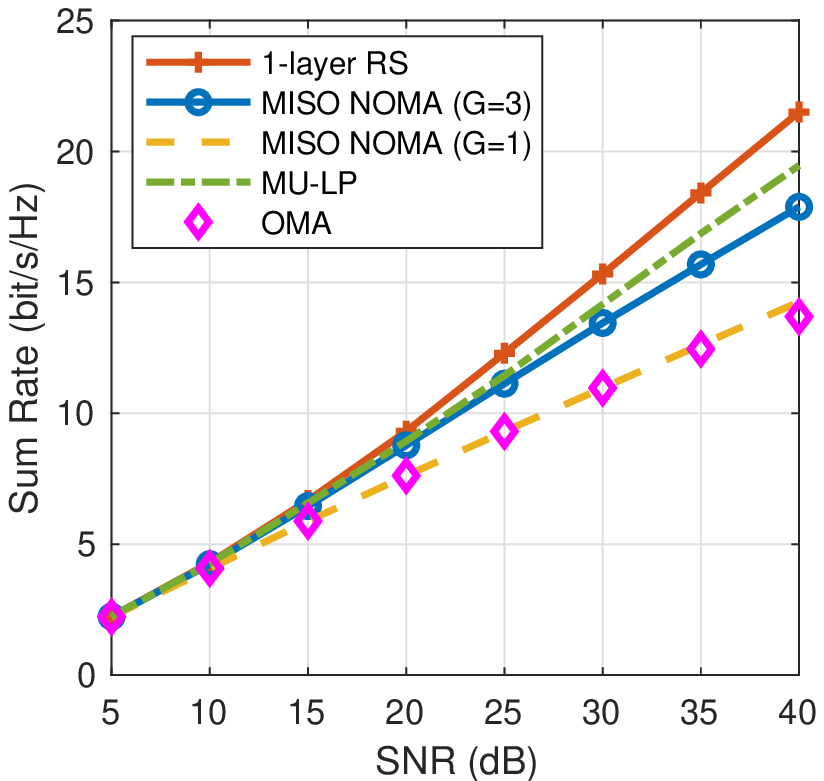}%
		\caption{$M=3, \sigma_k^2\in[0.1,1]$ }
	\end{subfigure}%
	~
	\begin{subfigure}[b]{0.38\columnwidth}
		\centering
		\includegraphics[width=\textwidth]{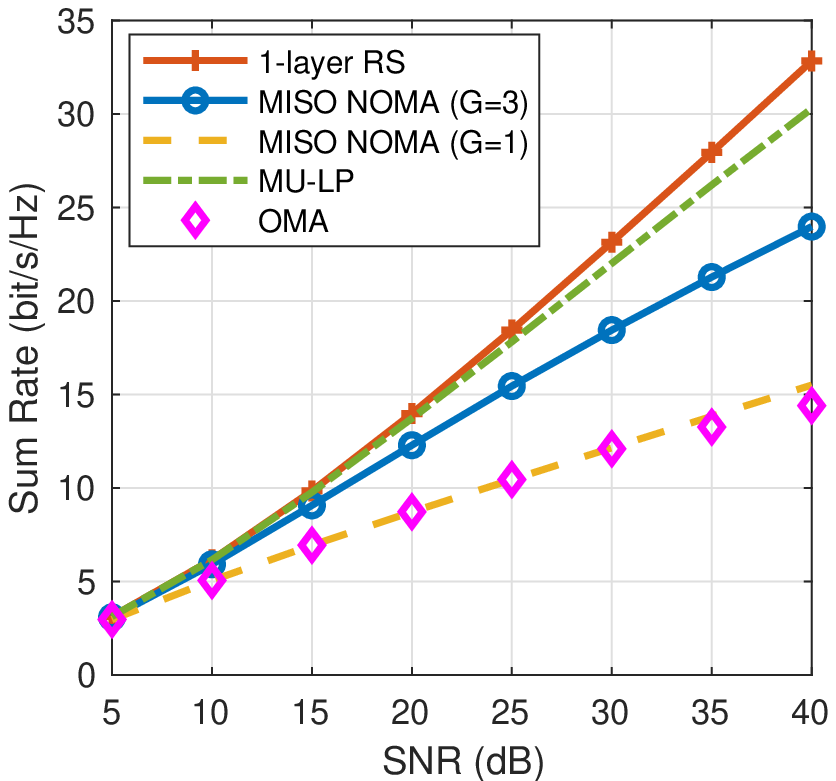}%
		\caption{$M=6, \sigma_k^2\in[0.1,1]$}
	\end{subfigure}%
	
	\caption{Sum-rate vs. SNR comparison of different strategies with imperfect CSIT, $\alpha=0.5$, $K=6$.}
	\label{fig: ESR vs SNR imperfect}
\end{figure}

\begin{figure}
	\centering
	\begin{subfigure}[b]{0.38\columnwidth}
		\centering
		\includegraphics[width=\textwidth]{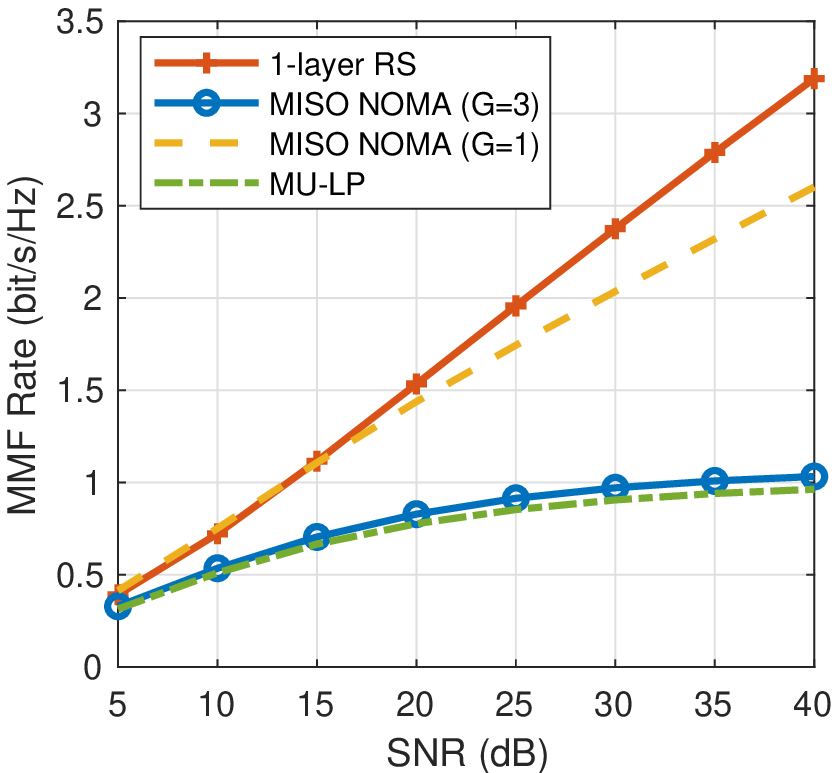}%
		\caption{$M=3$ }
	\end{subfigure}%
	~
	\begin{subfigure}[b]{0.38\columnwidth}
		\centering
		\includegraphics[width=\textwidth]{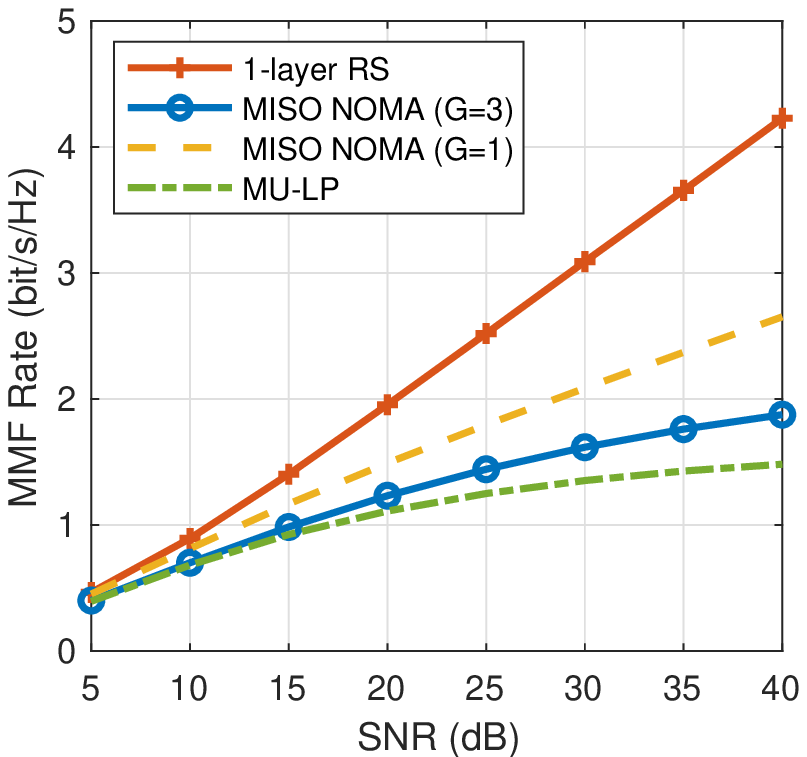}%
		\caption{$M=4$}
	\end{subfigure}%
	~\\
	\begin{subfigure}[b]{0.38\columnwidth}
		\centering
		\includegraphics[width=\textwidth]{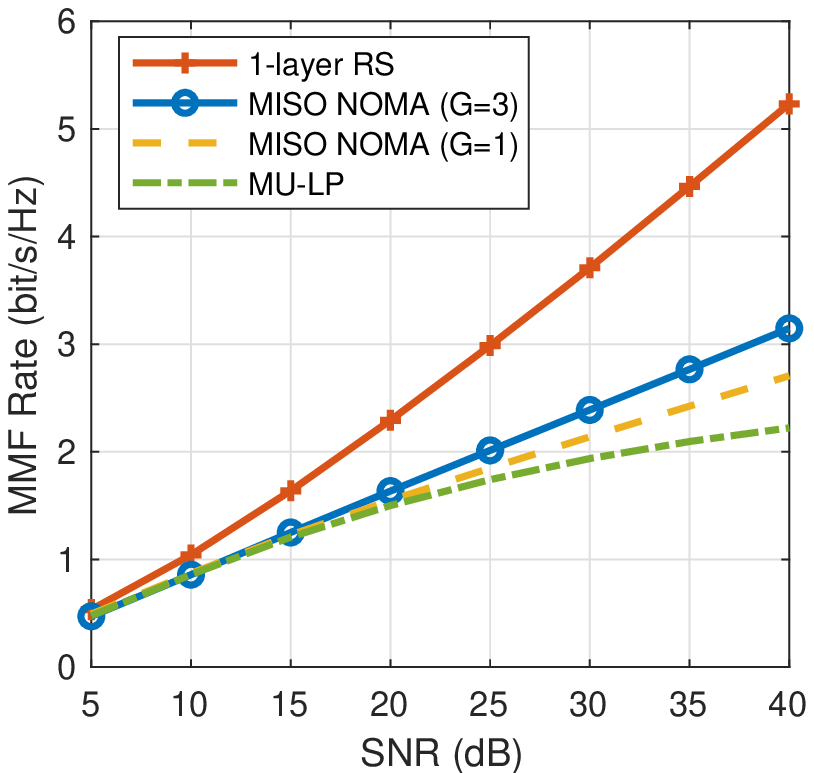}%
		\caption{$M=5$}
	\end{subfigure}%
	~
	\begin{subfigure}[b]{0.375\columnwidth}
		\centering
		\includegraphics[width=\textwidth]{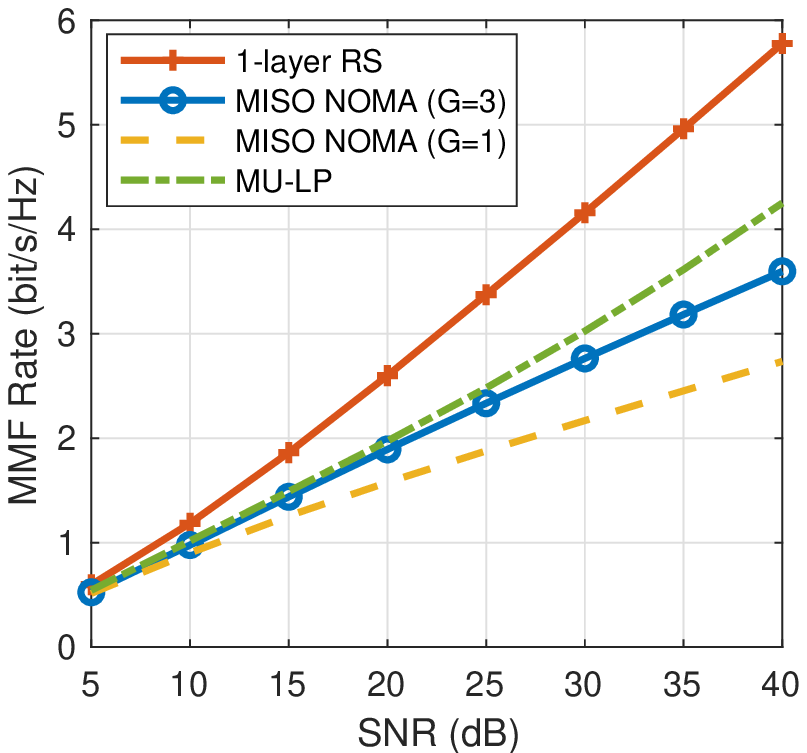}%
		\caption{$M=6$}
	\end{subfigure}%
	
	\caption{Max-min rate vs. SNR comparison of different strategies with imperfect CSIT, $\alpha=0.5$,  $K=6$, $\sigma_k^2=1, \forall k\in\mathcal{K}$.}
	\label{fig: maxMin vs SNR equalGain imperfect}
\end{figure}

\begin{figure}
	\centering
	\begin{subfigure}[b]{0.38\columnwidth}
		\centering
		\includegraphics[width=\textwidth]{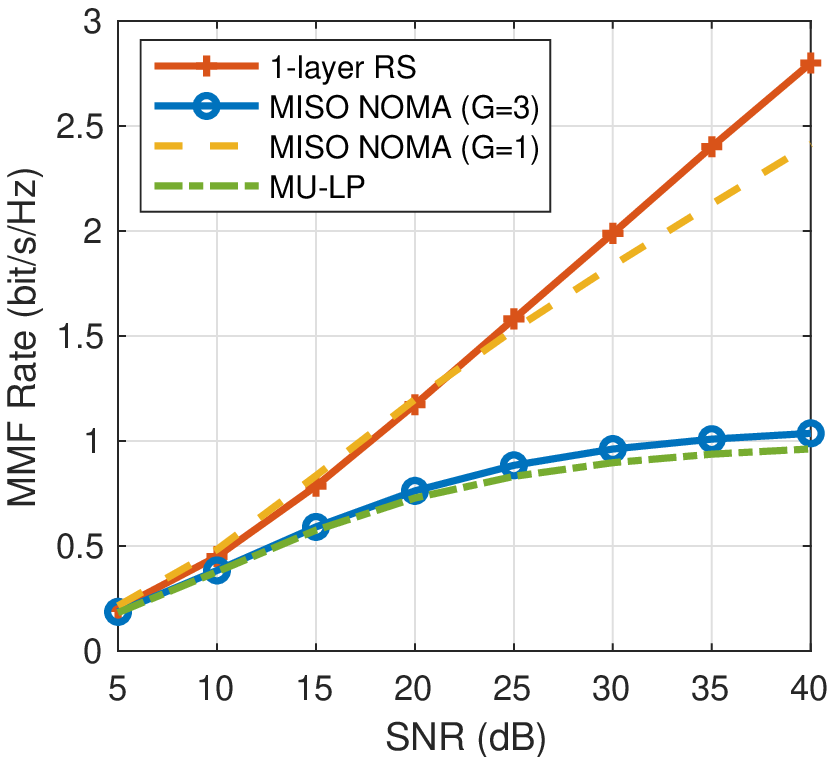}%
		\caption{$M=3$ }
	\end{subfigure}%
	~
	\begin{subfigure}[b]{0.38\columnwidth}
		\centering
		\includegraphics[width=\textwidth]{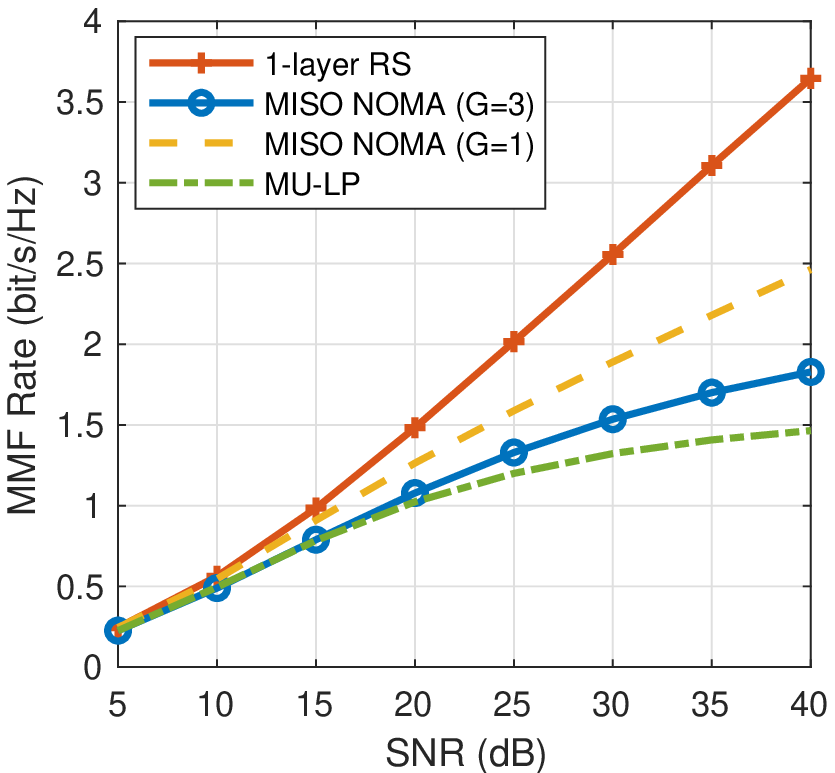}%
		\caption{$M=4$}
	\end{subfigure}%
	~\\
	\begin{subfigure}[b]{0.38\columnwidth}
		\centering
		\includegraphics[width=\textwidth]{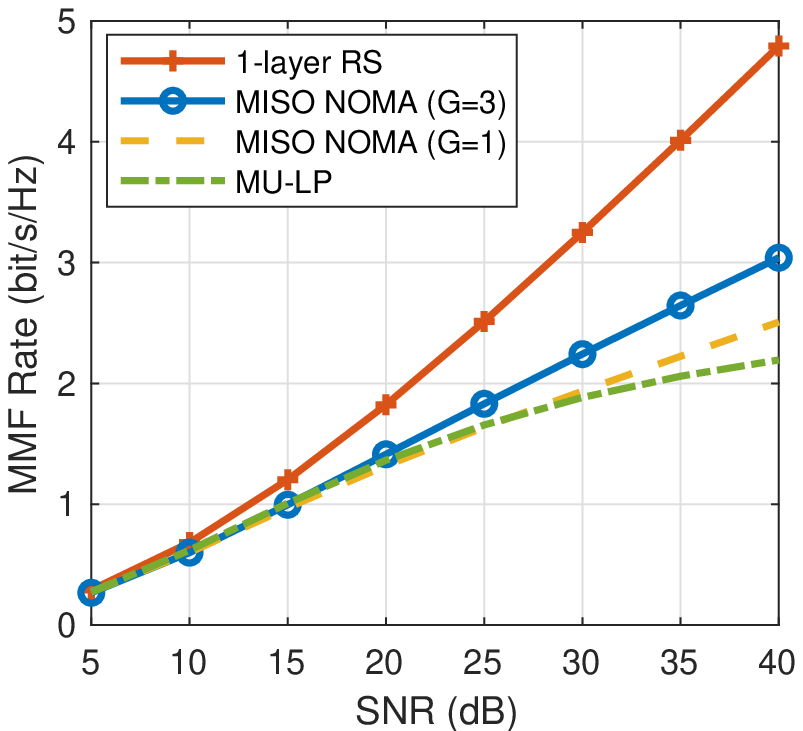}%
		\caption{$M=5$}
	\end{subfigure}%
	~
	\begin{subfigure}[b]{0.375\columnwidth}
		\centering
		\includegraphics[width=\textwidth]{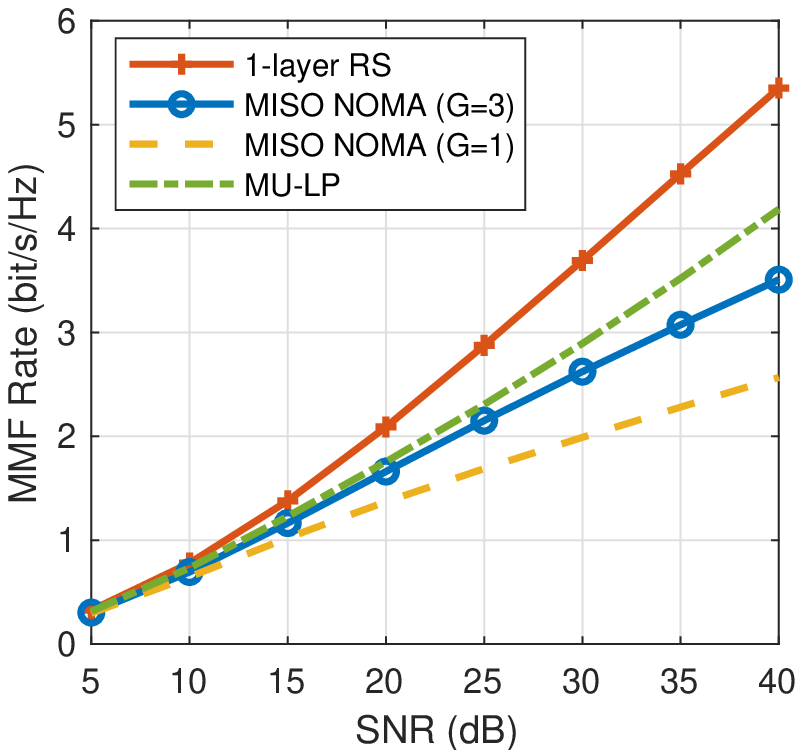}%
		\caption{$M=6$}
	\end{subfigure}%
	
	\caption{Max-min rate vs. SNR comparison of different strategies with imperfect CSIT, $\alpha=0.5$,   $K=6$, $\sigma_k^2\in[0.1, 1], \forall k\in\mathcal{K}$.}
	\label{fig: maxMin vs SNR diffGain imperfect}
\end{figure}

Let us now consider ergodic sum-rate  and minimum ergodic rate maximization problems when the CSIT is imperfect. The two problems are solved by extending the WMMSE algorithm specified in Section \ref{sec: precOpt} to the corresponding imperfect CSIT setting \cite{Joudeh:2016a}. This is achieved by using Sample Average Approximation (SAA) method \cite{Shapiro:2014} to transform the original ergodic problem to its deterministic counterpart and then using WMMSE to solve the corresponding deterministic problem. In  the following results, for a given channel estimate $\hat{\mathbf{h}}_k, k\in\mathcal{K}$,  $M=1000$ channel samples are generated. The ergodic sum-rate or max-min ergodic rate is obtained by averaging  over 100 channel estimates. The channel estimate $\hat{\mathbf{h}}_k$ and channel estimation error $\tilde{\mathbf{h}}_k$ have i.i.d. complex Gaussian entries respectively drawn from the distributions $\mathcal{CN}(0,\sigma_{k}^2-\sigma_{e,k}^2)$, $\mathcal{CN}(0,\sigma_{e,k}^2)$, where $\sigma_{e,k}^2=\sigma_k^2P_t^{-\alpha}$.
As only channel estimate $\hat{\mathbf{h}}_k, k\in\mathcal{K}$ is  known at the transmitter,  the precoders are initialized using the same method as in the perfect CSIT scenario but based on realistic channel estimates. Fig. \ref{fig: ESR vs SNR imperfect}, \ref{fig: maxMin vs SNR equalGain imperfect}, and \ref{fig: maxMin vs SNR diffGain imperfect} are the  imperfect CSIT results corresponding to  Fig. \ref{fig: ESR vs SNR}, \ref{fig: maxMin vs SNR equalGain}, and \ref{fig: maxMin vs SNR diffGain}, respectively. The unspecified parameters in this subsection remain the same as the corresponding ones used for perfect CSIT. 

Fig. \ref{fig: ESR vs SNR imperfect} illustrates the sum-rate vs. SNR comparison of the five strategies for imperfect CSIT. The sum multiplexing  gains  of  all  strategies in Fig. \ref{fig: ESR vs SNR imperfect}   match  the  theoretical  sum  multiplexing  gains   in  Table  \ref{tab: DoF compare}. When $\alpha=0.5$ and $M=3/6$, the sum multiplexing gains of the five strategies are $d_{\textnormal{s}}^{(\textnormal{R})}=2/3.5$ for 1-layer RS,  $d_{\textnormal{s}}^{(\textnormal{M})}=1.5/3$ for MU--LP, $d_{\textnormal{s}}^{(\textnormal{N,G=3})}=1.5/1.5$ for MISO NOMA ($G=3$), and $d_{\textnormal{s}}^{(\textnormal{N,G=1})}=d_{\textnormal{s}}^{(\textnormal{O})}=1/1$ for MISO NOMA ($G=1$) and OMA.
As suggested by the multiplexing gain results, where MISO NOMA ($G=1$) has the lowest multiplexing gain, we also observe from Fig. \ref{fig: ESR vs SNR imperfect} that  though MISO NOMA ($G=1$) has the highest receiver complexity, its ergodic sum rate performance  is the worst even in the  preferred NOMA overloaded setting when the users have channel strength disparities. MISO NOMA ($G=1$) always achieves a worse sum-rate than MU--LP. It is not beneficial for enhancing the sum-rate of multi-antenna scenarios regardless of whether perfect or imperfect CSIT is used. In comparison, 1-layer RS achieves explicit sum multiplexing gains and sum-rate improvement over all other strategies.  

Fig. \ref{fig: maxMin vs SNR equalGain imperfect} and \ref{fig: maxMin vs SNR diffGain imperfect} illustrate the MMF ergodic rate results.  In general, the MMF multiplexing gains of all strategies in both figures match the theoretical MMF multiplexing gain results specified in Table \ref{tab: DoF compare}. 
When $M=3/4/5/6$,  the corresponding MMF multiplexing gains of MISO NOMA ($G=3$) and MISO NOMA ($G=1$) when $\alpha=0.5$ are $d_{\textnormal{mmf}}^{(\textnormal{N,G=3})}=0/0/\frac{1}{4}/\frac{1}{4}$ and  $d_{\textnormal{mmf}}^{(\textnormal{N,G=1})}=\frac{1}{6}/\frac{1}{6}/\frac{1}{6}/\frac{1}{6}$, respectively, and the corresponding MMF multiplexing gain of MU--LP and RS are $d_{\textnormal{mmf}}^{(\textnormal{M})}=0/0/0/\frac{1}{2}$, and $d_{\textnormal{mmf}}^{(\textnormal{R})}=\frac{1}{4}/\frac{1}{3}/\frac{1}{2}/\frac{1}{2}$. We observe that 1-layer RS achieves significantly higher multiplexing gains, which is also reflected in the MMF ergodic rate performance in Fig. \ref{fig: maxMin vs SNR equalGain imperfect} and \ref{fig: maxMin vs SNR diffGain imperfect}. In both the perfect and imperfect CSIT settings,  user fairness cannot be improved  by MISO NOMA. The MMF ergodic rate performance of MISO NOMA is  much worse than that of 1-layer RS.

Therefore, the four misconceptions behind multi-antenna NOMA are further verified for imperfect CSIT. Higher sum-rate and MMF rate gaps between RS and MU--LP/multi-antenna NOMA are generally observed by comparing the corresponding perfect and imperfect CSIT results. By partially decoding the interference and treating the remaining interference as noise, 1-layer RS is more robust to CSIT inaccuracy. The large performance gain of RS makes it an appealing strategy for future communication networks.

\subsection{Discussions}
\label{evaluation_discussion}
\par The simulations so far fully validate the theoretical multiplexing gain analysis and confirm the inefficiency of MISO NOMA. We therefore conclude that the fundamental design principle of NOMA, namely forcing one user to decode the message(s) of other user(s), should be reconsidered
or very carefully used for multi-antenna settings.

\par Thanks to its ability to partially decode interference and partially treat interference as noise,  1-layer RS achieves equal or higher sum-rate and MMF rate performance than all other strategies in both underloaded and overloaded regimes, especially when it comes to metrics that favor user fairness (e.g., MMF rate) in an overloaded regime. This is due to the fact that the inter-user interference becomes stronger in the setting when all users are active and the number of transmit antenna is limited. The superiority of 1-layer RS   in managing multi-user interference becomes more pronounced when users suffer from stronger interference. Most importantly, 1-layer RS  requires no user grouping and decoding order optimization at the transmitter and only one layer of SIC at each user. Compared with MISO NOMA, the sum-rate and MMF rate performance gain of RS comes at a much reduced transmitter and receiver complexity. 1-layer RS enables a better trade-off between the rate performance gains and the number of SIC layers. Hence, we conclude that 1-layer RS is a more powerful and promising strategy for multi-antenna networks.

\par Though the evaluations have been limited to 1-layer RS as the basic RSMA scheme, further rate enhancements over 1-layer RS can be obtained with multi-layer RS where the message of a user is split multiple times and multiple SIC layers are implemented at the receivers, as demonstrated in \cite{Mao:2017,Dai:2016,Mao:2019,Mao:2020_TCOM,Mao:2020_Asilomar}.

\section{Conclusions and Future Research}\label{conclusions}
This paper provides a broad, different and useful perspective on multi-antenna NOMA and non-orthogonal transmission to the community working on NOMA and multiple access, and to the future generations of researchers working on multi-user multi-antenna communications. Although NOMA in single-antenna settings has been well understood for a long time, the paper shows that the design of non-orthogonal transmission strategies for multi-antenna settings should be done with care so as to benefit from the multi-antenna dimensions and SIC receivers.

\par The paper showed in Section \ref{twouser} that a two-user multi-antenna NOMA increases the receiver complexity and at the same time incurs a loss in multiplexing gain (and therefore rate at high SNR) compared to conventional multiuser precoding (as in used in 4G and 5G), therefore raising concerns on the efficiency of multi-antenna NOMA. Subsequently, a general $K$-user setting with perfect CSIT and imperfect CSIT were studied in Section \ref{kuser} and Section \ref{Imperfect_CSIT_section}, respectively and various multiplexing gains of multi-antenna NOMA were derived. Then we introduced two baseline schemes, namely $K$-user conventional multiuser precoding Section in \ref{MULP_section} and $K$-user multi-antenna rate-splitting in Section \ref{RS_section}, and studied the multiplexing gains of those schemes. Section \ref{sec: misconceptions NOMA} compares the multiplexing gains of all considered schemes and provides strong theoretical grounds for performance comparisons among all schemes. In particular it identifies the scenarios where NOMA incurs a gain and a loss compared to multiuser linear precoding and demonstrates how NOMA always leads to lower multiplexing gains than rate-splitting though it makes use of a larger number of SIC layers at the receivers. This section is instrumental and exposes various misconceptions and shortcomings of multi-antenna NOMA. Simulation results are then used in Section \ref{evaluations} to confirm our findings and prediction from the multiplexing gain analysis.

\par Our results show that NOMA is not an efficient solution to cope with the high throughput, reliability, heterogeneity of Quality-of-Service (QoS), and connectivity requirements of the downlink of future 5G and beyond multi-antenna wireless networks. This is due to the fact that the fundamental principle of NOMA consisting in forcing one user in each group to fully decode the messages of other co-scheduled users is an inefficient design in multi-antenna settings. Consequently, the benefits to the research community and future standards and networks of multi-antenna NOMA for downlink communications (e.g. MISO/MIMO techniques for NOMA, NOMA for massive MIMO and cell-free massive MIMO, multi-antenna NOMA for millimetre and terahertz communications, NOMA for multi-beam satellite communications, multi-antenna NOMA in reconfigurable intelligent surfaces, multi-antenna in Multiuser Superposition Transmission (MUST) in 3GPP, etc) are questionable and should be considered carefully in light of the results in this paper. 

\par Instead, non-orthogonal transmission strategies for multi-antenna settings should be designed such that interference is partially decoded and partially treated as noise based on the rate-splitting (multiple access) literature so as to truly benefit from multi-antenna transmitters (and potentially multi-antenna receivers) and SIC receivers.

\par In this paper, we limited the multiplexing gain analysis and the numerical evaluations to two metrics, namely sum multiplexing gain/sum-rate and MMF multiplexing gain/MMF rate, and to the MISO BC. Nevertheless, the results can be extended to other metrics such as the weighted sum-rate (WSR) and to other scenarios. Readers are invited to check \cite{Mao:2017} that confirms the inefficiency of NOMA and the superiority of RS from a WSR perspective, and are encouraged to consult the growing literature on RS (and RSMA) demonstrating the superiority of RS over NOMA and MU-LP in numerous scenarios and applications, namely multi-user multi-antenna communications \cite{Mao:2017,Clerckx:2020,Yang:2019}, multigroup multicast \cite{Joudeh:2017,Yalcin:2019}, energy efficiency \cite{Mao:2019,Mao:2018_iswcs,Rahmati:2019}, multi-cell joint transmission \cite{Mao:2019_icc}, non-orthogonal unicast and multicast transmission \cite{Mao:2019}, wireless information and power transfer \cite{Mao:2019_spawc}, cooperative communication with user relaying \cite{Zhang:2019}, imperfect CSIT \cite{Mao:2020_TCOM,Mao:2020_Asilomar}, link-level simulations \cite{Dizdar:2020}, C-RAN \cite{Yu:2019}, secrecy rate \cite{Li:2020,Fu:2020}, aerial networks \cite{Rahmati:2019,Jaafar:2020}, imperfect CSIT and CSIR \cite{An:2020}, visible light communications \cite{Tao:2020,Naser:2020}, network performance analysis \cite{Demarchou:2020}, reconfigurable intelligent surface \cite{Yang:2020}.

\par The emphasis of the paper was on downlink multi-user communications. Results suggest that future downlink multi-user multi-antenna communications would strongly benefits from RSMA. RSMA is a gold mine of research problems for academia and industry with issues spanning numerous areas: RSMA to achieve the fundamental limits of wireless networks; RSMA for multi-user/multi-cell multi-antenna networks; RSMA-based robust interference management; RSMA in MU-MIMO, CoMP, Massive MIMO, millimetre wave and higher frequency bands, relay, cognitive radio, caching, physical layer security, cooperative communications, cloud-enabled platforms (C-RAN, F-RAN), intelligent reflecting surfaces, etc; RSMA to unify, generalize and outperform SDMA and NOMA; physical layer design of RSMA-based network; coding and Modulation for RSMA; cross-layer design, optimization and performance analysis of RSMA; implementation and standardization of RSMA; RSMA in B5G services such as enhanced eMBB, enhanced URLLC, enhanced MTC, massive MTC, massive IoT, V2X, cellular, UAV and satellite networks, wireless powered communications, integrated communications and sensing, etc. RSMA can also be used in the uplink, as originally shown for single-antenna systems in \cite{Rimoldi:1996}, and much is left to be done to identify the benefits of RSMA for general uplink multi-user multi-antenna communications. The performance benefits of RSMA vs NOMA vs OMA vs other multiple accesses in the uplink, beyond the existing NOMA vs OMA comparison \cite{Wei:2020}, is also much worth investigating.

 \appendices

 \section{Proof of Proposition \ref{Theorem_NOMA_MMF_DoF_alpha}}
\label{app: prop4}
\par Let us first consider $G>1$ and $M\geq K-g+1$. Recalling from the proof of Proposition \ref{Theorem_NOMA_DoF_alpha} that the sum multiplexing gain of $G\alpha$ can be split equally among the $G$ groups so that each group gets a (group) sum multiplexing gain of $\alpha$, and following again the MAC argument, the (group) sum multiplexing gain of $\alpha$ in each group can then be further split equally among the $g$ users, which leads to an upper bound on the MMF multiplexing gain of $\frac{\alpha}{g}$. Achievability is obtained by designing precoders using ZFBF, and allocating power (consider group 1 for simplicity) to user $k=1,\ldots,g$ as $O(P^{1-\frac{g-k}{g}\alpha})$, which causes the SINR for user-$k$ to scale as $O(P^{\alpha/g})$ and an achievable MMF multiplexing gain of $\frac{\alpha}{g}$. 

\par To illustrate the achievability in more detail, we consider a simple example associated with $K=4$, $G=2$, $g=2$, and $M\geq 3$. First, we design the precoders $\mathbf{p}_1$ and $\mathbf{p}_2$ in group 1 to be orthogonal to the channel estimates $\hat{\mathbf{h}}_3$ and $\hat{\mathbf{h}}_4$ of users 3 and 4. Similarly, $\mathbf{p}_3$ and $\mathbf{p}_4$ in group 2 are made orthogonal to $\hat{\mathbf{h}}_1$ and $\hat{\mathbf{h}}_2$. Second, allocate power $O(P^{b})$ with $b=1-\alpha/2$ to users 1 and 3, and $O(P-P^{b})=O(P)$ to users 2 and 4. Using these precoders and power allocations, the received signals in group 1  can be written as
\begin{align}
    y_1&=\underbrace{\mathbf{h}_1^H\mathbf{p}_1 s_1}_{P^{b}}+\underbrace{\mathbf{h}_1^H\mathbf{p}_2 s_2}_{P}+\underbrace{\tilde{\mathbf{h}}_1^H\mathbf{p}_3 s_3}_{P^{b-\alpha}}+\underbrace{\tilde{\mathbf{h}}_1^H\mathbf{p}_4 s_4}_{P^{1-\alpha}}+\underbrace{n_1}_{P^0},\label{stream_1}\\
    y_2&=\underbrace{\mathbf{h}_2^H\mathbf{p}_1 s_1}_{P^{b}}+\underbrace{\mathbf{h}_2^H\mathbf{p}_2 s_2}_{P}+\underbrace{\tilde{\mathbf{h}}_2^H\mathbf{p}_3 s_3}_{P^{b-\alpha}}+\underbrace{\tilde{\mathbf{h}}_2^H\mathbf{p}_4 s_4}_{P^{1-\alpha}}+\underbrace{n_2}_{P^0},\label{stream_2}
\end{align}
where the quantities under the brackets refer to how the power level of each term scales. From \eqref{stream_1} and \eqref{stream_2}, $s_2$ can be decoded at an SINR level scaling as $\frac{P}{P^b+P^{1-\alpha}+P^{b-\alpha}+P^{0}}=\frac{P}{P^b}=P^{\alpha/2}$ (since $b\geq 1-\alpha\geq b-\alpha$ and $b\geq 0$). Using SIC, $s_2$ is cancelled in $\eqref{stream_1}$, and $s_1$ can be decoded at an SINR level scaling as $\frac{P^b}{P^{1-\alpha}+P^{b-\alpha}+P^{0}}=P^{\alpha/2}$. Similar expressions hold for group 2, and we note that all four streams have an SINR scaling as $P^{\alpha/2}$, therefore achieving an MMF multiplexing gain of $\frac{\alpha}{2}$.
\par Let us now consider $G>1$ and $M < K-g+1$. Since the MMF multiplexing gain collapses to 0 in the perfect CSIT setting, the same holds for imperfect CSIT.
\par Let us now consider $G=1$. The situation here is the same as in the perfect CSIT setting. There is no inter-group interference and the sum multiplexing gain of one in the single group can be split equally among the $K$ users, which leads to an upper bound on the MMF multiplexing gain of $\frac{1}{K}$. Achievability is obtained by choosing the powers of users $k=1,\ldots,K$ as $O(P^{k/K})$, which causes the SINR of user-$k$ to scale as $O(P^{1/K})$ and results in an achievable MMF multiplexing gain of $\frac{1}{K}$.

\section{WMMSE Optimization Framwork}
\label{sec: precOpt}
The WMMSE optimization framework  to solve both problems (\ref{eq: SR prob}) and (\ref{eq: maxmin prob}) is specified as follows.

At user-$j, j\in\mathcal{K}_i$, equalizer $g_{j,k}$ is employed to decode  stream $s_k, k\in \{k\mid k\geq j,k\in\mathcal{K}_i\}$.  
The estimate of $s_k$ at user-$j$ is obtained as $\widehat{s}_{j,k}=g_{j,k}y_{j,k}$, where $y_{j,k}=\sum_{m\leq k, m\in\mathcal{K}_i}\mathbf{h}_j^H\mathbf{p}_ms_m+\sum_{l\neq i, l\in\mathcal{G}}\sum_{m\in\mathcal{K}_l}\mathbf{h}_j^H\mathbf{p}_ms_m+n_j$ is the signal received at user-$j$ after removing the streams decoded before $s_k$.
The  corresponding Mean Square Error (MSE) is given by
\begin{equation}
\label{eq: MSE}
	\begin{aligned}
	\varepsilon_{j,k}&=\mathbb{E}\{|\widehat{s}_{j,k}-s_k|^2\}\\
	&=|g_{j,k}|^2T_{j,k}-2 \Re\{g_{j,k}\mathbf{h}_j^H\mathbf{p}_k\}+1,
	\end{aligned}
\end{equation}
where $T_{j,k}=|\mathbf{h}_j^H\mathbf{p}_k|^2+I_{j,k}^{(in)}+I_{j,k}^{(ou)}$ is the power received at user-$j$ when decoding $s_k$. Furthermore, $I_{j,k}^{(in)}$ and $I_{j,k}^{(ou)}$ are respectively the intra-group and inter-group interference power defined in (\ref{eq: intefe K MIMONOMA}).

By solving $\frac{\partial\varepsilon_{j,k}}{\partial g_{j,k}}=0$,  the optimal Minimum MSE (MMSE) equalizer is calculated as
\begin{equation}
\label{eq: MMSE equalizers}
g_{j,k}^{\textrm{MMSE}}=\mathbf{p}_{k}^{H}\mathbf{h}_{j}({T}_{j,k})^{-1}.
\end{equation}

Substituting (\ref{eq: MMSE equalizers}) back to (\ref{eq: MSE}), the corresponding MMSE is then obtained as 
\begin{equation}
\varepsilon_{j,k}^{\textrm{MMSE}}=\min_{g_{j,k}} \varepsilon_{j,k}=T_{j,k}^{-1}(I_{j,k}^{(in)}+I_{j,k}^{(ou)}).
\end{equation}
With the  introduced $\varepsilon_{j,k}^{\textrm{MMSE}}$, the rate at user-$j$ to decode the message of user-$k$ in (\ref{eq: rate K MIMONOMA}) is equivalently written as $R_{j,k}=-\log_2(\varepsilon_{j,k}^{\textrm{MMSE}})$. Defining the Weighted MSE (WMSE) of $\varepsilon_{j,k}$ with a weight $u_{j,k}>0$ as
\begin{equation}
\label{eq: WMSE}
		\xi_{j,k}= u_{j,k}\varepsilon_{j,k}-\log_{2}(u_{j,k}),
\end{equation}
and defining its Weighted MMSE (WMMSE) by minimizing $\xi_{j,k}$ over $u_{j,k}$ and $g_{j,k}$ as
\begin{equation}
\label{eq: WMMSE}
\xi_{j,k}^{\textrm{MMSE}}= \min_{u_{j,k},g_{j,k}}\xi_{j,k},
\end{equation}
we then establish the rate-WMMSE relationship, which is given by
\begin{equation}
\label{eq: rate-WMMSE}
\xi_{j,k}^{\textrm{MMSE}}=1-R_{j,k}.
\end{equation}
The rate-WMMSE relation in (\ref{eq: rate-WMMSE})  is obtained as follows. The optimum equalizer is calculated as $g_{j,k}^*=g_{j,k}^{\textrm{MMSE}}$ from $\frac{\partial\xi_{j,k}}{\partial g_{j,k}}=0$. Substituting $g_{j,k}^{\textrm{MMSE}}$ back to (\ref{eq: WMSE}) yields 
$\xi_{j,k}(g_{j,k}^{\textrm{MMSE}})=u_{j,k}\varepsilon_{j,k}^{\textrm{MMSE}}-\log_{2}(u_{j,k})$. By solving $\frac{\partial\xi_{j,k}(g_{j,k}^{\textrm{MMSE}})}{\partial g_{j,k}}=0$, we then obtain the optimal MMSE weight, which is given as 
\begin{equation}
	u_{j,k}^*=u_{j,k}^{\textrm{MMSE}}=(\varepsilon_{j,k}^{\textrm{MMSE}})^{-1}.
\end{equation}
Substituting $u_{j,k}^{\textrm{MMSE}}$ back to $\xi_{j,k}(g_{j,k}^{\textrm{MMSE}})$, we have $\min_{u_{j,k},g_{j,k}}\xi_{j,k}=1-R_{j,k}$. Following (\ref{eq: WMMSE}), we obtain (\ref{eq: rate-WMMSE}).

Motivated by the rate-WMMSE in (\ref{eq: rate-WMMSE}), we find that the achievable rate of user-$k$ in (\ref{eq: min rate K MIMONOMA}) is equal to
$R_k=1-\xi_{k}^{\textrm{MMSE}}$, where $\xi_{k}^{\textrm{MMSE}}=\max_{j\leq k, j\in\mathcal{K}_i}\xi_{j,k}^{\textrm{MMSE}}$.
By defining  the WMSE of user-$k$ as
\begin{equation}
\xi_{k}=\max_{j\leq k, j\in\mathcal{K}_i}\xi_{j,k},
\end{equation}
 and the respective set of equalizers and weights as $\mathbf{g}=\{g_{j,k}\mid j\leq k, k,j\in\mathcal{K}_i, i\in\mathcal{G}\}$, $\mathbf{u}=\{u_{j,k}\mid j\leq k, k,j\in\mathcal{K}_i, i\in\mathcal{G}\}$, 
the sum-rate WMMSE problem is formulated as
\begin{subequations}
	\label{eq: SR prob trasnformed}
	\begin{align}
	\min_{\mathbf{{P}}, \mathbf{{u}}, \mathbf{{g}}}\quad &\sum_{k\in\mathcal{K}}\xi_{k}\\
	\mbox{s.t.}\quad
	&	\text{tr}(\mathbf{P}\mathbf{P}^{H})\leq P.
	\end{align}
\end{subequations}
Following the proof of \cite{Joudeh:2016a}, we find that the MMSE solutions of the equalizers  $\mathbf{g}^{\textrm{MMSE}}=\{g_{j,k}^{\textrm{MMSE}}\mid j\leq k, k,j\in\mathcal{K}_i, i\in\mathcal{G}\}$ and weights $\mathbf{u}^{\textrm{MMSE}}=\{u_{j,k}^{\textrm{MMSE}}\mid j\leq k, k,j\in\mathcal{K}_i, i\in\mathcal{G}\}$  satisfy the KKT optimality conditions of (\ref{eq: SR prob trasnformed}).  Substituting ($\mathbf{g}^{\textrm{MMSE}},\mathbf{u}^{\textrm{MMSE}}$) back to (\ref{eq: SR prob trasnformed}) with affine transformations applied to the objective function, (\ref{eq: SR prob trasnformed}) boils down to (\ref{eq: SR prob}). In fact, for any point ($\mathbf{{P}}^*, \mathbf{{u}}^*, \mathbf{{g}}^*$) satisfying the KKT optimality conditions of (\ref{eq: SR prob trasnformed}), the solution $\mathbf{{P}}^*$ satisfies the KKT optimality conditions  of (\ref{eq: SR prob}). Hence, (\ref{eq: SR prob trasnformed}) yields a solution for (\ref{eq: SR prob}).

Although the transformed problem (\ref{eq: SR prob trasnformed})  is still non-convex, it is  block-wise convex with respect to $\mathbf{{P}}$ and ($\mathbf{{g}}, \mathbf{{u}}$).
For a given $\mathbf{{P}}$, the optimal solution of the weights and equalizers are $\mathbf{g}^{\textrm{MMSE}}(\mathbf{{P}}), \mathbf{u}^{\textrm{MMSE}}(\mathbf{{P}})$. When ($\mathbf{{g}}, \mathbf{{u}}$) are fixed,   problem (\ref{eq: SR prob trasnformed}) becomes convex and can be solved by interior-point methods. 
Motivated by the block-wise convexity, we use the Alternating Optimization (AO) algorithm as illustrated in Algorithm \ref{AO algorithm} to solve  (\ref{eq: SR prob trasnformed}). In each iteration, the equalizers and weights are first updated by ($\mathbf{g}^{\textrm{MMSE}}(\mathbf{{P}}), \mathbf{u}^{\textrm{MMSE}}(\mathbf{{P}})$) for a given $\mathbf{P}$. The updated equalizers and weights ($\mathbf{g}^{\textrm{MMSE}}(\mathbf{{P}}), \mathbf{u}^{\textrm{MMSE}}(\mathbf{{P}})$)  are substituted back to (\ref{eq: SR prob trasnformed}). Precoder $\mathbf{{P}}$ is then updated by solving (\ref{eq: SR prob trasnformed}). $\mathbf{{P}}$ and ($\mathbf{{g}}, \mathbf{{u}}$) are updated in an alternating manner until the convergence of the sum-rate.
Algorithm \ref{AO algorithm} is guaranteed to converge and it converges to the KKT solution of problem  (\ref{eq: SR prob}). Readers are referred to \cite{Joudeh:2016a} for the proof.

\begin{algorithm}[t!]
	\textbf{Initialize}: $t\leftarrow0$, $\mathbf{P}$\;
	\Repeat{convergence}{
		$t\leftarrow t+1$,	$\mathbf{P}^{[t-1]}\leftarrow \mathbf{P}$\;
		$\mathbf{g}\leftarrow\mathbf{g}^{\mathrm{MMSE}}(\mathbf{P}^{[t-1]})$; $\mathbf{u}\leftarrow\mathbf{u}^{\mathrm{MMSE}}(\mathbf{P}^{[t-1]})$\;
		Substitute $(\mathbf{g},\mathbf{u})$ back to (\ref{eq: SR prob trasnformed})  and 
		update $\mathbf{P}$ by solving (\ref{eq: SR prob trasnformed});	
	}	
	\caption{AO algorithm}
	\label{AO algorithm}				
\end{algorithm}

Following the same procedure, we are able to obtain the transformed WMMSE problem for max-min rate maximization, which is given by
 \begin{subequations}
 	\label{eq: maxmin prob trasnformed}
 	\begin{align}
 	\min_{\mathbf{{P}},\mathbf{{u}}, \mathbf{{g}}}\quad &\max_{k\in\mathcal{K}}\,\xi_{k}\\
 	\mbox{s.t.}\quad
 	&	\text{tr}(\mathbf{P}\mathbf{P}^{H})\leq P.
 	\end{align}
 \end{subequations}
By substituting problem (\ref{eq: SR prob trasnformed})  in Algorithm \ref{AO algorithm} with problem (\ref{eq: maxmin prob trasnformed}), we obtain the corresponding AO Algorithm to achieve the  KKT solution of the max-min rate problem (\ref{eq: maxmin prob}).


\end{document}